\documentclass[acmsmall]{acmart}
\usepackage[utf8]{inputenc}
\usepackage{newunicodechar}
\newunicodechar{♫}{\textmusicalnote}
\usepackage[ruled,vlined]{algorithm2e}
\usepackage{makecell}
\usepackage{times}
\usepackage{latexsym}
\usepackage{multirow}
\usepackage{amsmath}
\usepackage{booktabs}
\usepackage{xcolor}
\usepackage{tabularx}
\usepackage{graphicx}
\usepackage{pifont}
\usepackage{caption}
\usepackage{enumitem}
\usepackage{colortbl}
\usepackage[commandnameprefix=always]{changes}
\usepackage[T1]{fontenc}

\usepackage[utf8]{inputenc}

\usepackage{microtype}

\usepackage{listings}

\usepackage{algorithmic}
\usepackage{subfigure}
\usepackage{tcolorbox}
\usepackage{array}
\usepackage{lipsum} 
\usepackage{arydshln}
\usepackage[table]{xcolor}
\usepackage{xspace}
\usepackage{float}
\newcommand{\method}{TeaRAG\xspace}

\usepackage{ulem}
\definecolor{lightgray}{gray}{0.9}
\usepackage{indentfirst}
\AtBeginDocument{%
  }

\begin{document}

\setcopyright{cc}
\setcctype{by}
\acmJournal{TOIS}
\acmYear{2026} \acmVolume{1} \acmNumber{1} \acmArticle{}
\acmMonth{1} \acmDOI{10.1145/3818621}

\title{\method: A Token-Efficient Agentic Retrieval-Augmented Generation Framework}

\author{Chao Zhang}
\email{zclfe00@mail.ustc.edu.cn}
\affiliation{%
  \institution{University of Science and Technology of China}
  \country{China}
}
\affiliation{%
  \institution{ City University of Hong Kong}
  \country{China}}

\author{Yuhao Wang}
\affiliation{%
  \institution{ City University of Hong Kong}
  \country{China}}

\email{yhwang25-c@my.cityu.edu.hk}

\author{Derong Xu}
\email{derongxu@mail.ustc.edu.cn}
\affiliation{%
  \institution{University of Science and Technology of China}
  \country{China}
}
\affiliation{%
  \institution{ City University of Hong Kong}
  \country{China}
}

\author{Haoxin Zhang}
\affiliation{%
  \institution{Xiaohongshu Inc.}
  \country{China}}
\email{zhanghaoxin1994@gmail.com}

\author{Yuanjie Lyu}
\email{s1583050085@gmail.com}
\author{Yuhao Chen}
\email{isyuhaochen@mail.ustc.edu.cn}
\author{Shuochen Liu}
\email{shuochenliu@mail.ustc.edu.cn}
\author{Tong Xu}
\authornote{Corresponding authors.}
\email{tongxu@ustc.edu.cn}
\affiliation{%
  \institution{University of Science and Technology of China}
  \country{China}
}

\author{Xiangyu Zhao}
\authornotemark[1]
\email{xianzhao@cityu.edu.hk}
\affiliation{%
  \institution{ City University of Hong Kong}
  \country{China}
  }

\author{Yan Gao}
\email{yadun@xiaohongshu.com}
\author{Yao Hu}
\email{yaoohu@gmail.com}
\affiliation{%
  \institution{Xiaohongshu Inc.}
  \country{China}}

\author{Enhong Chen}
\email{cheneh@ustc.edu.cn}
\affiliation{%
  \institution{University of Science and Technology of China}
  \country{China}
}

\renewcommand{\shortauthors}{Zhang et al.}

\begin{abstract}
Retrieval-Augmented Generation (RAG) utilizes external knowledge to augment Large Language Models' (LLMs) reliability.
For flexibility, agentic RAG employs autonomous, multi-round retrieval and reasoning to resolve queries.
Although recent agentic RAG has improved via reinforcement learning, they often incur substantial token overhead from search and reasoning. This trade-off prioritizes accuracy over efficiency.
To address this issue, this work proposes \method, a \textbf{T}oken-\textbf{e}fficient \textbf{a}gentic \textbf{RAG} framework capable of compressing both retrieval content and reasoning steps. 
1) First, the retrieved content is compressed by augmenting chunk-based semantic retrieval with a graph retrieval using concise triplets. 
A knowledge association graph is then built from semantic similarity and co-occurrence. 
Finally, Personalized PageRank is leveraged to highlight key knowledge within this graph, reducing the number of tokens per retrieval.
2) Besides, to reduce reasoning steps, Iterative Process-aware Direct Preference Optimization (IP-DPO) is proposed. Specifically, our reward function evaluates the knowledge sufficiency by a knowledge matching mechanism, while penalizing excessive reasoning steps. 
This design can produce high-quality preference-pair datasets, supporting iterative DPO to improve reasoning conciseness.
Across six datasets, \method improves the average Exact Match by $4\%$ and $2\%$ while reducing output tokens by $61\%$ and $59\%$ on Llama3‑8B‑Instruct and Qwen2.5‑14B‑Instruct, respectively.
Code is available at \url{https://github.com/Applied-Machine-Learning-Lab/TeaRAG}.
\end{abstract}



\maketitle

\section{Introduction}

\label{intro}
Large Language Models (LLMs) have made substantial progress through learning from massive pre‑training corpora. 
Nevertheless, statistical inaccuracies in linguistic distribution modeling during pre‑training can lead to hallucination generation~\cite{kalai2025language}.
Retrieval-Augmented Generation (RAG) is an effective technique that mitigates hallucinations in LLMs by incorporating external retrieval information~\cite{gao2023retrieval,lyu2025crud}.
To improve the accuracy of RAG, conventional RAG systems typically adopt a predefined multi-step workflow consisting of planning~\cite{verma2024plan}, query rewriting~\cite{ma2023query,zhang2026personalize}, retrieval~\cite{lyu2024retrieve}, reranking~\cite{wang2025richrag}, refinement~\cite{jiang2023llmlingua}, and generation.
However, the fixed workflows constrain their capacity to address tasks that demand multi-step reasoning, adaptive decision-making, and dynamic integration of retrieved information~\cite{li2025search,lyu2024retrieve}.
To further enhance the flexibility and adaptiveness of the RAG framework, agentic RAG has recently been proposed.
Agentic RAG adopts agentic workflows that allow LLMs to autonomously control the workflow process to solve complex problems~\cite{lyu2024retrieve,song2025r1}. 
By leveraging the capabilities of LLMs in task decomposition and dynamic planning, agentic RAG can break down problems into a sequence of tractable steps. 
At each step, the framework proactively invokes retrieval modules to obtain relevant contextual information, thereby establishing an integration between reasoning and retrieval~\cite{trivedi2023interleaving,jin2025search}.

Recently, the performance of agentic RAG has been significantly enhanced through reinforcement learning (RL) optimization of agentic workflows for LLMs~\cite{song2025r1,sun2025zerosearch,shi2025search}. 
For example, Search-R1~\cite{jin2025search} employs PPO~\cite{schulman2017proximal} and GRPO~\cite{shao2024deepseekmath} to optimize agentic RAG based on an output-based reward.
However, the current paradigm still faces several challenges.
First, existing agentic RAG methods are token-inefficient.
This is because current systems predominantly prioritize maximizing the accuracy of the final outputs, while overlooking the substantial token overhead during the reasoning and retrieval processes~\cite{jin2025search,shi2025search}.
This optimization bias often leads models to overthink~\cite{cuadron2025danger,chen2024not} and perform redundant retrieval~\cite{wang2025stepsearch}, resulting in wasted computational resources and reduced system efficiency.
Second, current agentic RAG systems largely rely on chunk‑based semantic retrieval, which yields low information density and tends to introduce irrelevant noise~\cite{lyu2024retrieve,yu2024rankrag}.
While recent studies~\cite{yu2025graphrag,luo2025graph} have investigated the use of higher‑density knowledge graphs for retrieval, it remains unvalidated on large‑scale graphs and fails to leverage co‑occurrence relationships among chunks and knowledge triplets.
This prevents the system from leveraging both retrieval methods' strengths~\cite{gutierrez2025rag} and hinders retrieval systems in preserving critical information and removing irrelevant noise.
Finally, most agentic RAG systems train LLMs with outcome-based rewards via RL~\cite{song2025r1,jin2025search}. This reward design is typically sparse and noisy, which can hinder stable and efficient training~\cite{wang2025stepsearch}.
Besides, agentic RAG methods~\cite{wang2025stepsearch,shi2025search} are usually trained using RL algorithms such as PPO~\cite{schulman2017proximal} and GRPO~\cite{shao2024deepseekmath}. These approaches require multi-machine coordination and continuous external retrieval access during training~\cite{sun2025zerosearch}, which makes the training process prohibitively slow.
To address this limitation, recent approaches employ GPT‑4o for process‑based reward annotation~\cite{zhang2025process}, and then optimize using DPO~\cite{rafailov2023direct}. 
However, the reward annotation process is costly and depends on proprietary models.

\begin{figure}[!t]
    \centering
    \includegraphics[width=\textwidth]{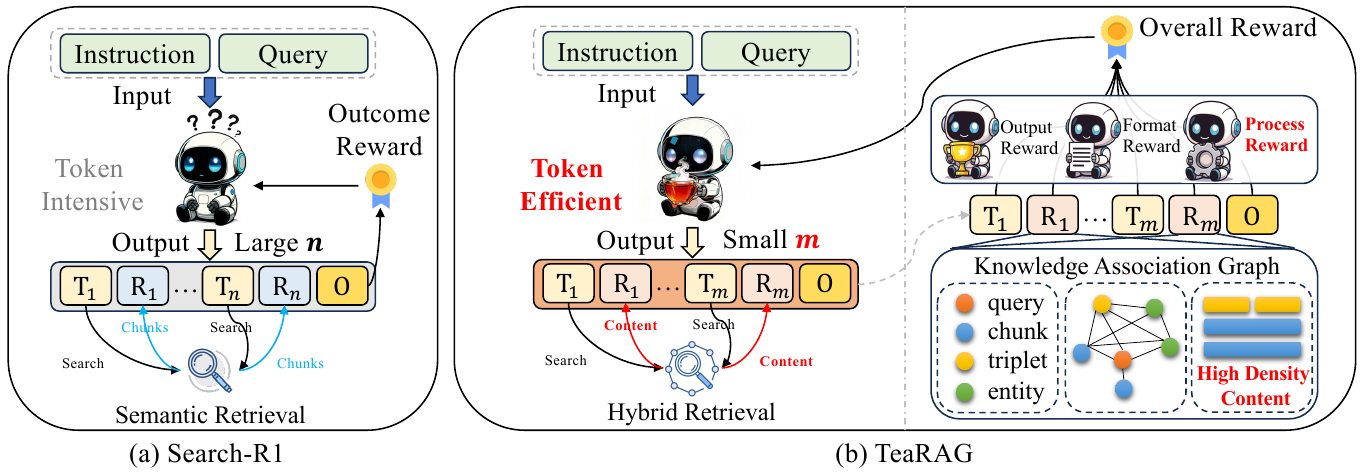}
    \caption{$T_i$ denotes the thinking tokens at the i-th step, $R_i$ denotes the retrieved context at the i-th step, and $O$ represents the final output. (a) illustrates Search-R1, a representative agentic RAG method optimized based on the final outcome. (b) shows our proposed method \method, which achieves a token-efficient agentic RAG by optimizing the retrieved content length with high-density triplets and controlling the number of LLM reasoning steps via a process-aware reward.}
    \label{fig:demo}
\end{figure}

\begin{figure*}[!t]
    \centering
    \includegraphics[width=\textwidth]{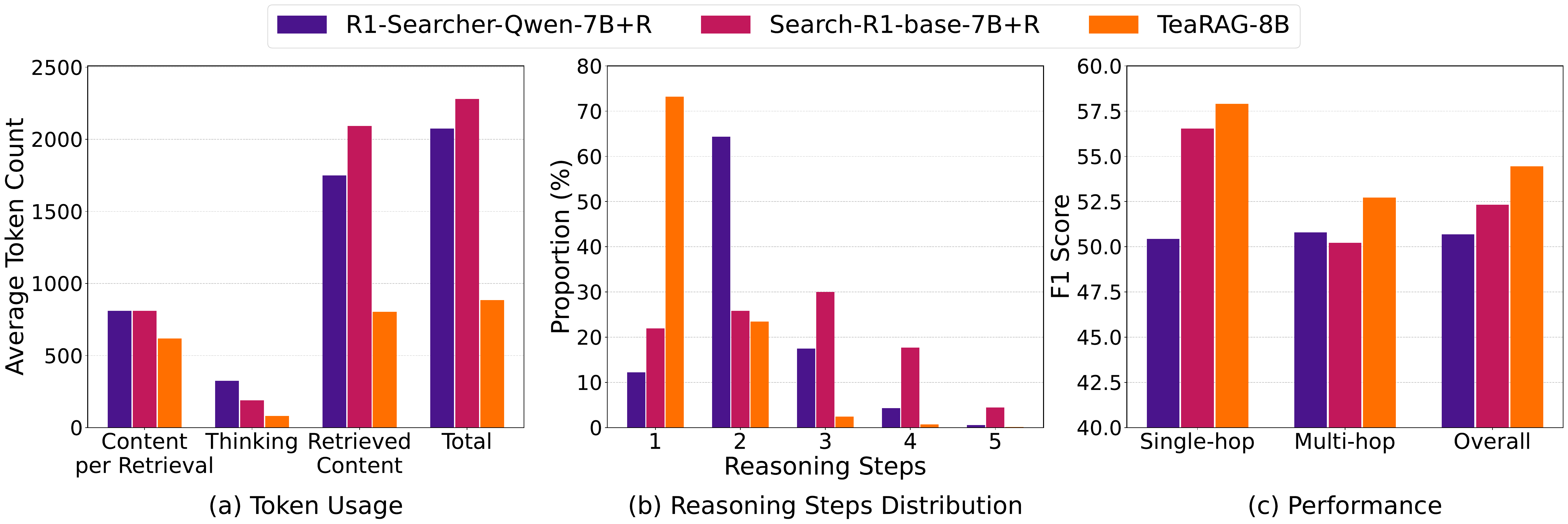}
    \caption{(a) shows the token usage. (b) shows the distribution of reasoning steps. (c) shows the F1 performance on single-hop, multi-hop, and overall QA benchmarks.}
    \label{fig:demo_exp}
\end{figure*}

To gain deeper insights into the token usage of existing agentic RAG systems, we analyze two representative methods: Search-R1~\cite{jin2025search} and R1-Searcher~\cite{song2025r1}.
As illustrated in Fig.~\ref{fig:demo} (a), the output of an agentic RAG system primarily consists of two types of tokens generated through multiple iterations.
The first is the LLM's thinking process, where these tokens are used for planning, problem decomposition, and reasoning over retrieved content. 
The second is the retrieved content that the LLM obtains by invoking a retriever to access external sources.
Through statistical analysis of different token types and iteration rounds, as observed in Fig.~\ref{fig:demo_exp} (a) and Fig.~\ref{fig:demo_exp} (b), we identify two critical inefficiencies that impact token utilization.
First,  the retrieved content constitutes the majority of the overall output.
This is because chunk-based retrieval methods typically return entire document segments as input to LLMs.
However, a substantial portion of these contents may consist of irrelevant or redundant background information that does not enhance the quality of the final answer~\cite{lyu2024retrieve}. Therefore, increasing the information density of the retrieved content is essential for improving the token efficiency in agentic RAG.
Second, agentic RAG methods generally adopt multi‑step reasoning, even when addressing single‑hop questions. As shown in Fig.~\ref{fig:demo_exp} (b), single‑hop questions account for $44\%$ of the test set, yet in most cases the number of reasoning steps exceeds one.
This is due to the lack of process supervision in outcome-based rewards, which leads to overthinking~\cite{chen2024not} and redundant retrieval~\cite{wang2025acting}.
Therefore, reducing unnecessary and repetitive steps in intermediate processes is key to improving the efficiency of token utilization.

To address these challenges, we introduce \method, a token-efficient agentic RAG framework that enhances token efficiency by simultaneously optimizing the conciseness of reasoning steps and the density of retrieved content, as shown in Fig.~\ref{fig:demo} (b).
Specifically, the workflow of this agentic RAG system is autonomously controlled by an LLM. 
First, the LLM identifies key entities in the question and breaks it down into sub-questions around these key entities. Subsequently, based on the sub-questions, the LLM calls a retriever to perform semantic retrieval at the chunk level.
To improve the density of the retrieved content, the LLM also uses graph retrieval to retrieve relevant and concise knowledge triplets.
Afterwards, based on the recalled chunks and triplet information, we construct a Knowledge Association Graph (KAG) using semantic similarity and co-occurrence. 
Using Personalized PageRank (PPR) on this graph, we filter out redundant and irrelevant chunk information while supplementing it with the higher-density triplet information. Finally, the LLM summarizes the retrieved content.
The LLM repeats this process until the final answer is obtained.

To effectively implement this pipeline and improve the conciseness of LLM reasoning, we propose a two‑stage training paradigm.
In the first stage, we construct supervised fine-tuning (SFT) data to train models to master the reasoning format and thinking process.
Specifically, we leverage the query decomposition processes from the MuSiQue~\cite{trivedi2022musique} dataset and employ Qwen2.5-72B-Instruct~\cite{yang2024qwen2} to transform structured question-answer pairs into natural language equivalents.
We then assemble these into complete reasoning processes following chain combinations.
Using this SFT dataset, we perform SFT on the models. In the second stage, we propose Iterative Process-aware Direct Preference Optimization (IP-DPO), which incorporates a novel process reward and iteratively optimizes LLMs. 
Specifically, we sample multiple reasoning paths for each query, assign rewards to them, and construct preference-pair datasets based on these rewards. For a given sampled reasoning path, the reward system integrates conventional output-based and format-based rewards with the proposed process reward to produce a comprehensive evaluation score.
The process reward employs a knowledge matching mechanism to assess evidence acquisition across three dimensions: subquery generation, context retrieval, and summarization. This mechanism measures the alignment of intermediate reasoning outputs in each dimension with ground truth knowledge.
Then, the process reward is computed by aggregating evidence acquisition scores from each dimension and normalizing by the total reasoning steps, resulting in the information gain per step.
Besides, the process reward also ensures consistency between extracted entities and subqueries to facilitate the PPR filtering.
Based on this reward framework, we can create high‑quality preference‑pair datasets, enabling iterative DPO to improve model generalization and yield more compact reasoning paths.
Comprehensive evaluations on six benchmark datasets demonstrate that \method attains superior accuracy while reducing both reasoning steps and output length as shown in Fig.~\ref{fig:demo_exp}. 
On Llama3‑8B‑Instruct, it improves the average Exact Match (EM) score by $4\%$ alongside a $61\%$ reduction in output tokens, whereas on Qwen2.5‑14B‑Instruct it yields a $2\%$ EM increase with a $59\%$ token reduction.

Our contributions can be summarized as follows:
\begin{itemize}[leftmargin=*, topsep=0pt, itemsep=0pt]
\item We conduct an in-depth analysis of token inefficiencies in agentic RAG and propose a token-efficient pipeline, \method, that simultaneously improves the information density of retrieved contents and reduces reasoning steps.
\item We propose a retrieval method that constructs KAGs via semantic similarity and co-occurrence, combining semantic and graph retrieval strengths. By employing PPR filtering to remove irrelevant contents, our approach retrieves more concise contents without sacrificing performance.
\item We propose a two-stage training paradigm that effectively activates agentic RAG reasoning capabilities while maintaining reasoning conciseness through process-aware training.
\item We conduct comprehensive experiments and detailed analyses across six benchmark datasets, validating the effectiveness and token efficiency of our pipeline and training methods.
\end{itemize}

\section{Related Work}
\subsection{Retrieval-Augmented Generation} 
LLMs are trained on massive datasets, possessing powerful capabilities in understanding~\cite{gao2023retrieval}, generation~\cite{yang2024qwen2}, and reasoning~\cite{shao2024deepseekmath,li2025search}. However, when faced with knowledge-intensive tasks~\cite{lewis2020retrieval}, LLMs that rely solely on their parametric memory are prone to hallucination~\cite{gao2023retrieval,jin2025search,webglm}. 
RAG addresses this by using a retriever to search a corpus for information snippets relevant to a query. These information are then provided to the LLM as evidence to generate more reliable answers. 
The field of RAG has attracted significant attention in recent years, spurring substantial research.
We will focus on introducing agentic RAG and graph-enhanced RAG, which are most relevant to our work.

\subsubsection{Agentic RAG} 
A comprehensive RAG system typically involves multiple sub-steps, including query planning~\cite{verma2024plan}, query rewriting~\cite{ma2023query,Agent4Ranking}, retrieval~\cite{lyu2024retrieve}, reranking~\cite{wang2025richrag}, refinement~\cite{jiang2023llmlingua}, and generation.
Traditional RAG systems usually follow pre-defined sub-steps that are manually designed by humans.
However, this approach fails to dynamically adapt to task requirements and performs poorly when facing complex multi-hop tasks~\cite{gutierrez2024hipporag,liu2026look,zhangevoking}.
To autonomously control the RAG process, agentic RAG applies the agent approach~\cite{wang2024survey,yao2023react,liu2026perma}.  
Through the planning and comprehension capabilities of LLMs, it iteratively breaks down complex problems into executable operations and queries until an answer is found.

Initial works use prompt-based approaches for preset process control~\cite{li2025search,trivedi2023interleaving,yueinference}, determining whether RAG should continue querying or provide an answer to the question.
However, due to the lack of training specifically for RAG scenarios, these methods rely on LLMs' instruction-following and reasoning capabilities.
To enhance the adaptability of LLMs to RAG tasks, some studies~\cite{lyu2024retrieve,asai2023self} employ SFT to train agentic RAG systems. However, such approaches often fall short in further stimulating the generalization capability of the models.
Recently, as RL methods have greatly enhanced the reasoning and generalization capabilities of LLMs~\cite{shao2024deepseekmath}, numerous studies have employed outcome-based RL to stimulate LLMs' thinking and tool-use abilities~\cite{jin2025search,song2025r1}.
These works typically use string matching methods to simply determine whether the LLM's output is consistent with the ground truth~\cite{jin2025search,song2025r1,jiang2025s3} or whether the LLM's reasoning process contains the ground truth~\cite{shi2025search}.
Nevertheless, such approaches often suffer from suboptimal reasoning processes due to the lack of supervision over intermediate steps~\cite{wang2025stepsearch}. 
Some recent work has explored process-reward methods to more precisely optimize agentic RAG systems~\cite{wang2025stepsearch,zhang2025process}. These approaches include employing predefined reasoning format rewards to supervise the model in learning correct reasoning formats~\cite{song2025r1plus,jin2025empirical} and designing reward mechanisms based on the number of retrieval calls~\cite{song2025r1plus,yu2025graphrag}. 
Further improvements involve leveraging more comprehensive evidence paragraphs to supervise the reasoning process~\cite{wang2025stepsearch}, or utilizing GPT-4o to annotate intermediate reasoning steps~\cite{zhang2025process}.

Although current agentic RAG has made significant progress, existing methods focus on training LLMs on how to invoke retrievers for better performance, often neglecting token efficiency.
Besides, these approaches either rely on high-resource RL methods such as PPO~\cite{schulman2017proximal} and GRPO~\cite{shao2024deepseekmath}, which require multi-machine collaboration and continuous retrieval resource requests during training~\cite{wang2025stepsearch}, or utilize GPT-4o for process scoring~\cite{zhang2025process}. 
Both approaches make it difficult to achieve efficient, low-resource implementation of agentic RAG.
Therefore, our work aims to improve the token efficiency of agentic RAG by reducing token usage while maintaining performance. Additionally, we propose an iterative process-aware DPO to implement agentic RAG.

\subsubsection{Graph-Enhanced RAG}
Most traditional RAG relies on chunk-based retrieval approaches~\cite{edge2024local}, which split complete document information into multiple chunks and use these chunks as the units for retrieval and input to LLMs. 
However, since question-related clues are embedded in complex chunk contexts, existing chunk-based approaches struggle to accurately capture subtle clues~\cite{lyu2024retrieve,guo2024lightrag} and multi-hop knowledge associations~\cite{gutierrez2024hipporag,gutierrez2025rag}.
Previous work organizes corpus knowledge into knowledge graph structures to capture fine-grained clues and knowledge associations more systematically~\cite{edge2024local,guo2024lightrag,xu2024multi,xu2022relation}.
Leveraging the topological structure of knowledge graphs, many works replace chunk-based retrieval methods with graph algorithms to achieve more accurate results~\cite{gutierrez2024hipporag,gutierrez2025rag,xu2025align,chenplan,fang2025kirag}.
These graph-enhanced RAG approaches can be categorized into three strategies. 
First, some methods employ classical graph algorithms such as Prize-Collecting Steiner Tree~\cite{he2024g} and PPR~\cite{gutierrez2024hipporag,gutierrez2025rag,alonso2024mixture} to retrieve all query-related information in one step, then feed the entire retrieved context to the LLM for direct answer generation~\cite{xu2025align,zhu-etal-2025-knowledge}.
Second, some approaches employ LLM prompts to select the optimal triplet path from all candidate reasoning paths based on topological relationships in the graph structure~\cite{chenplan,fang2025kirag,sunthink}.
Last, recent concurrent work has explored extending RL-based training approaches from agentic RAG to graph-enhanced RAG, aiming to fully leverage the reasoning capabilities of LLMs~\cite{luo2025graph,yu2025graphrag}.
Nevertheless, a key challenge in knowledge graph construction is the potential for information loss~\cite{xu2024large}, as the process simplifies complete text into structured triplets. To address this issue, some works dynamically construct knowledge graphs after receiving specific queries~\cite{jiang2025ras,xu2023recomp}, which enables the targeted capture of core information relevant to the query from the text. Nevertheless, these approaches incur additional online inference time and high computational costs. Alternatively, some methods adopt a hybrid approach that combines chunk-based semantic retrieval with graph-based retrieval mechanisms to leverage the advantages of both strategies~\cite{yu2025graphrag,gutierrez2025rag,sarmah2024hybridrag,ma2024think,liang2025kag}.

Current graph-enhanced RAG still has its limitations. First, these methods typically utilize existing small-scale knowledge graphs
~\cite{xu2025align,he2024g} or construct them from relatively small (sub-million scale) corpora~\cite{gutierrez2024hipporag,gutierrez2025rag,luo2025graph}, lacking validation on sufficiently large-scale graph data. Furthermore, graph-enhanced RAG systems typically select a fixed number of top-ranked triplets~\cite{xu2025align,fang2025kirag,chenplan} or their associated text chunks~\cite{gutierrez2024hipporag,gutierrez2025rag}, or directly concatenate the retrieved triplets and chunks as input~\cite{yu2025graphrag,sarmah2024hybridrag}. However, these approaches fail to fully exploit the co-occurrence of different representations of the same knowledge element. In our work, we conduct experimental validation by constructing a large-scale knowledge graph based on a common wiki corpus~\cite{karpukhin2020dense}. We use PPR on a KAG of retrieved chunks and knowledge triplets to select the most relevant information. This process boosts information density while preserving key context, achieving a token-efficient agentic RAG.

\subsection{Reinforcement Learning for LLMs} 

LLMs gain knowledge through pre-training on vast data~\cite{brown2020language,yang2024qwen2} and are then instruction-tuned via SFT to follow instructions~\cite{grattafiori2024llama,ouyang2022training,chen2025xiangqi}. However, this imitation-based learning limits their ability to generalize.
To address these limitations, Reinforcement Learning from Human Feedback~\cite{ouyang2022training} (RLHF) first trains a reward model to capture human preferences over model outputs. 
This reward model then guides the LLM optimization using PPO~\cite{schulman2017proximal}, enabling the generation of higher-quality, more aligned responses.
Beyond alignment with human values, RL techniques have proven effective in enhancing LLM reasoning capabilities~\cite{shao2024deepseekmath,jaech2024openai,lyu2025correction}. For example, methods like GRPO~\cite{shao2024deepseekmath} improve performance on mathematical reasoning tasks by optimizing the model's generation process using either outcome-based rewards~\cite{shao2024deepseekmath,yu2025dapo} or fine-grained process reward models that provide step-by-step supervision~\cite{zeng2024scaling}.
However, the above methods are typically online RL approaches, requiring LLMs to perform sampling simultaneously during training~\cite{shao2024deepseekmath,jin2025search}, which often leads to low training efficiency when LLM inference is lengthy or requires calling complex tools~\cite{rafailov2023direct,gao2025beyond}. By separating the sampling and training phases, easier and more efficient RL methods achieve higher training efficiency, sometimes at the expense of performance~\cite{ivison2024unpacking}. These include rejection sampling~\cite{yuan2023scaling,khaki2024rs,xiong2025minimalist}, which enhance model performance by filtering high-quality samples from generated data for subsequent SFT. DPO~\cite{rafailov2023direct,khaki2024rs} circumvents explicit reward model training by treating the LLM itself as an implicit reward model, enabling RL through preference pair construction. This approach has been successfully applied to reasoning tasks through iterative sampling and training cycles~\cite{pang2024iterative,tu2025enhancing}.
Building on this, our work utilizes RL algorithms to optimize the multi-turn retrieval and reasoning capabilities of agentic RAG systems. We design novel process-based rewards that employ knowledge matching to evaluate key components throughout the LLM's reasoning paths. These rewards effectively distinguish between positive and negative samples, and we employ multi-round iterative DPO to train LLMs for adaptation to agentic RAG tasks efficiently.

\section{Preliminaries}
Given a task instruction $I$, a question $q$, a retriever $\mathcal{R}$, a collection of document chunks $\mathcal{D} = \{d_1, \ldots, d_n\}$, and a knowledge graph $\mathcal G = \langle \mathcal V, \mathcal{E} \rangle$ where $\mathcal V = \{v_1, \ldots, v_m\}$ is the entity set and $\mathcal{E} = \left\{\, e_i\,|\, e_i = (v_i^{h}, rel, v_i^{t}),\, v_i^{h}, v_i^{t} \in \mathcal{V} \right\}$ is the triplet set, RAG utilizes the retriever $\mathcal{R}$ to retrieve relevant information based on question $q$ and inputs it into an LLM for reasoning and answering.

\begin{definition}
\textbf{Single-Round RAG.} This paradigm involves a single retrieval step for a given query $q$. A retrieval algorithm $\mathcal{A}_{\mathcal{R}}$, using a retriever $\mathcal{R}$, gathers information from a document corpus $\mathcal{D}$ and a knowledge graph $\mathcal{G}$. The resulting context, $\mathcal{C}_q = \mathcal{A}_{\mathcal{R}}(q, \mathcal{D}, \mathcal{G})$, is the union of retrieved document chunks ($\mathcal{D}_q \subseteq \mathcal{D}$) and knowledge graph triplets ($\mathcal{E}_q \subseteq \mathcal{G}$). This aggregated context is then combined with the instruction $I$ and query $q$ to form the final prompt for the LLM, which generates the answer $a$:
$a = \text{LLM}(I, q, \mathcal{C}_q)$.
\end{definition}

\begin{definition}
\textbf{Agentic RAG.} 
This paradigm empowers the LLM to act as an autonomous agent that constructs a dynamic workflow to answer complex questions. The agent iteratively builds a reasoning path, $\mathcal{P}_k$, by interleaving thought and retrieved context. Each step $i$ of the process unfolds in two parts. First, in the thinking phase, the LLM generates an internal reasoning step $T_i = \text{LLM}(I, q, \mathcal{P}_{i-1})$ based on the instruction $I$, question $q$, and its previous reasoning steps. Second, in the retrieval phase, a subquery $q_i$ is extracted from the thought $T_i$ and used to retrieve relevant context $\mathcal{C}_{q_i} = \mathcal{A}_\mathcal{R}(q_i, \mathcal{D}, \mathcal{G})$. This cycle continues until a complete reasoning path, $\mathcal{P}_k = [T_1, \mathcal{C}_{q_1}, \ldots, T_k, \mathcal{C}_{q_k}]$, is constructed. Finally, the LLM generates the final answer based on this complete path: $a = \text{LLM}(I, q, \mathcal{P}_k)$.
\end{definition}

This work aims to develop a token‑efficient agentic RAG system that preserves correctness while compressing $\mathcal{P}_k$, thereby enhancing token efficiency.
As analyzed in Section~\ref{intro}, \method achieves this goal via two key perspectives. (1) constructing higher-density contexts $\mathcal{C}_{q_i}$ by shortening the context per retrieval while retaining essential content. (2) reducing the number of reasoning steps $k$ thereby enhances the efficiency of reasoning.

\section{Methodology}

In this section, we present the overall implementation of \method.
First, we describe the knowledge graph construction in Section~\ref{kg_extract}.
Second, based on the existing corpus, we outline the \method\ pipeline in Section~\ref{pipeline}.
Finally, we introduce the training strategies employed to realize this pipeline in Section~\ref{training}, including SFT and IP-DPO.

\subsection{Knowledge Graph Construction}
\label{kg_extract}
Most graph-enhanced RAG~\cite{luo2025graph,yu2025graphrag} methods construct knowledge graphs from relatively small corpora (sub-million scale). 
Such settings cannot reliably evaluate model performance on large-scale data, as larger graphs inevitably introduce more noise and complexity.  
To investigate the performance of \method under a large-scale knowledge graph, we construct a knowledge graph based on the widely used Wikipedia corpus~\cite{karpukhin2020dense}.  
Following~\cite{zhu-etal-2025-knowledge}, we employ the Qwen2.5‑14B‑Instruct model~\cite{yang2024qwen2} to extract knowledge triplets from each document chunk. This ensures that the content of the graph is consistent with the underlying corpus.  
The extracted entities are added to the entity set $\mathcal{V}$, and the extracted triplets are incorporated into the edge set $\mathcal{E}$. All entities and triplets obtained from the corpus together constitute the knowledge graph $\mathcal{G}$.
Statistics for this graph are provided in Table~\ref{tab:kg_stats}.

The complete knowledge graph extraction from the Wikipedia corpus~\cite{karpukhin2020dense} requires approximately $389$ GPU hours on A100 80G GPUs. This one-time preprocessing cost is effectively amortized over the system's deployment lifetime, as the constructed graph is persistently cached and reused across all subsequent queries.
Furthermore, in online deployment settings, triplet extraction is required only for newly ingested documents. Since a single A100 80GB GPU processes approximately $15$ chunks per second, the incremental extraction process is both efficient and cost-effective.

\subsection{Pipeline of \method}
\label{pipeline}
\begin{figure}[!t]
    \centering
    \includegraphics[width=\textwidth]{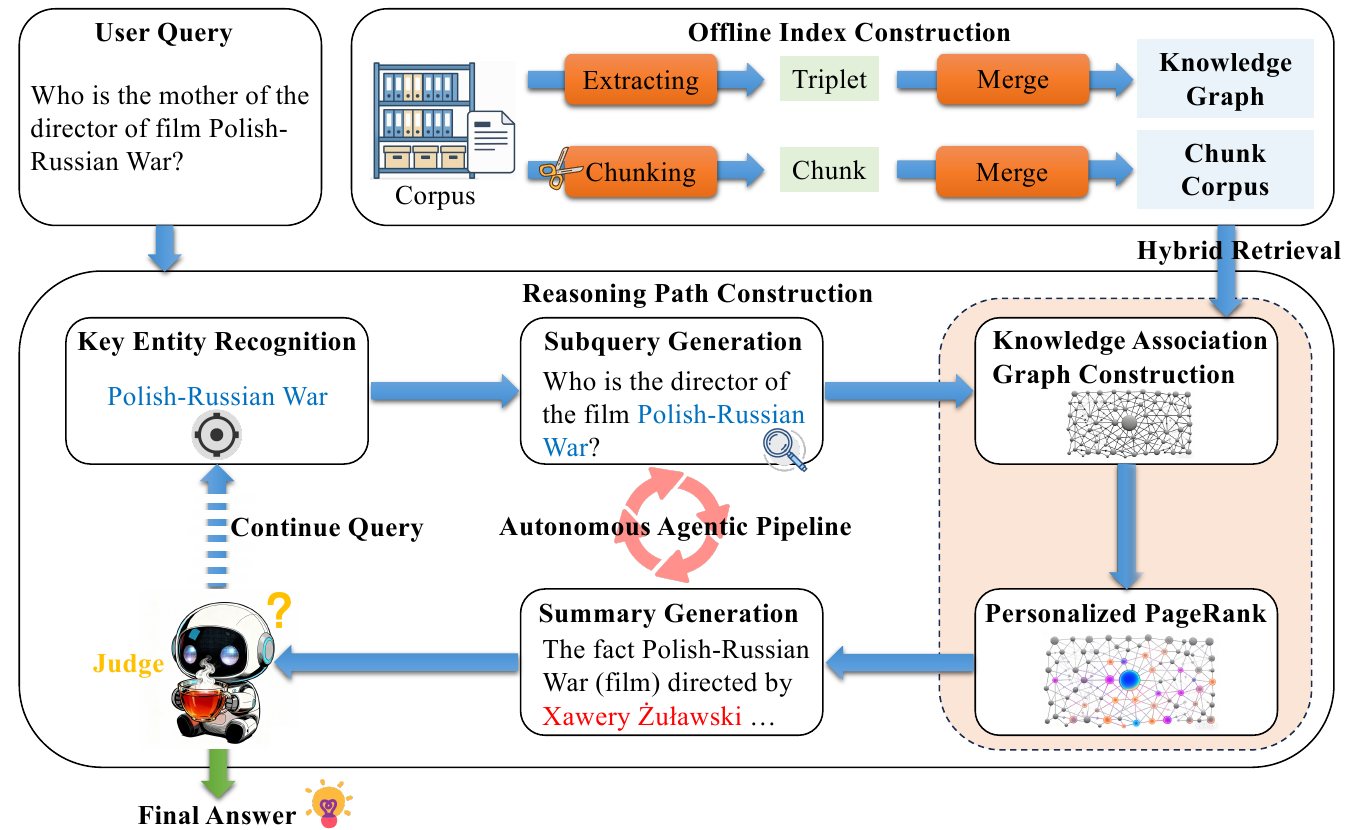}
    \caption{The overall pipeline of \method. Based on an offline-built knowledge graph and chunk corpus index, \method progressively constructs a reasoning path until the final answer is determined.}
    \label{fig:pipeline}
\end{figure}
\label{sec:reasoning_chain_structure}

Leveraging the Wikipedia corpus and its knowledge graph, the system can perform hybrid retrieval to autonomously address problems.
In agentic RAG, conventional retrieval methods are confined to semantic retrieval~\cite{song2025r1,jin2025search}, graph retrieval~\cite{luo2025graph,yu2025graphrag}, or their naive concatenation~\cite{yu2025graphrag,sarmah2024hybridrag,wen2025hybridrag}, overlooking both their complementary strengths and the strong relevance signal from their co-occurrence.
Semantic retrieval offers rich background and conceptual context but suffers from low information density, noise~\cite{lyu2024retrieve,asai2023self,xu2023recomp}, and inefficient token use. Conversely, graph retrieval yields high-density triplets containing precise facts but lacking contextual grounding.
Besides, when a chunk from semantic retrieval and a triplet from graph retrieval correspond to the same source data, their co-occurrence constitutes a robust relevance signal, functioning as a high-confidence filter.
Therefore, \method exploits this intrinsic relationship between chunks and triplets to reconcile the accuracy–efficiency trade-off. By prioritizing co-occurring chunk–triplet pairs over less certain chunks, \method can enhance contextual accuracy while improving token efficiency for the LLM.

\begin{table}[t]
\color{blue}
\centering
\caption{The instruction prompt utilized in \method. \textcolor{red!80!black}{\textless{}Reference\textgreater{}} and \textcolor{red!80!black}{\textless{}/Reference\textgreater{}} are special tokens for wrapping retrieved information.}
\label{tab:instruction}
\vspace{-1em}

\begin{tcolorbox}[
    colback=gray!10!white,   
    colframe=gray!60!black,
    boxrule=0.6pt,
    arc=4pt,
    width=\linewidth,
    left=8pt, right=8pt, top=8pt, bottom=8pt,
    nobeforeafter
]
\ttfamily\small
Follow these iterative steps to answer the overall question:\\
1. Important entity: Identify key entities to investigate for the current step. Multiple entities should be separated by \#\#\#.\\
2. Subquery: Formulate search queries based on important entities.\\
3. Search: Execute the search. All results MUST be placed within \textcolor{red!80!black}{\textless{}Reference\textgreater{}} and \textcolor{red!80!black}{\textless{}/Reference\textgreater{}} tags.\\
4. Summary: Analyze the search results (e.g., based on clues, obtain sub-answers).\\
5. Decide: If the information is sufficient, provide the "Final Answer:". Otherwise, return to step 1.
\end{tcolorbox}

\vspace{-1.5em}
\end{table}

Consequently, we implement the pipeline of \method as shown in Fig.~\ref{fig:pipeline}.  
We redesign the structure of the agent's reasoning path as $\mathcal{P}_k = [S_1, \ldots, S_k]$, where $S_i$ denotes the $i$-th reasoning step. The structure of a reasoning step is shown in Fig.~\ref{fig:KAG} (a).  
Specifically, after receiving the instruction prompt in Table~\ref{tab:instruction}, \method first identifies the key entities central to the current reasoning step. Based on these entities, it formulates the subquery required for the step. Subsequently, \method performs context retrieval using a hybrid approach, extracting relevant information from both the chunk corpus and the knowledge graph, and constructs a KAG to enable PPR filtering for precise and concise context. \method then summarizes the core content from this context with respect to the subquery. Finally, leveraging the previous reasoning steps, \method decides whether to proceed with further reasoning or generate the final answer.  
Next, we present the detailed design of a reasoning step.

\subsubsection{Important Entity Recognition}
Before question decomposition and generation, LLMs need to identify the current anchor entity. By focusing on the key entities in the question, they can generate more targeted questions~\cite{mackie2023grm} and facilitate subsequent PPR filtering of irrelevant information.
The key entities recognized by LLMs at step $i$ are denoted as $v^i_{1}, \ldots, v^i_{j}$.

\subsubsection{Subquery Generation} Based on current key entities and previous reasoning information, the LLM decomposes the original question, generating the current subquery $q_i$ to be solved. The retrieval algorithms $\mathcal{A}_{\mathcal{R}}$ then acquire relevant context centered around this subquery $q_i$.

\begin{figure}[!t]
    \centering
    \includegraphics[width=\textwidth]{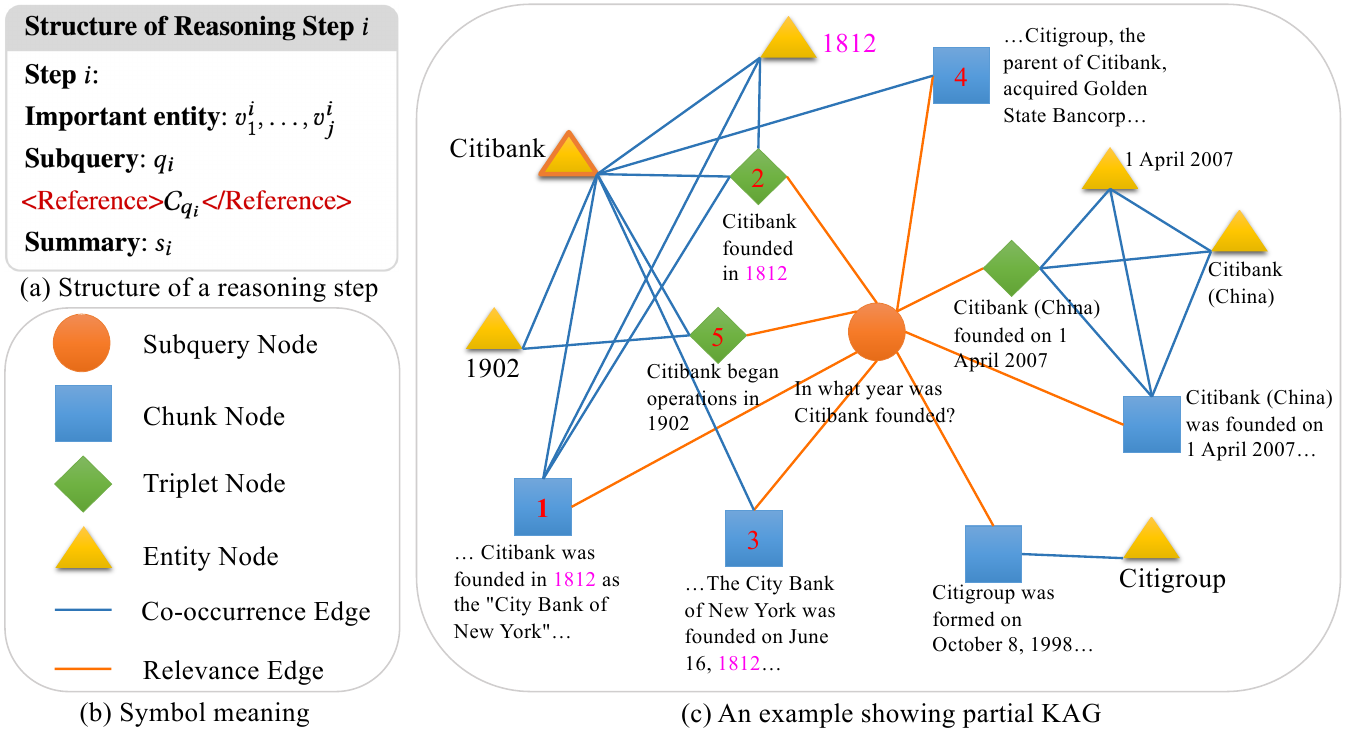}
    \caption{(a) shows the structure of a reasoning step. \textcolor{red!80!black}{\textless{}Reference\textgreater{}} and \textcolor{red!80!black}{\textless{}/Reference\textgreater{}} are special tokens for wrapping retrieved information.
    (b) shows the meaning of symbol in (c). (c) shows an example of partial KAG. The answer is highlighted by pink text. The red number on nodes is the ranking of the content selected in PPR.}
    \label{fig:KAG}
\end{figure}

\subsubsection{Context Retrieval}
\method's context retrieval combines semantic and graph retrieval.

The first component is semantic retrieval. The retriever, $\mathcal{R}$, retrieves relevant chunks $\mathcal{D}_{q_i}$ from the document corpus $\mathcal{D}$ based on their similarity to the subquery $q_i$. This is denoted as: $\mathcal{D}_{q_i} = \mathcal{R}(q_i,\mathcal{D})$.

The second component is graph retrieval, which retrieves a set of knowledge triplets $\mathcal{E}_{q_i}$ from the knowledge graph $\mathcal{G}$.
However, retrieving triplets based solely on identified key entities without considering detailed relationships risks collecting irrelevant triplets, whereas directly searching for relevant triplets from the entire set can be hampered by excessive noise from similar entries. 
To address these issues, we propose a two-stage graph retrieval method to perform graph retrieval in a more fine-grained manner.
The first stage retrieves relevant entities from the entity set $\mathcal{V}$ using the retriever $\mathcal{R}$. For the set of key entities $\{v^i_1, \ldots, v^i_j\}$ from the current reasoning step $i$, we generate entity queries by pairing each entity $v^i_t$ with the subquery $q_i$ in the format: ``Key entity: $v^i_t$. Query: $q_i$.'' The union of all entities returned by the retriever for these entity queries forms the set $\mathcal{V}_{q_i}$. We then collect all one-hop triplets connected to the entities in $\mathcal{V}_{q_i}$ from the knowledge graph, forming the set $\mathcal{E}_{entity}$. In the second stage, we retrieve the set of triplets most relevant to $q_i$ from this constrained triplet set, defined as $\mathcal{E}_{q_i}=\mathcal{R}(q_i,\mathcal{E}_{entity})$.

In the above two components, we employ a top-K strategy to directly select the top-K contexts with the highest retrieval scores. We denote the size of these sets as $|\mathcal{D}_{q_i}|=k_d$ and $|\mathcal{E}_{q_i}|=k_t$. 
Next, our aim is to select the final $k_f$ contexts for the LLM from the set $\mathcal{D}_{q_i}\cup\mathcal{E}_{q_i}$ by constructing a KAG based on the co-occurrence mechanism.

\subsubsection{Knowledge Association Graph Construction}
Based on the retrieved content, we construct a KAG, a heterogeneous graph designed to capture the co-occurrence relationships between different knowledge elements.
This graph comprises four distinct node types and two types of association edges that connect them, as shown in Fig.~\ref{fig:KAG} (b). The node types are defined as follows:
\begin{itemize}[topsep=0pt, itemsep=0pt]
    \item \textbf{Subquery node $n_{q_i}$:} A single node representing the current subquery $q_i$. This node serves as the central anchor of the graph, and all other information is retrieved based on this.
    
    \item \textbf{Chunk node $n_d$:} Each node of this type represents a specific textual chunk $d$ sourced from the retrieved document set $\mathcal{D}_{q_i}$.
    
    \item \textbf{Triplet node $n_e$:}  Represents a knowledge triplet $e$ from the triplet set $\mathcal{E}_{q_i}$.
    
    \item \textbf{Entity node $n_{v}$:} Each node of this type represents an entity $v$, which can be one of the entities $v^i_1, \ldots, v^i_j$ identified at reasoning step $i$, a source document title for a chunk, or a head or tail entity from a knowledge triplet.
\end{itemize}

The edges of the KAG capture either structural co-occurrence or semantic similarity, and are categorized as \textbf{co-occurrence edges} and \textbf{relevance edges}, respectively.
Besides, all edges are undirected.
Co-occurrence edges signify structural or contextual links between nodes. They are assigned a weight of $1$ and are defined by the following rules:
\begin{itemize}[topsep=0pt, itemsep=0pt]
    \item \textbf{Entity-Entity edge:} $(n_{v^h}, n_{v^t})$ connects the head and tail entities of a triplet $e=(v^h,rel,v^t)$.
    \item \textbf{Triplet-Entity edge:} $(n_e, n_{v^h})$ and $(n_e, n_{v^t})$ connect a triplet node to its constituent entities.
    \item \textbf{Chunk-Triplet edge:} If a triplet $e \in \mathcal{E}_{q_i}$ and its source chunk $d$ is also in $\mathcal{D}_{q_i}$, an edge $(n_e, n_d)$ is created, signifying a co-occurrence between the triplet and the chunk.
    \item \textbf{Chunk-Entity edge:} 
    When a triplet $(v^h,rel,v^t)$ co-occurs with a chunk $d$, edges are also added between the entity nodes of the triplet and the chunk node $(n_d, n_{v^h})$ and $(n_d, n_{v^t})$.
    Besides, an edge $(n_d,n_{title})$ connects a chunk node $n_d$ to the entity node of its document title $n_{title}$.
\end{itemize}
Relevance edges quantify the relevance of chunk and triplet nodes to the subquery node, defined as:
\begin{itemize}[topsep=0pt, itemsep=0pt]
    \item \textbf{Chunk-Subquery edge:} Weight of $(n_d, n_{q_i})$ for a chunk $d \in \mathcal{D}_{q_i}$ is based on their similarity:
    \begin{equation}
        w(n_d, n_{q_i}) = \operatorname{sigmoid}(\operatorname{sim_{\mathcal{R}}}(d, q_i)),
    \end{equation}
    where $\operatorname{sim_{\mathcal{R}}}$ is the relevance score from retriever $\mathcal{R}$, normalized via $\operatorname{sigmoid}$.
    \item \textbf{Triplet-Subquery edge:} To account for the contextual sparsity of triplets, which can lead to unreliable similarity scores, we apply a thresholding mechanism. The weight of the edge $(n_e, n_{q_i})$ for a triplet $e \in \mathcal{E}_{q_i}$ is:
    \begin{equation}
        w(n_e, n_{q_i}) = \max(\operatorname{sigmoid}(\operatorname{sim_{\mathcal{R}}}(e, q_i)) - \tau, 0),
    \end{equation}
    where $\tau$ is a threshold hyperparameter.
\end{itemize}

\subsubsection{Personalized PageRank Filtering}
After constructing the KAG, we employ PPR to identify the most important content nodes. This process allows us to dynamically select key information and compress tokens by replacing text chunks with high-density triplets, while ensuring relevance.
We denote the importance distribution across all nodes as $\pi$. The weighted adjacency matrix from the graph construction is normalized to produce the transition matrix $W$. Besides, we denote the personalization vector as $p$, which acts as a bias to steer the final importance distribution.
To ensure the relevance of the context to the subquery, the values for the subquery node $n_{q_i}$ in the personalized vector $p$ are set to 1. Meanwhile, the key entities identified by the LLM have their corresponding values in $p$ set to 0.5, serving as anchors for the question.
Subsequently, we iterate for $N$ rounds to obtain the updated importance distribution $\pi$:
\begin{equation}
\pi = \alpha W\pi + (1-\alpha) p,
\end{equation}
where $\alpha$ is a hyperparameter balancing the graph-derived importance with the personalization vector bias.
Finally, based on the ranking of the scores in $\pi$, we select the $k_f$ highest-scoring nodes from the chunk and triplet nodes to form the final context $\mathcal{C}_{q_i}$. This context is then wrapped in \textless{}Reference\textgreater{} and \textless{}/Reference\textgreater{} tags to signal to the LLM that it is an external resource.

We show a KAG example in Fig.~\ref{fig:KAG} (c). Core relevant knowledge forms dense graphs via co-occurrence edges. This indicates that using co-occurrence is usually a more precise way to filter noise and prevent the model's inference from being distracted by similar but irrelevant content.

\subsubsection{Summary Generation.} The LLM summarizes the collected external context to produce a summary, $s_i$, which enables subsequent steps to easily identify the key information for this step.

\subsubsection{Final Answer Generation.} The LLM autonomously determines whether the information collected in the reasoning path is sufficient. If not, it continues to generate the next reasoning step $S_{i+1} = \text{LLM}(I, q, \mathcal{P}_i)$. Conversely, when the information is sufficient, the LLM generates the final answer $a = \text{LLM}(I, q, \mathcal{P}_i)$. And the final answer $a$ starts with the format ``Final answer: '', which facilitates the format check.

\subsection{Model Training}
\label{training}

\begin{figure}[!t]
    \centering
    \includegraphics[width=\textwidth]{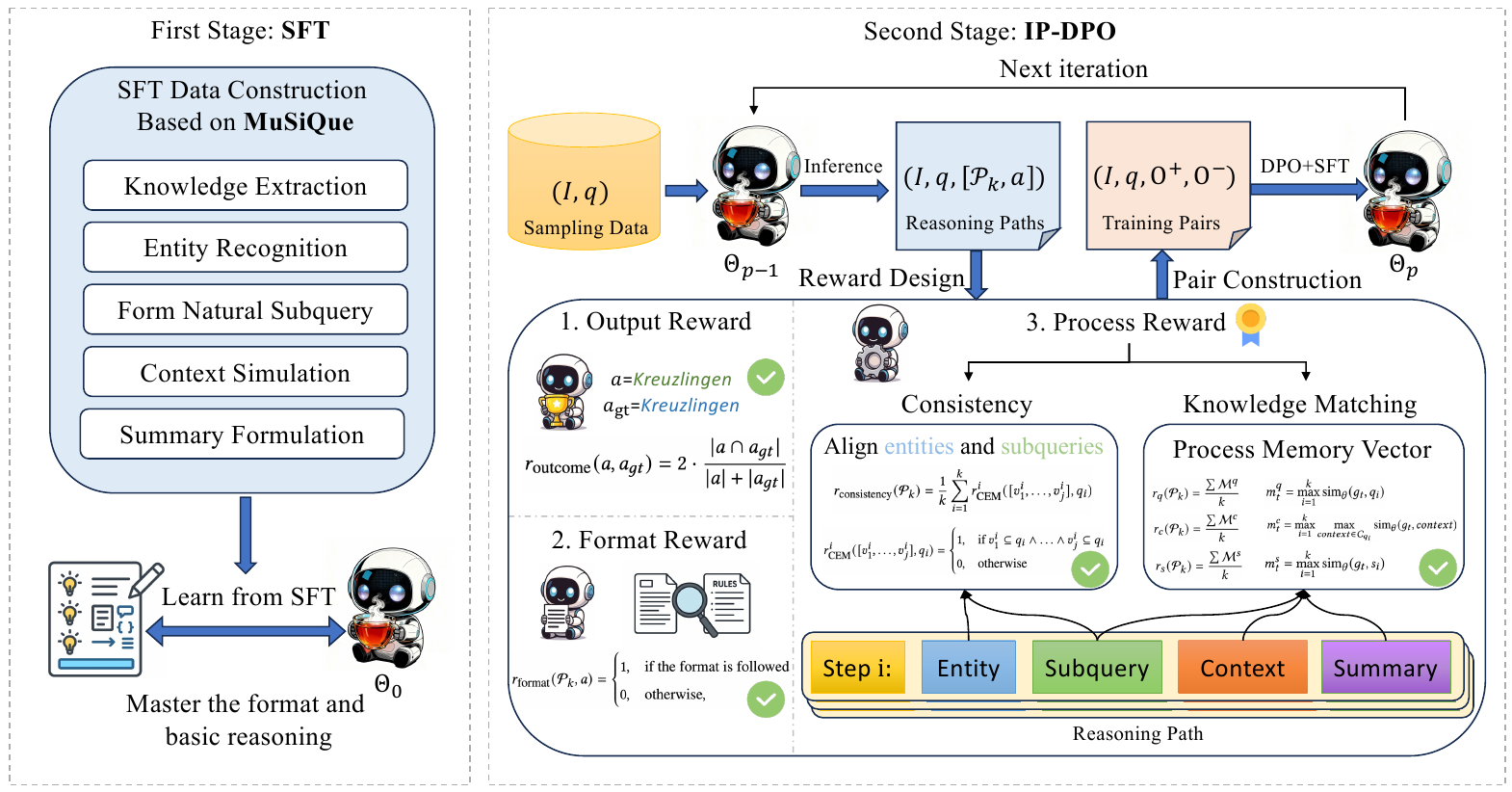}
    \caption{The overall training framework for \method follows a two-stage paradigm. First, we conduct SFT on the preprocessed MuSiQue dataset to help the LLM learn the required format and develop basic reasoning skills. In the second stage, we apply IP-DPO with a process-aware reward to further improve the model while preventing overthinking.}
    \label{fig:training_framework}
\end{figure}

To enable LLMs to autonomously execute the aforementioned pipeline, current advanced agentic RAG mainly adopts outcome-based RL~\cite{song2025r1,jin2025search}. However, these methods rely on high-resource RL methods such as PPO~\cite{schulman2017proximal} and GRPO~\cite{shao2024deepseekmath}, which require continuous inference and retrieval during training, and these methods only focus on the final results, making it difficult to effectively control the reasoning steps.
Although some methods utilize GPT-4o to annotate intermediate reasoning steps~\cite{zhang2025process} or consider complex rewards in the PPO framework~\cite{wang2025stepsearch}, these approaches make it difficult to achieve efficient, low-resource training of agentic RAG.
Therefore, to control the number of reasoning steps $k$ in $\mathcal{P}_k$ and enable efficient training, we propose a two-stage training paradigm that improves the efficiency and conciseness of reasoning by introducing process-level rewards.
The overall training framework is shown in Fig.~\ref{fig:training_framework}.
In the first stage, SFT trains the model on basic reasoning patterns.
In the second stage, IP-DPO is applied to further enhance the model’s generalization.

\subsubsection{First Stage: SFT}

To build reasoning SFT data, we use the Musique dataset~\cite{trivedi2022musique}, which provides structured query decomposition and supporting golden evidence paragraphs.
Each reasoning step is constructed as follows:
\begin{enumerate}[leftmargin=*, topsep=0pt, itemsep=0pt]
    \item \textbf{Knowledge extraction:} We first process the golden evidence paragraphs of the current subquery with Qwen2.5-14B-Instruct to extract a set of knowledge triples, which are aligned with our knowledge graph construction. From this set, Qwen2.5-72B-Instruct then identifies the supporting triplets that directly answer each intermediate subquery.
    \item \textbf{Important entity recognition:} We view the entities in supporting triplets as anchor entities.
    \item \textbf{Subquery generation:} We employ Qwen2.5-72B-Instruct to transform the structured intermediate questions from Musique into fluent, natural language subqueries.
    \item \textbf{Context simulation:} To improve fact localization, we synthetically construct a context for training. We create this context by taking the golden evidence paragraphs and randomly inserting the evidence triplets. This process helps the LLM learn to adapt to diverse and imperfect contexts.
    \item \textbf{Summary formulation:} We use diverse templates to construct target summaries. Every summary follows a two-part structure. It first presents the key supporting triplets and then concludes with the answer to the subquery.
\end{enumerate}
After constructing each reasoning step, we concatenate them in the order specified by the Musique dataset to form a complete reasoning path. Alternatively, each individual reasoning step can be treated as a single-hop question.
We denote the SFT dataset as $\mathcal{S}_{SFT} = \{ (I, q, O) \} = \{ (I, q, [\mathcal{P}_k, a]) \}$, where $O$ represents either the direct output of the LLM or context autonomously retrieved by the LLM.
The SFT dataset statistics are shown in Table~\ref{tab:sft_data}.

To avoid interfering with the LLM's ability to learn reasoning patterns due to the retrieved context, following~\cite{jin2025search}, we mask the context between the two special tokens <Reference> and </Reference>.
We fine-tune the model on the SFT dataset, using the following objective function:
\begin{equation}
\mathcal{L}_{SFT} = - \mathbb{E}_{(I, q, O)  \sim\mathcal{S}_{SFT}}
  \sum_{t=1}^{|O|} \mathbb{I}[O_t \text{ not masked}] \log P(O_t \mid I, q, O_{<t}; \Theta),
\end{equation}
where $\mathbb{I}$ is the indicator function, and $\Theta$ denotes the parameters of the LLM.
Through this SFT training phase, the model can generate outputs following the constructed reasoning path structure and acquire preliminary reasoning capabilities.

\subsubsection{Second Stage: IP-DPO}
\label{sec:ipaDPO}
The models trained with SFT are not optimized for real-world retrieval environments, where evidence is often noisy and incomplete. Consequently, their reasoning abilities are poorly suited to realistic retrieval scenarios, and they exhibit limited generalization~\cite{kirkunderstanding}.
To address this issue and control the number of reasoning steps $k$, we propose IP-DPO, which introduces a process-aware reward and uses iterative DPO~\cite{rafailov2023direct} to optimize the LLMs for better and more concise reasoning paths.
First, we collect a sampling dataset containing a diverse set of tasks and use the LLM to generate outputs for this dataset. Next, we score each sampled instance with a process-aware reward and the final outcome. These scored data are then used to construct data pairs, which serve as the training inputs for IP-DPO. After each training cycle, we resample using the updated model and repeat this entire process iteratively to improve performance further.

\textbf{Dataset collection.}
To enhance data diversity, we collect a dataset by sampling from the training sets of NQ~\cite{kwiatkowski2019natural}, HotpotQA~\cite{yang2018hotpotqa}, and Musique~\cite{trivedi2022musique}.
From the NQ dataset, we randomly sample 4,000 single-hop questions.
From the HotpotQA dataset, we sample questions from the hard level that contain as many evidence sentences as possible, drawing 2,500 questions each from the bridge and comparison types.
From the Musique dataset, we sample a total of 1,000 questions, distributed evenly across the 2-hop, 3-hop, and 4-hop data.
The data statistic is shown in Table~\ref{tab:dpo_data}.
The model then performs inference using the overall pipeline on each question in this dataset to repeatedly generate $R$ reasoning paths.
For each reasoning path, both the reasoning process and the final answer are recorded for subsequent evaluation.

\textbf{Reward design.}
To better guide the LLM's reasoning process and encourage more concise reasoning steps, we design a process-aware reward based on knowledge matching. This enables a more fine-grained evaluation of the quality of different generated outputs.
Our total reward system consists of three components: the outcome reward, the format reward, and the process reward. 

The first part is the outcome reward, which is based on a comparison between the final answer generated by the LLM and the ground truth.
Because rewards based on an exact match are too sparse to effectively compare the accuracy of LLM-generated answers, we follow~\cite{song2025r1} and use the F1 score as the metric for the outcome reward.
Specifically, we define the outcome reward as follows:
\begin{equation}
r_{\text{outcome}}(a,a_{gt}) = 2 \cdot \frac{|a \cap a_{gt}|}{|a|+|a_{gt}|},
\label{eq:outcome}
\end{equation}
where $a_{gt}$ represents the ground truth, $|a|$ is the word count of the LLM-generated answer $a$, and $|a \cap a_{gt}|$ is the number of overlapping words between $a$ and $a_{gt}$.

The second part is the format reward, which evaluates whether the structure of the model’s reasoning path conforms to the structure constructed in Section~\ref{sec:reasoning_chain_structure}. 
The format reward of reasoning path $\mathcal{P}_k$ and final answer $a$ is defined as follows:
\begin{equation}
r_{\text{format}}(\mathcal{P}_k,a) = 
\begin{cases}
1, & \text{if the format is followed} \\
0, & \text{otherwise},
\end{cases}
\label{eq:format}
\end{equation}

The third part is the process reward, which assesses the rationality of the LLM’s reasoning process. 
The process rewards include two components: entity-subquery consistency reward and knowledge matching rewards.
The first component is the entity-subquery consistency reward, intended to align the entities identified by the LLM with the anchor of the subsequent subquery. This alignment enables more targeted question decomposition and more precise PPR filtering.
We implement a cover exact match (CEM) mechanism that mandates complete inclusion of key entities within the subquery. The corresponding reward function is defined as follows:
\begin{equation}
r_{\text{consistency}}(\mathcal{P}_k)=\frac{1}{k}\sum_{i=1}^{k}r_{\text{CEM}}^i([v_{1}^i, \ldots, v^i_{j}],q_i),
\label{eq:consistency1}
\end{equation}
\begin{equation}
r_{\text{CEM}}^i([v_{1}^i, \ldots, v^i_{j}],q_i) = 
\begin{cases}
1, & \text{if } v^i_{1} \subseteq q_i \wedge \ldots \wedge v^i_{j} \subseteq q_i \\
0, & \text{otherwise}
\end{cases}.
\end{equation}

The second component is the knowledge matching rewards.
These rewards evaluate the correctness of the LLM's reasoning process by assessing whether its intermediate steps successfully capture the golden evidence.
We evaluate three types of intermediate steps: subqueries, retrieved contexts, and summaries.
We denote the set of golden evidence for an overall question as $G=[g_1,\ldots,g_l]$, where $l$ is the number of golden evidence pieces.
The non-deterministic nature of the reasoning process makes it difficult to pre-assign which specific step should capture a particular piece of golden evidence. 
To address this, following~\cite{wang2025stepsearch}, we create a memory vector for each type of intermediate step. 
This vector stores the maximum similarity score achieved between each piece of evidence and any step in the reasoning path $\mathcal{P}_k$. This score serves as a metric to evaluate whether the model successfully integrated the required evidence.
Specifically, we denote the subquery memory vector as $\mathcal{M}^q=[m_1^q,\ldots,m_l^q]$.
Here, $m_t^q$ records the maximum similarity between the golden evidence $g_t$ and any subquery $q_i$ in the reasoning path (where $1\le t\le l$):
\begin{equation}
m_t^q = \max_{i=1}^k \text{sim}_{\theta}(g_t,q_i),
\end{equation}
where $\text{sim}_{\theta}()$ is a similarity function based on a reranker model $\theta$, which uses a sigmoid function to normalize the score to the range $[0,1]$.
Similarly, we define the context memory vector $\mathcal{M}^c$ and the summary memory vector $\mathcal{M}^s$. The memory elements, $m_t^c$ and $m_t^s$, are calculated as follows:
\begin{equation}
m_t^c = \max_{i=1}^k \max_{context\in \mathcal{C}_{q_i}} \text{sim}_{\theta}(g_t,context),
\end{equation}
\begin{equation}
m_t^s = \max_{i=1}^k \text{sim}_{\theta}(g_t,s_i).
\end{equation}

These calculations yield three memory vectors, $\mathcal{M}^q, \mathcal{M}^c, \text{ and } \mathcal{M}^s$, for the LLM's reasoning process. For a multi-hop question with $l$ pieces of golden evidence, an ideal path might decompose the problem into approximately $l$ subqueries. For a reasoning path $\mathcal{P}_k$ with $k$ steps, a value of $k \le l$ suggests a concise reasoning process without redundant retrievals. Conversely, when $k > l$, it may indicate inefficiencies such as overthinking or repetitive searches. To promote conciseness and penalize inefficiency, we normalize the summed memory scores by the number of steps, $k$. The final rewards are calculated as follows:
\begin{equation}
r_q(\mathcal{P}_k) = \frac{\sum\mathcal{M}^q}{k}, \quad r_c(\mathcal{P}_k) = \frac{\sum\mathcal{M}^c}{k}, \quad r_s(\mathcal{P}_k) = \frac{\sum\mathcal{M}^s}{k},
\end{equation}
where $r_q(\mathcal{P}_k)$, $r_c(\mathcal{P}_k)$, and $r_s(\mathcal{P}_k)$ are the process rewards for subquery generation, the retrieved context, and the summary, respectively. Based on these, the overall process reward is as follows:
\begin{equation}
r_{process}(\mathcal{P}_k)=0.1r_{\text{consistency}}(\mathcal{P}_k) + 0.3r_q(\mathcal{P}_k)+0.3r_c(\mathcal{P}_k)+0.3r_s(\mathcal{P}_k).
\end{equation}

\textbf{DPO pair construction.}
Based on the above reward design, each reasoning path can obtain three rewards: $r_{format}$, $r_{outcome}$, and $r_{process}$. The core challenge in DPO~\cite{rafailov2023direct} is constructing positive-negative preference pairs. To this end, we design the following pair construction algorithm.

First, we select the chosen response set, which exhibits a more accurate reasoning process. 
Our approach varies based on the question type.
For single-hop questions, which are typically simple and whose answers can be found with a single retrieval, we do not apply a process reward. 
This is also because our single-hop dataset lacks golden evidence.
Instead, we directly scale the outcome reward by dividing it by the length of the reasoning path $k$, setting $r_{outcome} = \frac{r_{outcome}}{k}$ and $r_{process} = 0$. This approach incentivizes the LLM to solve single-hop questions in a single step.
For more complex multi-hop questions, we utilize the process reward to differentiate and evaluate the quality of the reasoning steps. When considering format and outcome rewards, we must also ensure that the process reward is sufficiently high. Furthermore, many reasoning paths in multi-hop questions demonstrate relatively good reasoning processes, but due to variations in the presentation of the output, there are cases where small differences exist between the outcome and ground truth. Therefore, we have relaxed the conditions.
Given a question $q$, the chosen response set $\mathcal{W}$ is formally defined as:
\begin{equation}
\resizebox{0.9\textwidth}{!}{$
\mathcal{W} = \left\{(q,\mathcal{P}_k, a, r_{format}, r_{outcome}, r_{process}) \middle|
\begin{cases} 
r_{format}=1 \land r_{outcome}=1, & \text{if } q \text{ is single-hop} \\ 
r_{format}=1 \land \Big((r_{outcome}=1 \land r_{process} \geq 0.7) \lor \\ 
\quad (r_{outcome} \geq 0.8 \land r_{process} \geq 0.8)\Big), & \text{if } q \text{ is multi-hop}
\end{cases} 
\right\}.
$}
\label{eq:chosen_set_criteria} 
\end{equation}

Next, for each selected response in $\mathcal{W}$, we choose a corresponding rejected response. Given a chosen response $(q,\mathcal{P}_k^+, a^+, r_{\text{format}}^+, r_{\text{outcome}}^+, r_{\text{process}}^+)$, we rank sampled reasoning paths for the question $q$ in ascending order based on the sum of the three reward scores. 
While traversing this ranked list sequentially, we define three types of rejected samples:
\begin{itemize}[leftmargin=*, topsep=0pt, itemsep=0pt]
    \item \textbf{Format rejection:} $r_{\text{format}}^- = 0$. The reasoning path violates the required format.
    
    \item \textbf{Easy rejection:} $r_{\text{format}}^- = 1$ and $r_{\text{outcome}}^- \le 0.3$. The generated reasoning path adheres to the format but fails to reach the correct answer.
    
    \item \textbf{Hard rejection:} $r_{\text{format}}^- = 1$, $0.3 < r_{\text{outcome}}^- \le r_{\text{outcome}}^+ - 0.3$, and $r_{\text{process}}^- \le r_{\text{process}}^+$. The path largely progresses toward an answer but still contains specific errors.
\end{itemize}
To maintain the data diversity of DPO pairs, once a rejected response is obtained for a chosen sample, the algorithm continues traversing the ranked list from the next position to find the next rejected sample for the next chosen sample of the question $q$.
This approach prevents the selection of duplicate rejected samples.
Additionally, for single-hop questions, due to their simplicity, rejected samples are limited to format and easy rejection.
Following this procedure, we construct the DPO pair dataset.
We denote the dataset at iteration $p$ by $\mathcal{S}_{DPO-p}= \{(I, q, O^+, O^-)\}$. 

\textbf{Iterative DPO training.}
Given that the quality of samples generated by the preliminary SFT model remains suboptimal, the performance gains from a single round of DPO training are inherently limited. To address this, we adopt an iterative DPO procedure designed to enhance the model's comprehension of preference patterns and enable progressively stronger alignment.
Following~\cite{pang2024iterative,pal2024smaug}, we jointly optimize the DPO objective together with the SFT objective to mitigate distribution shift in the model’s chosen responses. The training losses are defined as follows:
\begin{equation}
\mathcal{L}_{SFT} = - \mathbb{E}_{(I, q, O^+,O^-)  \sim\mathcal{S}_{DPO-p}}
  \sum_{t=1}^{|O^+|} \mathbb{I}[O_t^+ \text{ not masked}] \log P(O_t^+ \mid I, q, O_{<t}^+; \Theta_p),
\end{equation}
\begin{align}
\mathcal{L}_{DPO} &= - \mathbb{E}_{(I, q, O^+,O^-)  \sim\mathcal{S}_{DPO-p}} \left[ \log \sigma \left( \beta \sum_{t=1}^{|O^+|} \mathbb{I}[O_t^+ \text{ not masked}] \log \frac{P(O_t^+ \mid I, q, O_{<t}^+; \Theta_p)}{P(O_t^+ \mid I, q, O_{<t}^+; \Theta_{p-1})} \right. \right. \nonumber \\
&\quad \quad\quad\quad\quad\quad\quad\quad\quad\quad\quad\quad \left. \left. - \beta \sum_{t=1}^{|O^-|} \mathbb{I}[O_t^- \text{ not masked}] \log \frac{P(O_t^- \mid I, q, O_{<t}^-; \Theta_p)}{P(O_t^- \mid I, q, O_{<t}^-; \Theta_{p-1})} \right) \right],
\end{align}
\begin{equation}
\mathcal{L}_{ALL} = \eta \mathcal{L}_{SFT} + \mathcal{L}_{DPO},
\end{equation}
where $\beta$ and $\eta$ are hyperparameters, and $\Theta_p$ denotes the model parameters in the $p$-th round, initialized from $\Theta_{p-1}$. $\Theta_0$ represents the model parameters after the first-stage SFT. 

Once the model completes training with the aforementioned loss function, the next round of data sampling, sample reward scoring, pair construction, and training can be performed, implementing iterative DPO training to continuously optimize the model's capabilities.

\section{Experiments}
\subsection{Experiment Setup}

\subsubsection{Datasets and Evaluation Metrics.}
\label{dataset}
We use the training datasets from Section~\ref{training} to perform SFT and IP-DPO.
The statistics of these datasets are shown in Table~\ref{tab:sft_data} and Table~\ref{tab:dpo_data}.
Besides, we focus on question-answer (QA) datasets. For single-hop QA datasets, we use \textbf{NQ}~\cite{kwiatkowski2019natural} and \textbf{PopQA}~\cite{mallen2023not}. For multi-hop QA datasets, we use \textbf{HotpotQA}~\cite{yang2018hotpotqa}, \textbf{2WikiMultiHopQA}~\cite{ho2020constructing}, \textbf{Musique}~\cite{trivedi2022musique}, and \textbf{Bamboogle}~\cite{press2023measuring}. For datasets without a test set, we use the development set for testing. The statistical details of these datasets are shown in Table~\ref{tab:qa_datasets}. Since we use some training data from NQ, HotpotQA, and Musique, the tests on NQ, HotpotQA, and Musique datasets are in-domain tests, and the tests on PopQA, 2WikiMultiHopQA, and Bamboogle are out-of-domain tests. 
For the retrieval corpus, following~\cite{jin2024flashrag}, we adopt the Wikipedia corpus~\cite{karpukhin2020dense}, as the chunk corpus. The statistics of the chunk corpus are shown in Table~\ref{tab:chunk_stats}. Based on the knowledge graph construction described in Section~\ref{kg_extract}, we constructed a large-scale knowledge graph, and the statistical information of this knowledge graph is shown in Table~\ref{tab:kg_stats}.
For all datasets, we adopt EM and F1 as the evaluation metrics.

\renewcommand{\arraystretch}{1.2}
\begin{table}[ht]
\centering

\begin{minipage}{0.45\textwidth}
\caption{SFT data distribution.}
\label{tab:sft_data}
\centering
\begin{tabular}{l|rrrr|r}
\Xhline{1.2pt}
\textbf{Dataset} & \textbf{1-hop} & \textbf{2-hop} & \textbf{3-hop} & \textbf{4-hop} & \textbf{Total} \\
\Xhline{0.6pt}
Train & 2,700 & 1,800 & 1,800 & 1,057 & 7,357 \\
Test  &   300 &   200 &   200 &   118 &   818 \\
\Xhline{1.2pt}
\end{tabular}
\end{minipage}%
\hspace{0.04\textwidth}
\begin{minipage}{0.45\textwidth}
\caption{IP-DPO dataset composition.}
\label{tab:dpo_data}
\centering
\begin{tabular}{lr}
\Xhline{1.2pt}
\textbf{Source Dataset} & \textbf{Number} \\
\Xhline{0.6pt}
NQ         & 4,000 \\
HotpotQA   & 5,000 \\
Musique    & 1,000 \\
\Xhline{0.6pt}
Total      & 10,000 \\
\Xhline{1.2pt}
\end{tabular}
\end{minipage}

\vspace{1em}

\begin{minipage}{0.45\textwidth}
\caption{QA datasets for testing.}
\label{tab:qa_datasets}
\centering
\resizebox{\textwidth}{!}{
\begin{tabular}{l|l|rr}
\Xhline{1.2pt}
\textbf{Task} & \textbf{Dataset} & \textbf{\# Dev} & \textbf{\# Test} \\
\Xhline{0.6pt}
\multirow{2}{*}{Single-hop QA} & NQ & 8,757 & 3,610 \\
                               & PopQA & -- & 14,267 \\
\Xhline{0.6pt}
\multirow{4}{*}{Multi-hop QA}  & HotpotQA & 7,405 & -- \\
                               & 2WikiMultiHopQA & 12,576 & -- \\
                               & Musique & 2,417 & -- \\
                               & Bamboogle & -- & 125 \\
\Xhline{1.2pt}
\end{tabular}
}
\end{minipage}%
\hspace{0.04\textwidth}
\begin{minipage}{0.45\textwidth}
\caption{Chunk corpus statistics.}
\label{tab:chunk_stats}
\centering
\begin{tabular}{lr}
\Xhline{1.2pt}
\textbf{Metric} & \textbf{Value} \\
\Xhline{0.6pt}
Size of document set & 3,232,908 \\
Size of chunk set $\mathcal{D}$ & 21,015,324 \\
Average length of chunk & 100 \\
\Xhline{1.2pt}
\end{tabular}
\end{minipage}

\vspace{1em}

\begin{minipage}{0.5\textwidth}
\caption{Knowledge graph statistics.}
\label{tab:kg_stats}
\centering
\resizebox{0.99\linewidth}{!}{
\begin{tabular}{lr}
\Xhline{1.2pt}
\textbf{Metric} & \textbf{Value} \\
\Xhline{0.6pt}
Size of entity set $\mathcal{V}$ & 51,063,765 \\
Size of relation set $\mathcal{E}$ & 130,931,564 \\
Average out-degree per head entity & 9.24 \\
Average in-degree per tail entity & 3.05 \\
Average degree per entity & 5.13 \\
\Xhline{1.2pt}
\end{tabular}
}
\end{minipage}

\end{table}

\subsubsection{Baselines.}
We compare our \method with various baselines, which are divided into the following two categories: one-turn generation and iterative RAG.

One-turn generation prepares all the required information and utilize the LLM in a single pass to generate the final answer.
The methods in this category are as follows:
\begin{itemize}[leftmargin=*, topsep=0pt, itemsep=0pt]
\item \textbf{Zero-shot:} This method uses the LLM to generate answers without retrieving relevant information.
\item \textbf{RAG}~\cite{lewis2020retrieval}\textbf{:} This is the standard RAG method that uses semantic retrieval to search for relevant text chunks from the corpus, which are then input as evidence to the LLM to directly generate answers.
\item \textbf{Rerank}~\cite{yu2024rankrag}\textbf{:} This method uses semantic retrieval to find relevant chunks, then applies reranking to filter noise, before inputting them as evidence to the LLM to directly generate answers.
\item \textbf{RECOMP$_{abs}$}~\cite{xu2023recomp}\textbf{:} This method uses a compressor to extract core information related to the question from retrieved chunks before inputting them to the LLM.
\item \textbf{MixPR-RAG}~\cite{alonso2024mixture}\textbf{:} This method constructs a semantic graph from each sentence in the chunks and uses PPR to filter out important sentences to input to the LLM.
\end{itemize}

Iterative RAG utilizes LLMs to autonomously determine whether retrieval is needed, making multiple calls to retrieval tools and integrating multiple retrieval contexts to derive the final answer. The methods in this category are as follows:
\begin{itemize}[leftmargin=*, topsep=0pt, itemsep=0pt]
\item \textbf{IRCoT}~\cite{trivedi2023interleaving}\textbf{:} This method iteratively utilizes Chain-of-Thought (CoT) to guide retrieval and uses retrieval results to enhance CoT to derive the final answer.
\item \textbf{SelfRAG}~\cite{asai2023self}\textbf{:} This method utilizes LLMs to generate special tokens to help the model decide when to retrieve and whether to use the retrieved content.
\item \textbf{R1-Searcher}~\cite{song2025r1}\textbf{:} This method employs two-stage training. The first stage trains the model to learn generating retrieval queries, and the second stage uses outcome-based and format rewards to train LLMs to automatically invoke retrieval during reasoning. Based on Reinforce++~\cite{hu2025reinforce++}, this method is trained using data from HotpotQA and 2WikiMultiHopQA. We use two versions of models trained by this method: R1-Searcher-Llama-8B and R1-Searcher-Qwen-7B.
\item \textbf{Search-R1}~\cite{jin2025search}\textbf{:} This method utilizes PPO~\cite{schulman2017proximal} with complete NQ and HotpotQA training data to optimize LLMs based on outcome rewards. We test four versions of models trained by this method: Search-R1-instruct-7B, Search-R1-base-7B, Search-R1-instruct-14B, Search-R1-base-14B.
\end{itemize}

For all baselines, the retrieval systems use E5-base-V2~\cite{wang2022text} as the retriever unless otherwise specified. The ``\textbf{+R}'' symbol indicates that the retrieval system incorporates both retrieval and reranking stages, with BGE-reranker-v2~\cite{bge_m3} serving as the reranker model. This configuration ensures a fair comparison with our \method.

\subsubsection{Implementation Details.}
To ensure robustness, we employ two widely used LLMs for training and inference: Llama3-8B-Instruct~\cite{grattafiori2024llama} and Qwen2.5-14B-Instruct~\cite{yang2024qwen2}. For the retrieval stage, following~\cite{jin2024flashrag}, we utilize E5-base-V2~\cite{wang2022text} as the retriever and BGE-reranker-v2~\cite{bge_m3} as the reranker, which together constitute our retrieval system $\mathcal{R}$.
For semantic retrieval, we retrieve the top-$20$ chunks, which are then reranked to a final set of $k_d=5$ chunks. 
For graph retrieval, we initially retrieve the top-$10$ entities and $20$ edges. After reranking, these are narrowed down to $5$ entities and $k_t=10$ edges. The final number of items composing the context is set to $k_f=5$.
During the PPR process, we set the hyperparameters $\tau=0.2$ and $\alpha=0.5$.
The iteration $N$ of PPR is set to $200$.
We configure the LLM to perform a maximum of $5$ reasoning steps, i.e., $k\le5$.

Our experiments are conducted using 8 NVIDIA A100 (80G) GPUs. In the SFT phase, the learning rate is $5 \times 10^{-4}$, with an overall batch size of $128$ and a per-device batch size of $16$. We train for $1$ epoch with a weight decay of $0.001$. 
In the IP-DPO phase, $R = 8$ reasoning paths are repeatedly sampled for each question.
And the number of epochs for the first two DPO rounds is $2$, while the final round is trained for only $1$ epoch to prevent overfitting from data repetition. 
The hyperparameter $\beta$ is set to $0.5$. 
The SFT weighting parameter $\eta$ is tuned within the range $[0.25, 0.5, 1]$ and is progressively reduced as DPO iterations increase to prevent overfitting. 
Besides, the learning rate is set to $1 \times 10^{-4}$, the overall batch size is $16$, the per-device batch size is $2$, and the weight decay is $0.001$.
In both training stages, we adopt LoRA~\cite{hu2022lora} for parameter-efficient fine-tuning. We set the LoRA rank to $8$, $\alpha$ to $32$, and dropout to $0.1$, and apply it to the \texttt{q\_proj}, \texttt{k\_proj}, \texttt{v\_proj}, and \texttt{o\_proj} layers.

\begin{table*}
\centering
\caption{Performance comparison of various methods on six QA benchmarks. ``\textbf{*}'' indicates testing using officially released LLMs. ``\textbf{+R}'' indicates that the retrieval system includes retrieval and rerank stages. The best results are highlighted in bold. ``\textbf{{$\dag$}}'' indicates the statistically significant improvements in average results (i.e., two-sided t-test with $p<0.05$) over all baselines.}
\label{tab:overall_performance}
\renewcommand\arraystretch{1.2}
\resizebox{\textwidth}{!}{
\begin{tabular}{l|l|cccc|cccccccc|cc}
\hline
\multirow{2}{*}{\textbf{LLM}}&\multirow{2}{*}{\textbf{Method}} & \multicolumn{2}{c}{\textbf{NQ}} & \multicolumn{2}{c|}{\textbf{PopQA}} & \multicolumn{2}{c}{\textbf{HotpotQA}}& \multicolumn{2}{c}{\textbf{2Wiki}}& \multicolumn{2}{c}{\textbf{Musique}}&\multicolumn{2}{c|}{\textbf{ Bamboogle}}&\multicolumn{2}{c}{\textbf{ Avg.}} \\
\cmidrule(lr){3-4} \cmidrule(lr){5-6} \cmidrule(lr){7-8} \cmidrule(lr){9-10} \cmidrule(lr){11-12} \cmidrule(lr){13-14} \cmidrule(lr){15-16}
& & \textbf{EM} & \textbf{F1} & \textbf{EM} & \textbf{F1}& \textbf{EM} & \textbf{F1} & \textbf{EM} & \textbf{F1} & \textbf{EM} & \textbf{F1} & \textbf{EM} & \textbf{F1} & \textbf{EM} & \textbf{F1} \\
\hline
\multirow{16}{*}{\textbf{Llama3-8B}}& \textbf{zero-shot} & 22.49 & 31.62 & 22.56 & 26.84 & 19.14 & 26.91 & 17.81 & 26.14 & 4.30 & 10.00 & 12.80 & 18.91 & 16.41 & 23.40\\
& \textbf{RAG} & 36.03 & 47.19 & 37.73 & 46.43 &25.29&35.69 & 8.19 & 20.18 & 4.71 & 10.26 & 8.80 & 17.82 & 20.12 & 29.59\\
& \textbf{Rerank} & 37.28 & 48.48 & 41.08 & 50.27 & 29.53 & 40.80 & 9.64 & 22.12 & 5.54 & 12.08 & 11.20 & 19.94 & 22.37 & 32.28\\
& \textbf{RECOMP$_{abs}$} & 35.76 & 45.72 & 42.79 & 49.08 & 27.77 & 37.71 & 23.17 & 29.86 & 5.00 & 10.56 & 10.40 & 19.69 & 24.14 & 32.10\\
& \textbf{MixPR-RAG} & 26.73 & 38.88 & 35.22 & 44.96 & 23.07 & 34.47 & 7.96 & 21.02 & 3.72 & 10.08 & 8.80 & 18.85 & 17.58 & 28.04 \\
\cdashline{2-16}
& \textbf{IRCoT} & 35.31 & 46.57 & 39.72 & 47.42 & 33.50 & 44.67 & 25.81 & 34.58 & 9.43 & 15.87 & 32.80 & 43.21 & 29.42 & 38.72 \\
& \textbf{SelfRAG*} & 33.90 & 41.86 & 19.21 & 31.27 & 16.12 & 28.36 & 11.72 & 23.24 & 4.38 & 12.49 & 4.00 & 14.47 & 14.88 & 25.28 \\
& \textbf{R1-Searcher-Llama-8B*} & 39.92 & 49.89 & 42.01 & 47.31 & 43.46 & 55.12 & 44.92 & 50.12 & 17.54 & 25.46 & 44.00 & 55.57 & 38.64 & 47.25 \\
& \textbf{R1-Searcher-Llama-8B*+R} & 39.92 & 50.27 & 44.31 & 49.79 & 45.77 & 57.73 & 46.39 & 51.41 & 18.58 & 26.53 & 40.80 & 54.10 & 39.29 & 48.30  \\
& \textbf{R1-Searcher-Qwen-7B*} & 40.33 & 50.91 & 39.03 & 45.33 & 45.01 & 57.51 & 47.64 & 53.91 & 21.97 & 30.86 & 45.60 & 55.39 & 39.93 & 48.98 \\
& \textbf{R1-Searcher-Qwen-7B*+R} & 42.08 & 52.95 & 41.48 & 47.94 & 47.56 & \textbf{60.40} & \textbf{49.16} & \textbf{55.55} & 22.59 & 31.80 & 40.80 & 55.46 & 40.61 & 50.68 \\
&\textbf{Search-R1-instruct-7B*} & 40.25 & 50.41 & 40.32 & 45.88 & 38.47 & 49.23 & 35.67 & 41.88 & 16.71 & 24.53 & 41.60 & 53.33 & 35.50 & 44.21 \\
& \textbf{Search-R1-base-7B*} & 49.11 & 57.86 & 47.14 & 51.23 & 44.89 & 56.81 & 37.02 & 43.03 & 20.60 & 29.28 & 48.80 & 59.73 & 41.26 & 49.66 \\
& \textbf{Search-R1-base-7B*+R} & \textbf{50.39} & 59.09 & 49.86 & 54.00 & \textbf{47.66} & 60.14 & 39.49 & 45.65 & 21.64 & 30.57 & \textbf{51.20} & \textbf{64.51} & 43.37 & 52.33 \\
\cdashline{2-16}    
\rowcolor{blue!10}
\cellcolor{white}& \textbf{\method-SFT-8B} & 38.56 & 47.46 & 46.56 & 50.67 & 28.86 & 38.23 & 26.39 & 31.99 & 18.99 & 27.57 & 36.00 & 47.96 & 32.56 & 40.65 \\
\rowcolor{blue!10}
\cellcolor{white}& \textbf{\method-8B} & 50.06 & \textbf{59.71} & \textbf{51.98} & \textbf{56.08} & 46.59 & 59.48 & 47.89 & 54.21 & \textbf{26.98} & \textbf{37.34} & 47.20 & 59.89 & \textbf{45.12}$^\dag$ & \textbf{54.45}$^\dag$ \\
\hline
\multirow{16}{*}{\textbf{Qwen2.5-14B}}& \textbf{zero-shot} & 21.57 & 31.16 & 20.40 & 24.42 & 22.61 & 31.38 & 26.78 & 32.58 & 5.13 & 13.79 & 15.20 & 25.06 & 18.61 & 26.39 \\
& \textbf{RAG}  & 32.77 & 46.03 & 37.97 & 46.51 & 28.14 & 39.55 & 18.94 & 29.13 & 6.04 & 12.76 & 16.00 & 29.81 & 23.31 & 33.96 \\
& \textbf{Rerank} & 33.29 & 46.71 & 40.87 & 49.75 & 32.47 & 44.80 & 21.66 & 31.88 & 7.86 & 15.77 & 22.40 & 32.03 & 26.42 & 36.82 \\
& \textbf{RECOMP$_{abs}$} & 34.87 & 45.25 & 43.64 & 49.42 & 27.42 & 38.25 & 23.17 & 30.54 & 5.33 & 11.18 & 14.40 & 23.55 & 24.80 & 33.03 \\
& \textbf{MixPR-RAG} & 33.40 & 46.42 & 38.57 & 46.76 & 31.51 & 43.73 & 21.77 & 32.16 & 6.74 & 14.74 & 18.40 & 29.87 & 25.06 & 35.61 \\
\cdashline{2-16}
& \textbf{IRCoT} & 28.31 & 39.95 & 34.25 & 41.43 & 23.76 & 34.43 & 10.95 & 19.29 & 7.36 & 12.12 & 24.00 & 34.67 & 21.43 & 30.31 \\
& \textbf{SelfRAG*} & 33.90 & 41.86 & 19.21 & 31.27 & 16.12 & 28.36 & 11.72 & 23.24 & 4.38 & 12.49 & 4.00 & 14.47 & 14.88 & 25.28  \\
& \textbf{R1-Searcher-Llama-8B*} & 39.92 & 49.89 & 42.01 & 47.31 & 43.46 & 55.12 & 44.92 & 50.12 & 17.54 & 25.46 & 44.00 & 55.57 & 38.64 & 47.25 \\
& \textbf{R1-Searcher-Llama-8B*+R} & 39.92 & 50.27 & 44.31 & 49.79 & 45.77 & 57.73 & 46.39 & 51.41 & 18.58 & 26.53 & 40.80 & 54.10 & 39.29 & 48.30  \\
& \textbf{R1-Searcher-Qwen-7B*} & 40.33 & 50.91 & 39.03 & 45.33 & 45.01 & 57.51 & 47.64 & 53.91 & 21.97 & 30.86 & 45.60 & 55.39 & 39.93 & 48.98 \\
& \textbf{R1-Searcher-Qwen-7B*+R} & 42.08 & 52.95 & 41.48 & 47.94 & 47.56 & 60.40 & 49.16 & 55.55 & 22.59 & 31.80 & 40.80 & 55.46 & 40.61 & 50.68 \\
& \textbf{Search-R1-instruct-14B*} & 43.24 & 53.49 & 45.37 & 51.24 & 45.55 & 56.96 & 42.69 & 48.15 & 21.68 & 29.70 & 48.00 & 60.20 & 41.09 & 49.96 \\

& \textbf{Search-R1-base-14B*} & 49.67 & 58.36 & 48.10 & 51.69 & 47.78 & 59.97 & 46.33 & 52.22 & 24.95 & 33.40 & \textbf{51.20} & 64.78 & 44.67 & 53.40 \\
& \textbf{Search-R1-base-14B*+R} & \textbf{51.27} & 60.07 & 50.33 & 53.89 & \textbf{49.93} & 62.31 & 48.88 & 54.51 & 27.18 & 35.46 & \textbf{51.20} & \textbf{65.62} & 46.47 & 55.31 \\
\cdashline{2-16}
\rowcolor{blue!10}
\cellcolor{white}& \textbf{\method-SFT-14B} & 43.52 & 52.77 & 49.60 & 53.88 & 39.42 & 50.40 & 36.34 & 42.39 & 21.56 & 32.23 & 40.00 & 52.79 & 38.41 & 47.41 \\
\rowcolor{blue!10}
\cellcolor{white}& \textbf{\method-14B} & 50.33 & \textbf{60.31} & \textbf{53.59} & \textbf{58.02} & \textbf{49.93} & \textbf{63.14} & \textbf{52.29} & \textbf{57.91} & \textbf{27.93} & \textbf{39.22} & 50.40 & 64.07 & \textbf{47.41}$^\dag$ & \textbf{57.11}$^\dag$ \\
\hline
\end{tabular}
}
\end{table*}

\subsection{Overall Performance}

The overall performance of the model is shown in Table~\ref{tab:overall_performance}. We have the following observations:

\begin{enumerate}[leftmargin=*, topsep=0pt, itemsep=0pt]
    \item \textbf{Our \method outperforms existing agentic RAG methods on most datasets and achieves the best average performance across all datasets.} This empirical result demonstrates the effectiveness of \method, which successfully boosts the performance of LLMs on RAG tasks by constructing a KAG-based agentic workflow combined with process supervision, outperforming previous outcome-based agentic RAG approaches.
    
    \item \textbf{The two-stage training paradigm significantly boosts the performance of \method.}
    In the first training stage, SFT yields substantial performance improvements for \method-SFT, surpassing both the prompt-based agentic RAG approach, IRCOT, and the SFT-trained agentic RAG system, SelfRAG. Based on \method-SFT, the second stage, IP-DPO, delivers a pronounced enhancement. Specifically, IP-DPO achieves a relative gain in average EM score of $38.57\%$ on Llama3‑8B‑Instruct and $23.43\%$ on Qwen2.5‑14B‑Instruct. These improvements highlight the effectiveness of IP-DPO in strengthening the model’s generalization capability.
    
    \item \textbf{\method demonstrates strong out-of-domain generalizability.} Across both single-hop and multi-hop out-of-domain datasets, including PopQA, 2WikiMultiHopQA, and Bamboogle, our method delivers excellent performance. On the 2WikiMultiHopQA dataset, using the Llama3-8B-Instruct base model, \method-8B not only surpasses Search-R1-base-7B+R in the out-of-domain test by a significant margin but also achieves results on par with those of R1-Searcher-Qwen-7B+R in the in-domain evaluation.

    \item \textbf{More powerful LLMs achieve superior results.} This shows that \method leverages the core capabilities of LLMs. As the LLM's power increases, the agentic RAG's information-gathering abilities are correspondingly enhanced.

    \item \textbf{More advanced retrieval systems yield better results.} For a fair comparison with \method, we enhanced the original Search-R1 and R1-Searcher with a reranking-integrated retrieval system, which led to a significant performance increase. The continued superiority of \method over these augmented baselines highlights the effectiveness of its context retrieval strategy.
\end{enumerate}

\subsection{Analysis of Reasoning Paths}
In this section, we analyze the reasoning paths of test instances across all datasets, investigating the distribution of the number of reasoning steps and the token usage of these reasoning paths.

\subsubsection{Distribution of the Number of Reasoning Steps}

\begin{figure*}[!t]
    \centering
    \subfigure{
    \includegraphics[width=0.47\textwidth]{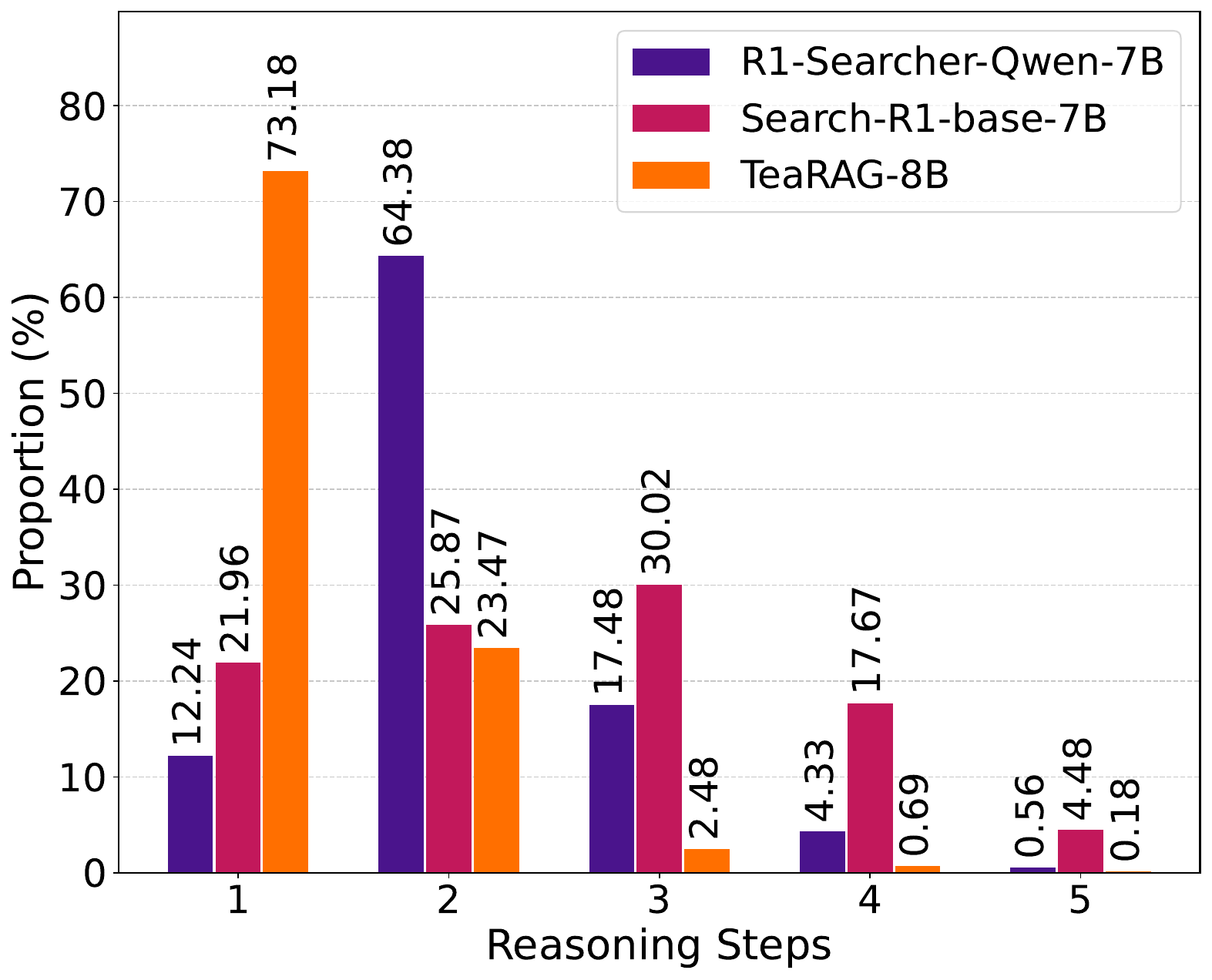}}
    \subfigure{
    \includegraphics[width=0.47\textwidth]{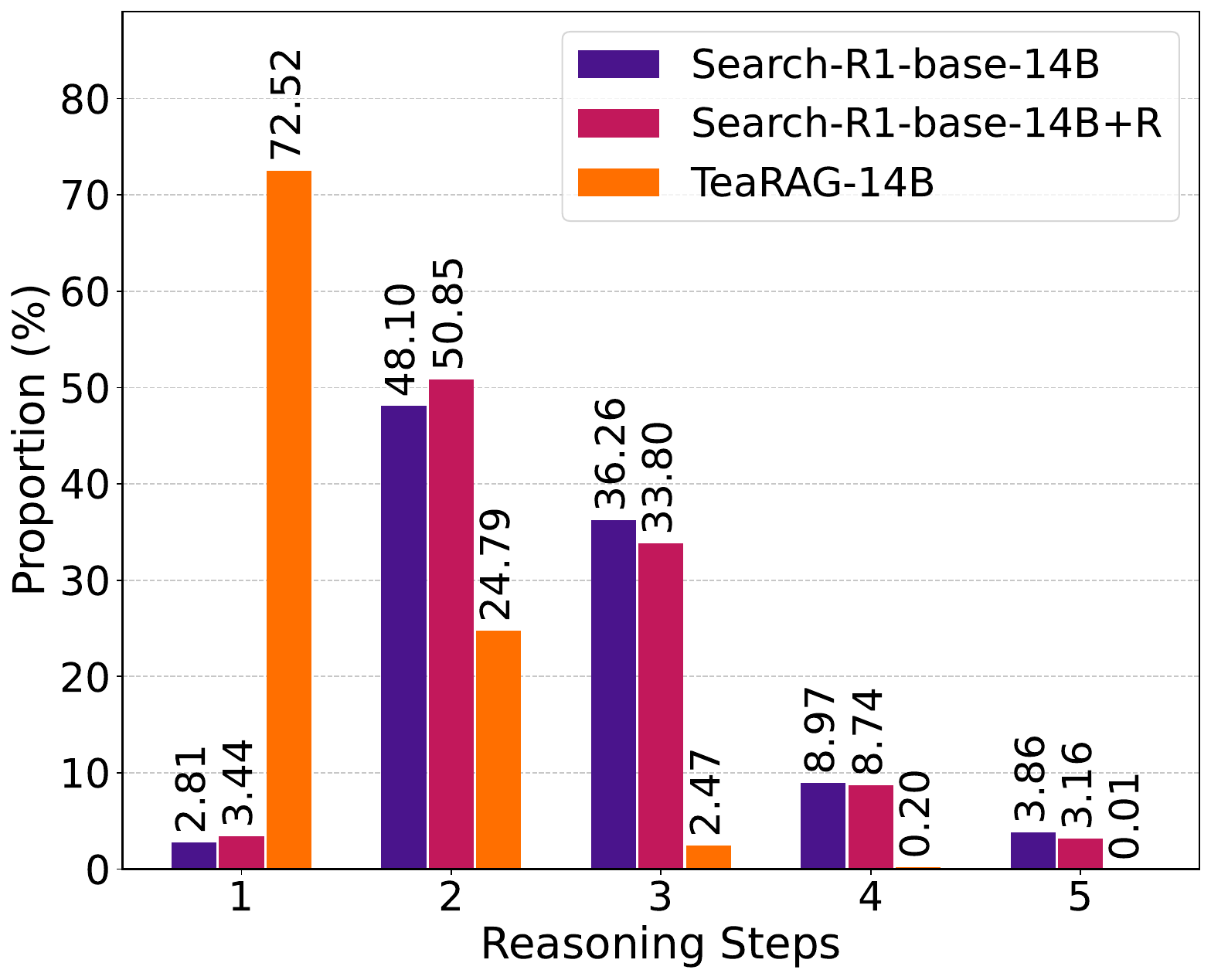}}
    \vspace{-2.1mm}
    \caption{Comparison of reasoning step distributions between \method and other baselines on six QA benchmarks. The figure on the left depicts the comparative results for \method-8B, while the figure on the right depicts those for \method-14B.}
    \label{fig:distribution}
\end{figure*} 

We present the detailed reasoning step distribution of \method and other similarly-scaled agentic RAG baselines in Fig.~\ref{fig:distribution}.
Since most methods require at least one reasoning step, we show the distribution of reasoning steps ranging from $1$ to $5$.
We obtain the following observations:
\begin{enumerate}[leftmargin=*, topsep=0pt, itemsep=0pt] 
\item \textbf{\method requires shorter reasoning steps compared to agentic RAG baseline methods.} This indicates that by incorporating process supervision signals, we effectively guide the LLM's reasoning process, reducing indirection and improving overall efficiency.

\item \textbf{Stronger information retrieval capabilities slightly improve reasoning efficiency.} For instance, by enhancing retrieval capabilities in the reranking stage, Search-R1-base-14B+R shows an increase in the proportion of reasoning completed in one or two steps to $54.3\%$, up from $50.9\%$ for Search-R1-base-14B. This finding indirectly validates the effectiveness of \method's retrieval strategy, which leverages KAG construction to improve information accuracy and prevent repetitive searches that occur when critical information is not retrieved.

\item \textbf{The distribution of reasoning path lengths for \method is stable across different base models.} In contrast, for the Search-R1 baseline, the proportion of one-step reasoning drops from $22.0\%$ with Search-R1-base-7B+R to just $3.4\%$ with Search-R1-base-14B+R. This suggests that the Search-R1 algorithm is highly sensitive to model scale. Conversely, the reasoning path lengths of \method remain remarkably stable across both 8B and 14B model scales, demonstrating our method's robustness and generalizability. 

\end{enumerate}

\begin{figure*}[!t]
    \centering
    \subfigure{
    \includegraphics[width=0.47\textwidth]{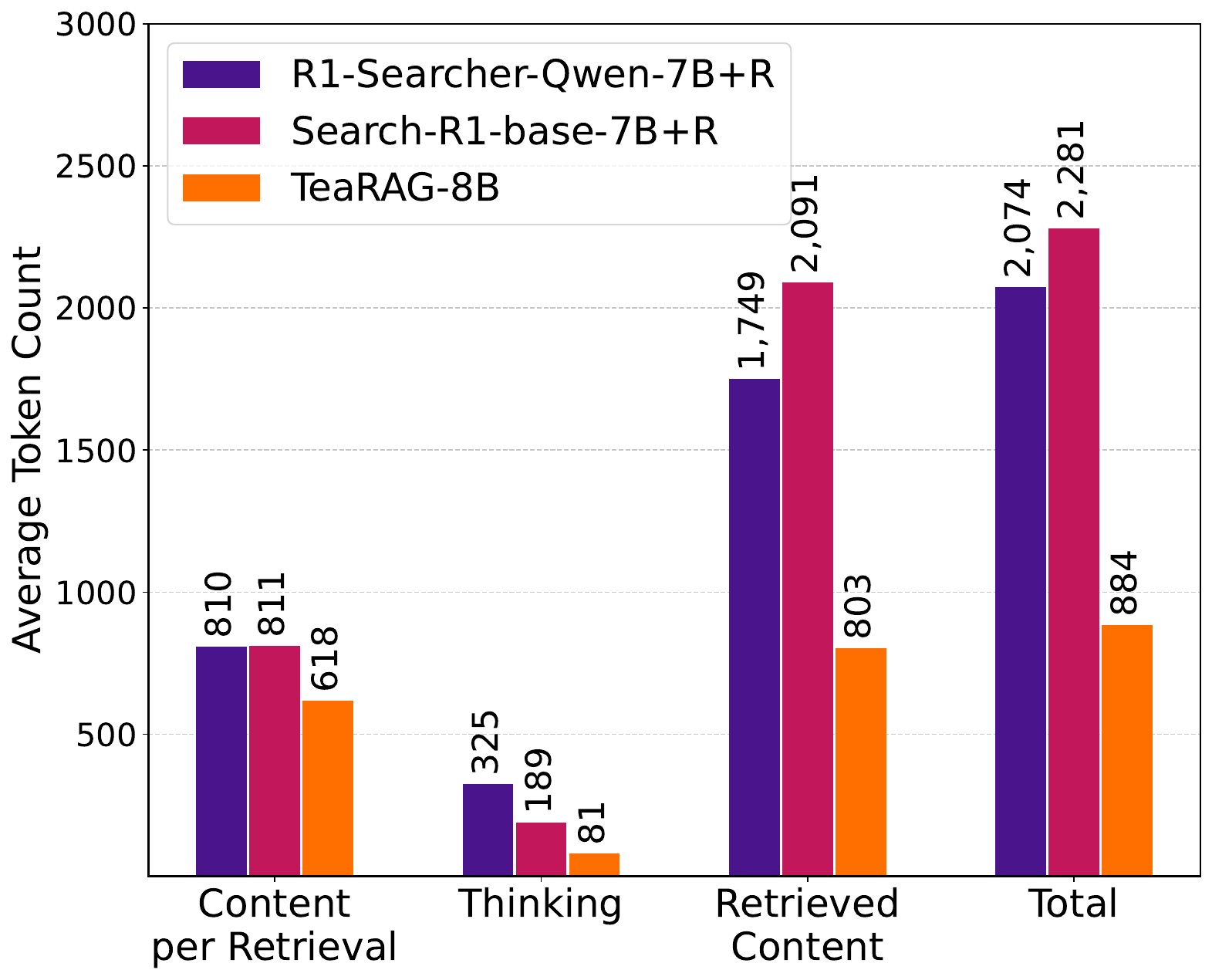}}
    \subfigure{
    \includegraphics[width=0.47\textwidth]{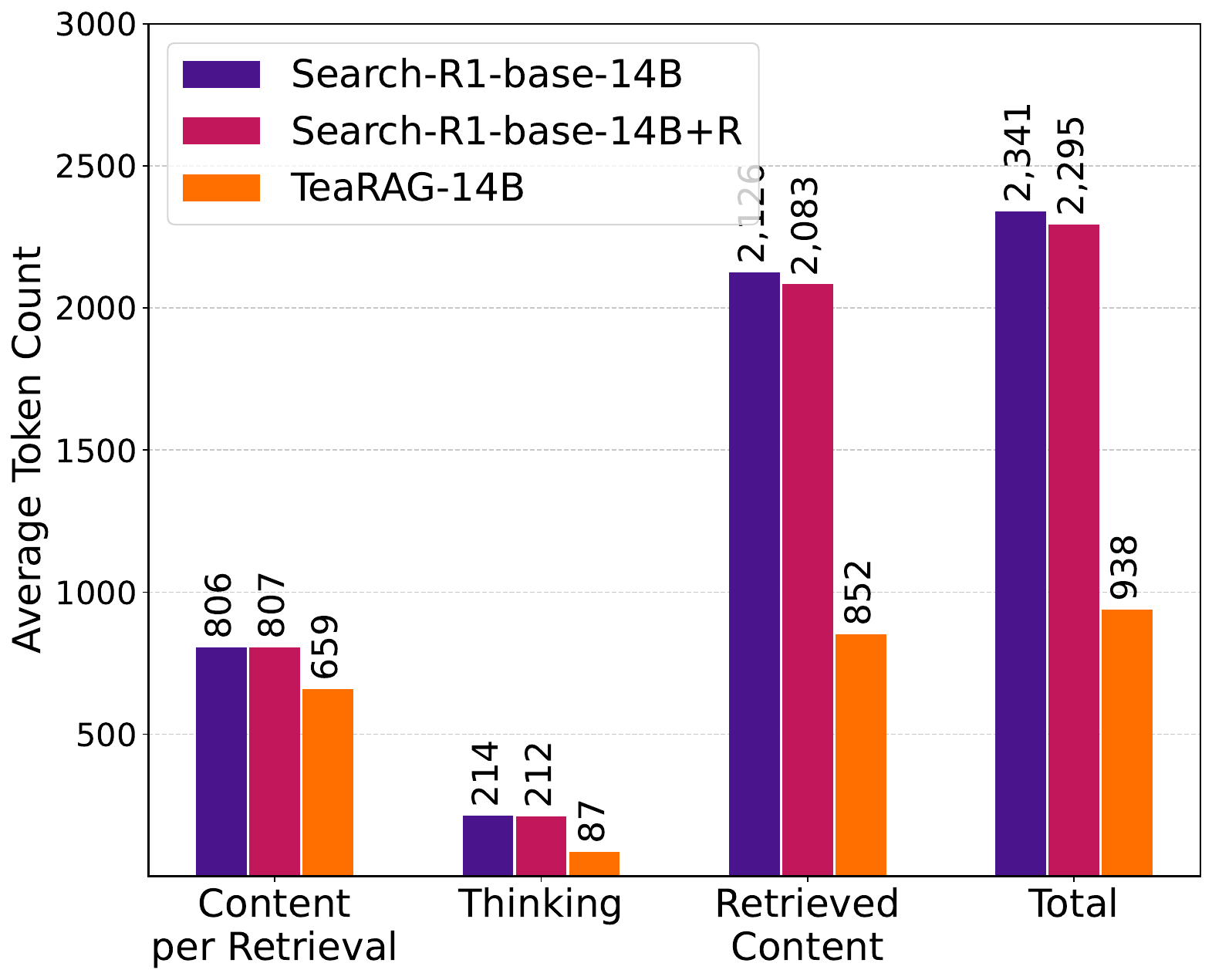}}
    \vspace{-2.1mm}
    \caption{Comparison of output token usage  between \method and other baselines on six QA benchmarks. The figure on the left depicts the comparative results for \method-8B, while the figure on the right depicts those for \method-14B.}
    \label{fig:token}
\end{figure*} 

\subsubsection{Analysis of Token Usage of Reasoning Paths}

In agentic RAG, the input prompt serves as static instructions, whereas the output contains the model's complete reasoning paths. Since input prompts are typically brief and fixed static instructions, they cannot adequately reflect the dynamic reasoning processes of LLMs. This is evidenced by the relatively modest instruction prompt token counts observed across agentic RAG methods, namely R1-Searcher-Qwen-7B (169), Search-R1-base-7B (119), TeaRAG-8B (136), Search-R1-base-14B (119), and TeaRAG-14B (134). Therefore, the analysis in this section focuses on the reasoning paths in the output tokens.

To analyze the token usage of the LLM's reasoning path more clearly, we measure four key metrics. (1) \textbf{Thinking tokens}, which are used for the LLM's thought processes like planning, problem decomposition, and summarization. (2) \textbf{Retrieved Content tokens}, which represent the token count for all externally retrieved information in the reasoning path. (3) \textbf{Total tokens}, which denote the tokens for the complete reasoning path including both Thinking tokens and Retrieved Content tokens. (4) \textbf{Content per Retrieval}, which is the average number of content tokens input to the LLM per retrieval. The results are shown in Fig.~\ref{fig:token}. 

Based on these results, we can draw the following conclusions:
\begin{enumerate}[leftmargin=*, topsep=0pt, itemsep=0pt] 
\item \textbf{\method demonstrates superior token efficiency compared to the other baselines across all four metrics.} This is primarily because introducing signals from process supervision enhances the LLM's reasoning efficiency, which in turn reduces ineffective reasoning and redundant retrievals, thereby significantly improving token efficiency. Furthermore, our adoption of the KAG retrieval method also reduces the number of tokens required per retrieval.

\item \textbf{\method's retrieval method effectively reduces the number of external content tokens per retrieval.} This is achieved by PPR to replace redundant and irrelevant chunks with high-information-density triplets. As a result, \method-8B reduces the average tokens per retrieval by $23.80\%$ compared to Search-R1-base-7B+R.

\end{enumerate}

\subsection{Ablation Study}

\subsubsection{Effect of Context Retrieval Methods}

We conduct ablation experiments on the context retrieval methods. We primarily focus on two core retrieval paradigms: iterative and hybrid search. To further understand the role of these paradigms, we design the following context retrieval variants:
\begin{itemize}[leftmargin=*, topsep=0pt, itemsep=0pt]  
\item \textbf{Single-G} performs single-round graph retrieval to obtain knowledge triplets pertinent to the query, which are subsequently provided to an LLM for answer generation.  
\item \textbf{Single-S} conducts single-round semantic retrieval, then feeds chunks to an LLM for output.
\item \textbf{Single-Con} adopts a single-round hybrid strategy combining semantic and graph retrieval without KAG construction and PPR filtering. The top-5 chunks and top-5 triplets are concatenated and passed to an LLM for answer generation.  
\item \textbf{Single-PPR} applies a hybrid retrieval approach incorporating semantic and graph retrieval. A KAG is constructed and refined using PPR, after which the top-5 ranked chunks and triplets are input to an LLM for answer generation.  
\item \textbf{\method-G} employs iterative question decomposition via an LLM, retrieving relevant knowledge triplets exclusively through graph retrieval until the final answer is produced.  
\item \textbf{\method-S} employs iterative question decomposition via an LLM, retrieving relevant chunks solely through semantic retrieval until the final answer is produced.  
\item \textbf{\method-Con} employs a hybrid retrieval approach combining semantic and graph retrieval. For each step, it directly concatenates the top-5 chunks and top-5 triplets without KAG construction and PPR filtering to form the context. An LLM then iteratively invokes this hybrid retrieval and decomposes the question until the final answer is obtained. 
\end{itemize}

It is worth noting that the Single-Con and \method-Con methods use $10$ information units per retrieval, while all other methods use $5$. We conduct an evaluation of context retrieval methods on two base models, with the results presented in Table~\ref{tab:retrieval}. From this, we derive the following observations:
\begin{enumerate}[leftmargin=*, topsep=0pt, itemsep=0pt]
    \item \textbf{Agentic iterative retrieval excels by decomposing complex problems.} This shows that by breaking down complex problems into simpler ones and solving them step by step, the retrieval system can better focus its information retrieval on a specific problem.
    
    \item \textbf{Hybrid retrieval achieves superior performance compared to using a single retrieval method.} This indicates that hybrid retrieval leverages the strengths of both semantic and graph retrieval. The fusion of content retrieved by the two methods further enhances the LLM's ability to answer questions, especially for non-iterative methods.

    \item \textbf{\method constructs a KAG and applies PPR-based filtering to increase the information density of retrieved content while preserving strong performance.} This is because PPR's co-occurrence-based mechanism removes irrelevant or verbose chunks and thereby further enhances accuracy. In contrast, although \method-Con utilizes hybrid retrieval, its direct concatenation of chunks and triplets increases input length and introduces extraneous noise.

    \item \textbf{Semantic retrieval outperforms graph retrieval by preserving information integrity.} This is primarily because important information within the chunks can be lost during graph construction. Additionally, the triplets' lack of sufficient context can lead to ambiguity and increase the difficulty for the LLM to comprehend.
    
\end{enumerate}

\begin{table*}
\centering
\caption{Performance comparison of context retrieval methods on QA benchmarks. CpR represents the average number of content tokens per retrieval. The best results are highlighted in bold.}
\label{tab:retrieval}
\renewcommand\arraystretch{1.5}
\resizebox{\textwidth}{!}{
\begin{tabular}{l|l|cc|c|cccc|cccccccc|cc}
\hline
\multirow{2}{*}{\textbf{LLM}}&\multirow{2}{*}{\textbf{Method}}&\multirow{2}{*}{\textbf{Retrieval}} & \multirow{2}{*}{\textbf{Iterative}} & \multirow{2}{*}{\textbf{CpR}} & \multicolumn{2}{c}{\textbf{NQ}} & \multicolumn{2}{c|}{\textbf{PopQA}} & \multicolumn{2}{c}{\textbf{HotpotQA}}& \multicolumn{2}{c}{\textbf{2Wiki}}& \multicolumn{2}{c}{\textbf{Musique}} & \multicolumn{2}{c|}{\textbf{Bamboogle}} & \multicolumn{2}{c}{\textbf{Avg.}} \\
\cmidrule(lr){6-7} \cmidrule(lr){8-9} \cmidrule(lr){10-11} \cmidrule(lr){12-13} \cmidrule(lr){14-15} \cmidrule(lr){16-17} \cmidrule(lr){18-19}
& & & & & \textbf{EM} & \textbf{F1} & \textbf{EM} & \textbf{F1}& \textbf{EM} & \textbf{F1} & \textbf{EM} & \textbf{F1} & \textbf{EM} & \textbf{F1} & \textbf{EM} & \textbf{F1} & \textbf{EM} & \textbf{F1} \\
\hline
\multirow{8}{*}{\textbf{Llama3-8B}}& \textbf{Single-G}& \textbf{Graph} & \textbf{\ding{56}} & 79 & 25.76 & 32.83 & 41.49 & 46.53 & 21.74 & 29.07 & 14.65 & 19.82 & 4.26 & 9.65 & 12.00 & 18.65 & 19.98 & 26.09 \\
& \textbf{Single-S}&\textbf{Semantic} & \textbf{\ding{56}} & 779 & 37.28 & 48.48 & 41.08 & 50.27 & 29.53 & 40.80 & 9.64 & 22.12 & 5.54 & 12.08 & 11.20 & 19.94 & 22.37 & 32.28 \\
& \textbf{Single-Con}& \textbf{Hybrid} & \textbf{\ding{56}} & 859 & 40.61 & 50.24 & 44.69 & 52.10 & 29.71 & 40.32 & 10.50 & 20.73 & 5.92 & 11.94 & 16.00 & 23.82 & 24.57 & 33.19   \\
& \textbf{Single-PPR}& \textbf{Hybrid} & \textbf{\ding{56}} & 650 & 38.89 & 49.52 & 44.72 & 52.84 & 31.42 & 41.69 & 12.59 & 20.58 & 7.65 & 13.72 & 16.80 & 24.06 & 25.35 & 33.73   \\
\cdashline{2-19}
& \textbf{\method-G}& \textbf{Graph} & \textbf{\ding{52}} & 88 & 34.27 & 43.20 & 45.78 & 49.94 & 31.71 & 42.87 & 40.35 & 46.04 & 15.85 & 25.46 & 35.20 & 44.15 & 33.86 & 41.94 \\
& \textbf{\method-S}& \textbf{Semantic} & \textbf{\ding{52}} & 790 & \textbf{50.36} & \textbf{59.98} & 51.66 & 55.85 & 46.47 & 59.25 & 46.88 & 53.47 & 26.44 & 36.55 & \textbf{47.20} & 59.32 & 44.84 & 54.07 \\
& \textbf{\method-Con}& \textbf{Hybrid} & \textbf{\ding{52}} & 879 & 50.08 & 59.79 & 51.97 & \textbf{56.25} & \textbf{46.66} & \textbf{59.68} & 47.38 & 53.88 & 26.73 & 37.11 & 45.60 & 58.92 & 44.74 & 54.27 \\
& \textbf{\method}& \textbf{Hybrid} & \textbf{\ding{52}} & 618 & 50.06 & 59.71 & \textbf{51.98} & 56.08 & 46.59 & 59.48 & \textbf{47.89} & \textbf{54.21} & \textbf{26.98} & \textbf{37.34} & \textbf{47.20} & \textbf{59.89} & \textbf{45.12} & \textbf{54.45}  \\
\hline
\multirow{8}{*}{\textbf{Qwen2.5-14B}}& \textbf{Single-G}& \textbf{Graph} & \textbf{\ding{56}} & 82 & 22.69 & 30.72 & 37.82 & 44.41 & 22.90 & 30.94 & 22.97 & 28.13 & 3.97 & 9.71 & 10.40 & 17.82 & 20.12 & 26.95  \\
& \textbf{Single-S}&\textbf{Semantic} & \textbf{\ding{56}} & 814 & 33.29 & 46.71 & 40.87 & 49.75 & 32.47 & 44.80 & 21.66 & 31.88 & 7.86 & 15.77 & 22.40 & 32.03 & 26.42 & 36.82 \\
& \textbf{Single-Con}& \textbf{Hybrid} & \textbf{\ding{56}} & 895 & 39.81 & 51.57 & 44.48 & 52.42 & 37.37 & 49.26 & 28.77 & 35.63 & 8.48 & 16.94 & 18.40 & 31.64 & 29.55 & 39.58 \\
& \textbf{Single-PPR}& \textbf{Hybrid} & \textbf{\ding{56}} & 678 & 40.91 & 52.95 & 46.29 & 53.48 & 37.84 & 49.68 & 29.87 & 36.57 & 9.10 & 17.42 & 22.40 & 34.18 & 31.07 & 40.71   \\
\cdashline{2-19}
& \textbf{\method-G}& \textbf{Graph} & \textbf{\ding{52}} & 90 & 35.84 & 44.64 & 46.58 & 50.62 & 34.27 & 45.33 & 40.67 & 46.13 & 16.22 & 27.61 & 38.40 & 52.11 & 35.33 & 44.41 \\
& \textbf{\method-S}& \textbf{Semantic} & \textbf{\ding{52}} & 823 & \textbf{51.30} & \textbf{60.98} & 53.66 & 58.09 & 50.43 & 63.41 & 51.77 & 57.64 & \textbf{28.22} & 39.37 & \textbf{50.40} & \textbf{65.30} & \textbf{47.63} & \textbf{57.47} \\
& \textbf{\method-Con}& \textbf{Hybrid} & \textbf{\ding{52}} & 902 & 50.33 & 60.31 & \textbf{54.13} & \textbf{58.65} & \textbf{50.71} & \textbf{63.73} & \textbf{52.36} & \textbf{58.13} & 27.68 & \textbf{39.41} & 48.00 & 63.99 & 47.20 & 57.37 \\
& \textbf{\method}& \textbf{Hybrid} & \textbf{\ding{52}} & 659 & 50.33 & 60.31 & 53.59 & 58.02 & 49.93 & 63.14 & 52.29 & 57.91 & 27.93 & 39.22 & \textbf{50.40} & 64.07 & 47.41 & 57.11  \\
\hline
\end{tabular}
}
\end{table*}

\begin{table*}
\centering
\caption{The impact of knowledge graph quality on performance in QA benchmarks. CpR represents the average number of content tokens per retrieval. The best results are highlighted in bold.}
\label{tab:KG_quality}
\renewcommand\arraystretch{1.5}
\resizebox{\textwidth}{!}{
\begin{tabular}{l|c|c|cccc|cccccccc|cc}
\hline
\multirow{2}{*}{\textbf{LLM}}&\multirow{2}{*}{\textbf{Pruning Ratio}} & \multirow{2}{*}{\textbf{CpR}} & \multicolumn{2}{c}{\textbf{NQ}} & \multicolumn{2}{c|}{\textbf{PopQA}} & \multicolumn{2}{c}{\textbf{HotpotQA}}& \multicolumn{2}{c}{\textbf{2Wiki}}& \multicolumn{2}{c}{\textbf{Musique}} & \multicolumn{2}{c|}{\textbf{Bamboogle}} & \multicolumn{2}{c}{\textbf{Avg.}} \\
\cmidrule(lr){4-5} \cmidrule(lr){6-7} \cmidrule(lr){8-9} \cmidrule(lr){10-11} \cmidrule(lr){12-13} \cmidrule(lr){14-15} \cmidrule(lr){16-17}
& & & \textbf{EM} & \textbf{F1} & \textbf{EM} & \textbf{F1}& \textbf{EM} & \textbf{F1} & \textbf{EM} & \textbf{F1} & \textbf{EM} & \textbf{F1} & \textbf{EM} & \textbf{F1} & \textbf{EM} & \textbf{F1} \\
\hline
\multirow{5}{*}{\textbf{Llama3-8B}}& \textbf{0\%} & 618 & 50.06 & 59.71 & 51.98 & 56.08 & 46.59 & 59.48 & \textbf{47.89} & \textbf{54.21} & \textbf{26.98} & \textbf{37.34} & 47.20 & 59.89 & 45.12 & \textbf{54.45}  \\
& \textbf{25\%} & 630 & 49.70 & 59.51 & 51.98 & 56.11 & \textbf{46.86} & \textbf{59.80} & 47.79 & 53.98 & 26.81 & 37.23 & 48.00 & 59.89 & \textbf{45.19} & 54.42 \\
& \textbf{50\%} & 649 & 49.78 & 59.45 & \textbf{52.02} & \textbf{56.18} & 46.37 & 59.37 & 47.42 & 53.69 & 26.60 & 37.10 & \textbf{48.80} & \textbf{60.36} & 45.16 & 54.36   \\
& \textbf{75\%} & 681 & 50.06 & 59.52 & 51.73 & 55.86 & 46.51 & 59.47 & 47.59 & 53.90 & 26.60 & 37.02 & 46.40 & 59.14 & 44.81 & 54.15   \\
& \textbf{100\%} & 790 & \textbf{50.36} & \textbf{59.98} & 51.66 & 55.85 & 46.47 & 59.25 & 46.88 & 53.47 & 26.44 & 36.55 & 47.20 & 59.32 & 44.84 & 54.07 \\
\hline
\multirow{5}{*}{\textbf{Qwen2.5-14B}}& \textbf{0\%} & 659 & 50.33 & 60.31 & 53.59 & 58.02 & 49.93 & 63.14 & \textbf{52.29} & 57.91 & 27.93 & 39.22 & 50.40 & 64.07 & 47.41 & 57.11 \\
& \textbf{25\%} & 660 & 50.47 & 60.50 & 53.45 & 57.88 & 50.06 & 63.24 & 52.19 & \textbf{57.96} & 27.55 & 39.09 & 50.40 & 64.26 & 47.35 & 57.16 \\
& \textbf{50\%} & 678 & 51.05 & 60.97 & 53.49 & 57.94 & 50.11 & 63.06 & 52.08 & 57.77 & 27.31 & 39.12 & 51.20 & 65.70 & 47.54 & 57.43   \\
& \textbf{75\%} & 710 & 51.00 & 60.76 & 53.26 & 57.71 & 50.40 & 63.29 & 51.94 & 57.81 & 27.97 & \textbf{39.57} & \textbf{52.00} & \textbf{65.80} & \textbf{47.76} & \textbf{57.49}   \\
& \textbf{100\%} & 823 & \textbf{51.30} & \textbf{60.98} & \textbf{53.66} & \textbf{58.09} & \textbf{50.43} & \textbf{63.41} & 51.77 & 57.64 & \textbf{28.22} & 39.37 & 50.40 & 65.30 & 47.63 & 57.47 \\
\hline
\end{tabular}
}
\end{table*}

\subsubsection{Effect of Knowledge Graph Quality}

To investigate the robustness of \method to knowledge graph quality, we simulate degradation by randomly pruning the triplet set $\mathcal{E}$ at varying ratios. The rationale is that extraction quality is inherently tied to LLM capability. Less capable models tend to produce sparser and less complete triplet sets, which can be approximated by progressively removing triplets from the original graph. A higher pruning ratio corresponds to a lower knowledge graph quality. The results are presented in Table~\ref{tab:KG_quality}, from which we make the following observations:

\begin{enumerate}[leftmargin=*, topsep=0pt, itemsep=0pt]
    \item \textbf{\method exhibits high robustness to knowledge graph degradation.} The PPR filtering mechanism effectively mitigates the impact of distracting noise from low-quality graphs, precluding the integration of irrelevant triplets into the LLM.

    \item \textbf{Lower knowledge graph quality correlates with an increased average token count per retrieval.} As triplet quality diminishes, the PPR mechanism filters out more graph-based results, leading to a higher retention of chunks. This confirms that \method dynamically prioritizes higher-quality retrieved content to maintain information density.

\end{enumerate}

\begin{table*}
\centering
\caption{Performance comparison of models trained with different rewards. \# Pairs indicates the number of preference pairs used for training. \# Steps indicates the number of reasoning path steps. The best results are highlighted in bold.}
\label{tab:process_reward}
\renewcommand\arraystretch{1.5}
\resizebox{\textwidth}{!}{
\begin{tabular}{c|ccc|c|c|cccc|cccccccc|cc}
\hline
\multirow{2}{*}{\textbf{Iteration}}&\multirow{2}{*}{\textbf{Outcome}}&\multirow{2}{*}{\textbf{Format}} &\multirow{2}{*}{\textbf{Process}} & \multirow{2}{*}{\textbf{\# Pairs}} & \multirow{2}{*}{\textbf{\# Steps}} & \multicolumn{2}{c}{\textbf{NQ}} & \multicolumn{2}{c|}{\textbf{PopQA}} & \multicolumn{2}{c}{\textbf{HotpotQA}}& \multicolumn{2}{c}{\textbf{2Wiki}}& \multicolumn{2}{c}{\textbf{Musique}}&\multicolumn{2}{c|}{\textbf{ Bamboogle}}&\multicolumn{2}{c}{\textbf{ Avg.}} \\
\cmidrule(lr){7-8} \cmidrule(lr){9-10} \cmidrule(lr){11-12} \cmidrule(lr){13-14} \cmidrule(lr){15-16} \cmidrule(lr){17-18} \cmidrule(lr){19-20}
& & & & & & \textbf{EM} & \textbf{F1} & \textbf{EM} & \textbf{F1}& \textbf{EM} & \textbf{F1} & \textbf{EM} & \textbf{F1} & \textbf{EM} & \textbf{F1} & \textbf{EM} & \textbf{F1} & \textbf{EM} & \textbf{F1} \\
\hline
\multirow{1}{*}{\textbf{SFT}} & \textbf{\ding{56}} & \textbf{\ding{56}} & \textbf{\ding{56}} & - & 1.47 & 38.56 & 47.46 & 46.56 & 50.67 & 28.86 & 38.23 & 26.39 & 31.99 & 18.99 & 27.57 & 36.00 & 47.96 & 32.56 & 40.65 \\
\hline
\multirow{3}{*}{\textbf{1}} & \textbf{\ding{52}} & \textbf{\ding{56}} & \textbf{\ding{56}} & 8,886 & 1.48 & 48.39 & 57.57 & 50.92 & 55.37 & 43.97 & 55.82 & 45.52 & 51.93 & 24.33 & 33.69 & \textbf{44.00} & 56.91 & 42.85 & 51.88 \\
& \textbf{\ding{52}} & \textbf{\ding{52}} & \textbf{\ding{56}} & 9,198 & 1.48 & 48.42 & 57.49 & 51.22 & 55.43 & 43.58 & 55.79 & \textbf{46.09} & \textbf{51.97} & 23.91 & 33.30 & \textbf{44.00} & \textbf{58.23} & 42.87 & 52.03 \\
& \textbf{\ding{52}} & \textbf{\ding{52}} & \textbf{\ding{52}} & 8,760 & 1.38 & \textbf{49.22} & \textbf{58.13} & \textbf{51.97} & \textbf{56.27} & \textbf{44.02} & \textbf{56.25} & 45.28 & 51.34 & \textbf{25.69} & \textbf{35.22} & 41.60 & 56.18 & \textbf{42.96} & \textbf{52.23} \\
\hline
\multirow{3}{*}{\textbf{2}} & \textbf{\ding{52}} & \textbf{\ding{56}} & \textbf{\ding{56}} & 5,024 & 1.57 & 42.11 & 49.42 & 46.30 & 49.76 & 39.47 & 49.18 & 37.76 & 40.86 & 17.91 & 22.62 & 39.20 & 46.37 & 37.12 & 43.03 \\
& \textbf{\ding{52}} & \textbf{\ding{52}} & \textbf{\ding{56}} & 4,576 & 1.76 & 48.64 & 57.94 & 50.75 & 55.02 & 45.25 & 57.19 & 44.78 & 49.99 & 23.13 & 32.51 & \textbf{47.20} & \textbf{59.07} & 43.29 & 51.95 \\
& \textbf{\ding{52}} & \textbf{\ding{52}} & \textbf{\ding{52}} & 4,760 & 1.35 & \textbf{48.89} & \textbf{58.98} & \textbf{51.28} & \textbf{55.84} & \textbf{45.62} & \textbf{58.88} & \textbf{46.72} & \textbf{53.48} & \textbf{26.02} & \textbf{36.69} & 43.20 & 54.58 & \textbf{43.62} & \textbf{53.07}  \\
\hline
\multirow{2}{*}{\textbf{3}} & \textbf{\ding{52}} & \textbf{\ding{52}} & \textbf{\ding{56}} & 3,832 & 2.13 & 48.42 & 57.61 & 50.19 & 54.30 & 39.91 & 49.61 & 39.42 & 43.14 & 18.82 & 25.28 & 41.60 & 50.45 & 39.73 & 46.73  \\
& \textbf{\ding{52}} & \textbf{\ding{52}} & \textbf{\ding{52}} & 2,941 & 1.31 & \textbf{50.06} & \textbf{59.71} & \textbf{51.98} & \textbf{56.08} & \textbf{46.59} & \textbf{59.48} & \textbf{47.89} & \textbf{54.21} & \textbf{26.98} & \textbf{37.34} & \textbf{47.20} & \textbf{59.89} & \textbf{45.12} & \textbf{54.45}  \\
\hline
\end{tabular}
}
\end{table*}

\subsubsection{Effect of Process Rewards}

To analyze the advantages of process rewards, we perform an ablation study specifically on process rewards. We validate our approach on Llama3-8B-Instruct, considering three distinct schemes: using only \textbf{Outcome} reward, using \textbf{Outcome-Format} reward, and using \textbf{Outcome-Format-Process} reward. We train models with each reward type for multiple rounds. The results are shown in Table~\ref{tab:process_reward}.
We observe the following phenomena: 
\begin{enumerate}[leftmargin=*, topsep=0pt, itemsep=0pt] 

\item \textbf{Process rewards effectively enhance model performance.} By providing rewards for each intermediate step of the reasoning path, the LLM can learn from high-quality reasoning examples.

\item \textbf{Process rewards reduce reasoning steps and increase reasoning efficiency.} The reward mechanism penalizes unnecessary reasoning steps that are not instrumental to problem-solving. In contrast, models trained without process rewards exhibit a tendency to gradually increase their reasoning step count. This phenomenon is attributed to the lack of intermediate supervision, which can reinforce redundant reasoning paths that ultimately lead to the correct result.

\item \textbf{Process rewards make training more stable.} With only outcome rewards, performance collapses by the second training round. The model prefers correct answers via incorrectly formatted paths, undermining its ability to follow the intended reasoning structure.
Similarly, when using both outcome and format rewards, collapse occurs by the third round. The model still rewards correct answers produced through flawed logic. This failure mode is especially common in binary-choice comparison tasks, leading it to internalize misguided preferences for certain reasoning paths.
Process rewards prevent this by constructing preference pairs under stricter criteria, blocking these spurious preferences and stabilizing training.

\end{enumerate}

\subsubsection{Effect of IP-DPO Iterations}

\begin{figure}[t]
    \centering
    \includegraphics[width=\textwidth]{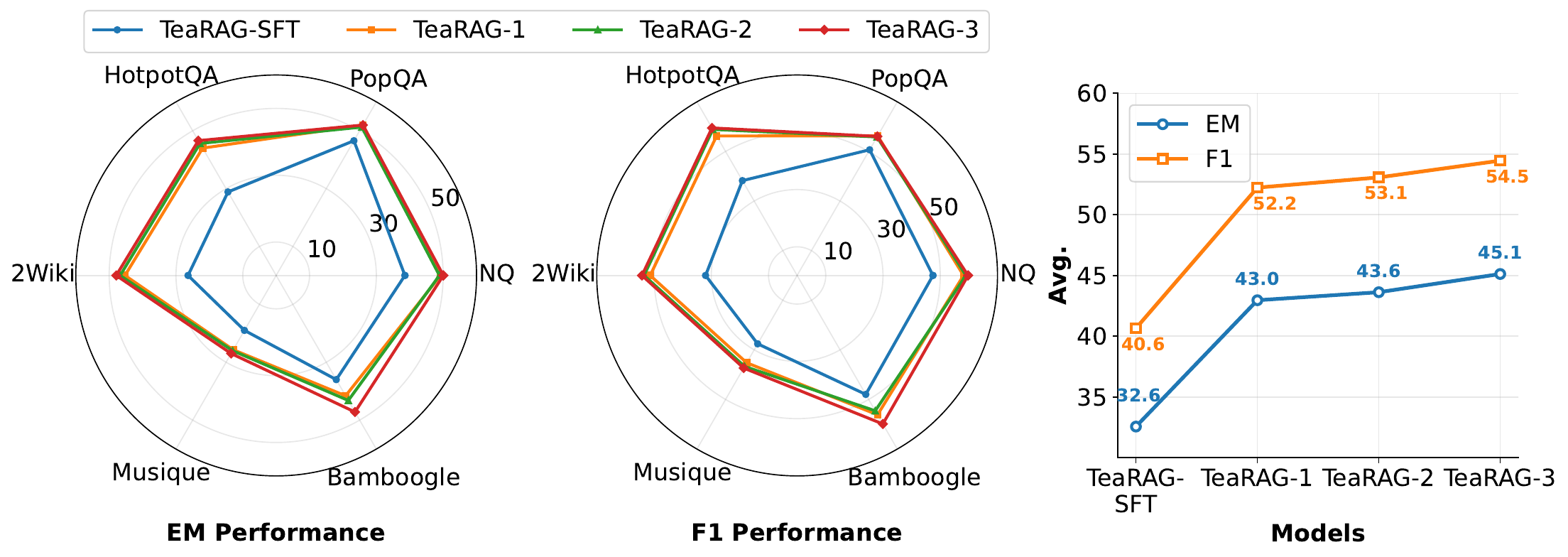}
    \caption{Performance of \method with Llama3-8B-Instruct across IP-DPO iterations. The left and middle figures show the EM and F1 scores of \method on six datasets, respectively. The right figure displays the average EM and F1 scores of \method across the six datasets.}
    \label{fig:dpo_iteration_llama}
\end{figure}

\begin{figure}[t]
    \centering
    \includegraphics[width=\textwidth]{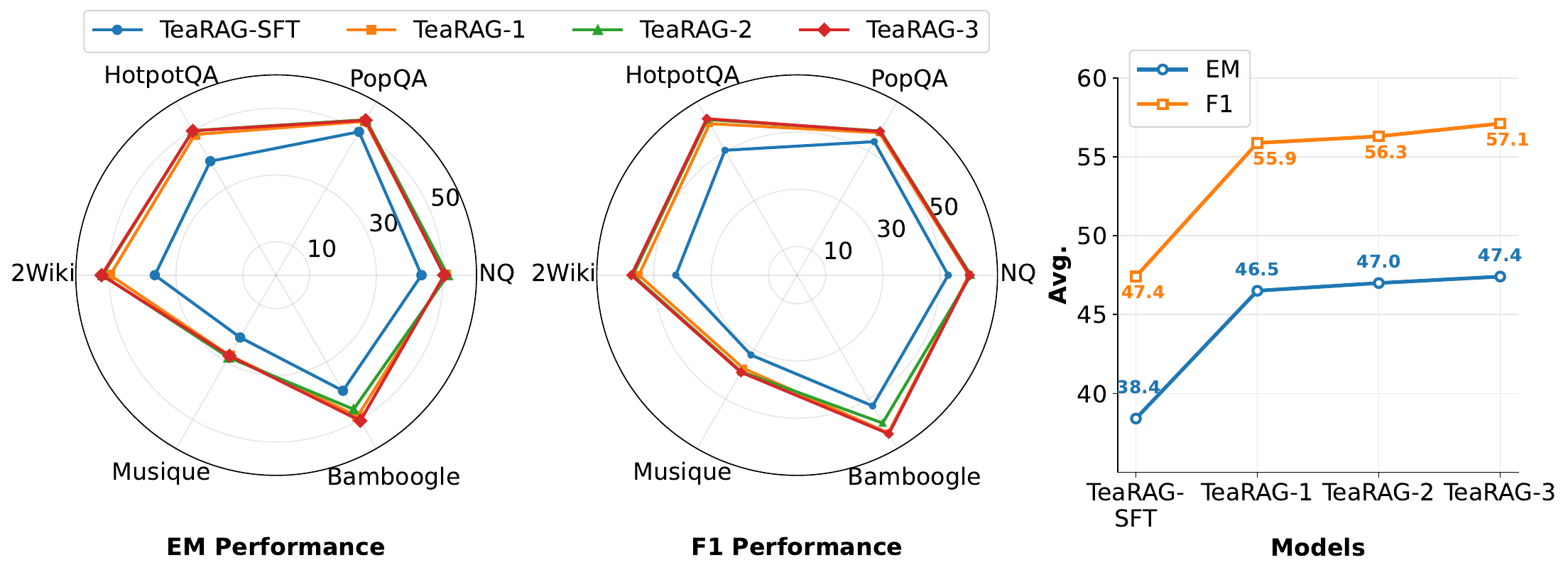}
    \caption{Performance of \method with Qwen2.5-14B-Instruct across IP-DPO iterations. The left and middle figures show the EM and F1 scores of \method on six datasets, respectively. The right figure displays the average EM and F1 scores of \method across the six datasets.}
    \label{fig:dpo_iteration_qwen}
\end{figure}

We investigate the impact of IP-DPO iterations by evaluating models across all training stages.
\method-SFT denotes the model trained only with SFT.
\method-1, \method-2, and \method-3 refer to the models after 1, 2, and 3 rounds of DPO training, respectively.
The results are presented in Figure~\ref{fig:dpo_iteration_llama} and Figure~\ref{fig:dpo_iteration_qwen}.
We can draw the following conclusions:
\begin{enumerate}[leftmargin=*, topsep=0pt, itemsep=0pt]
    \item \textbf{The performance of \method exhibits continuous enhancement as the number of IP-DPO rounds increases.} The underlying mechanism is that each round of optimization leverages the outputs from the model of the preceding round. This enables the model to sequentially master and refine its reasoning paradigms.
    
    \item \textbf{The performance gains from IP-DPO exhibit diminishing returns with additional rounds.} This occurs because the model's reasoning capabilities begin to converge, establishing a stable inference pattern. After many rounds of iteration, the model consistently produces correct responses for many questions, causing the preference data to provide a diminishing learning signal and thus limiting further improvement.
    
    \item \textbf{IP-DPO demonstrates consistent and stable improvements across models of varying scales and families.} This consistent success demonstrates IP-DPO's broad applicability for enhancing reasoning across diverse LLMs.
\end{enumerate}

\subsection{Hyper-Parameter Analysis}

\subsubsection{Number of Input Contents per Retrieval}

\begin{figure*}[t]
    \centering
    \includegraphics[width=\textwidth]{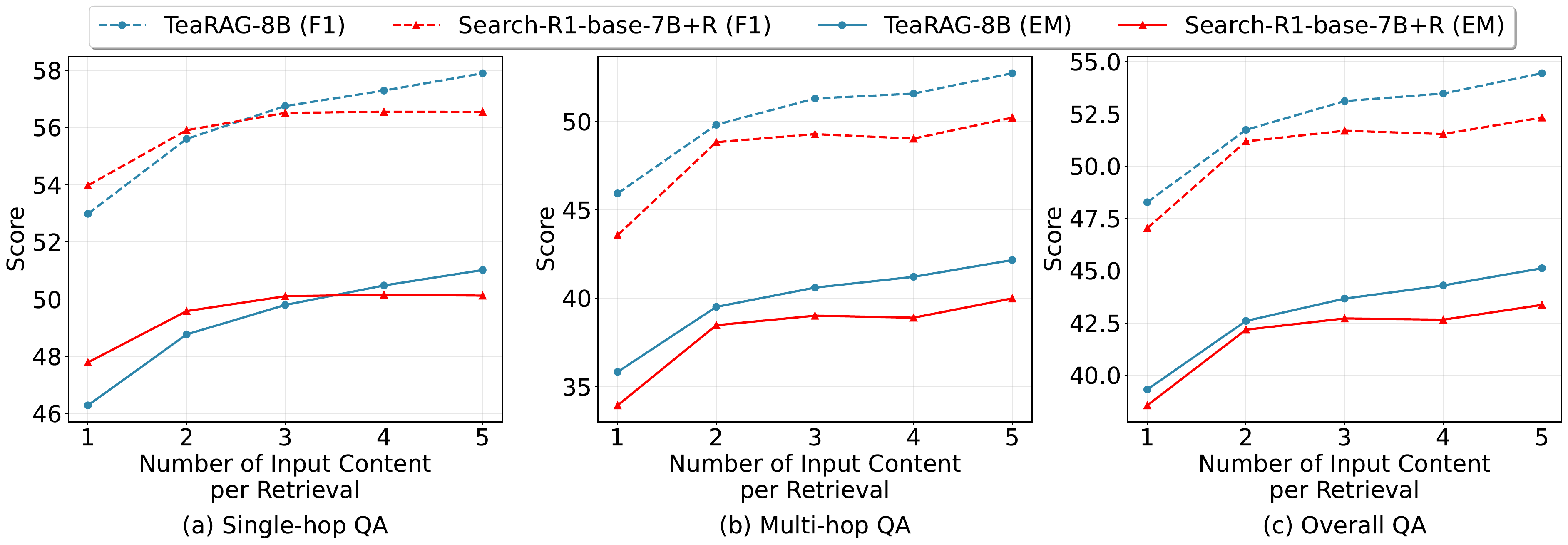}
    \caption{Comparative performance of \method-8B and Search-R1-base-7B+R across varying numbers of input content per retrieval.}
    \label{fig:8b_inputdoc}
\end{figure*} 

\begin{figure*}[t]
    \centering
    \includegraphics[width=\textwidth]{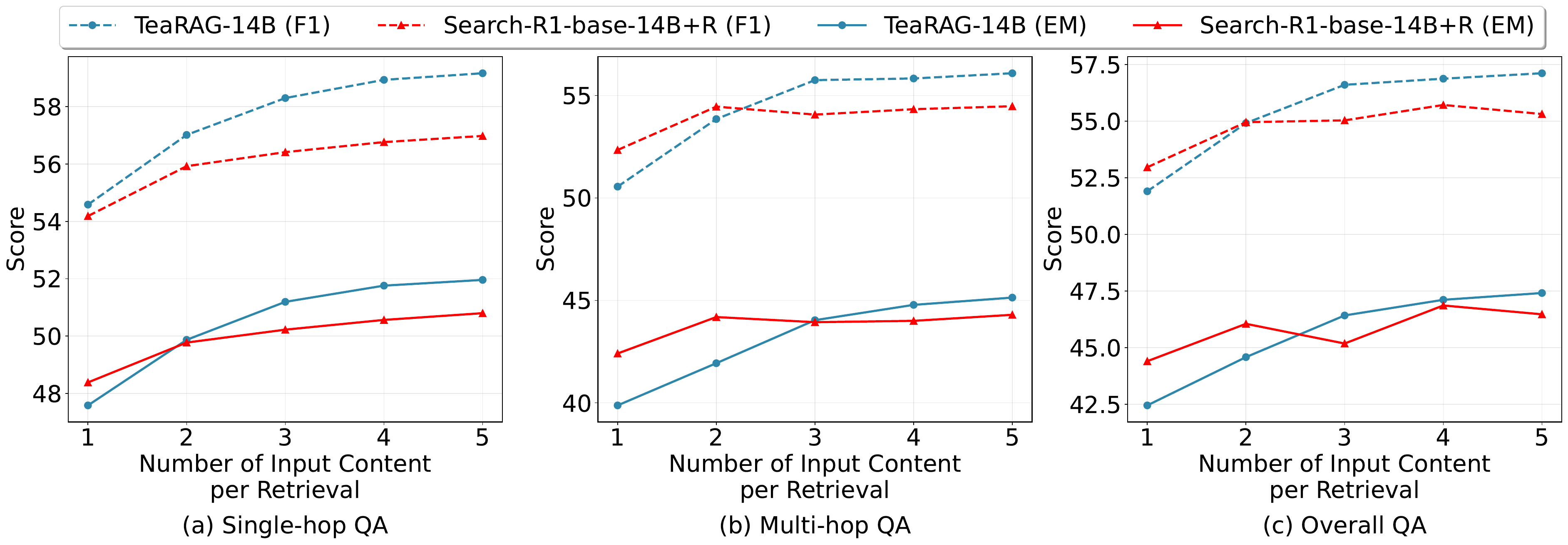}
    \caption{Comparative performance of \method-14B and Search-R1-base-14B+R across varying numbers of input content per retrieval.}
    \label{fig:14b_inputdoc}
\end{figure*} 

In this section, we investigate the impact of the number of input contents per retrieval on method performance. We compare the EM and F1 scores of \method-8B and Search-R1-base-7B+R, as well as \method-14B and Search-R1-base-14B+R.
The experimental results are shown in Fig.~\ref{fig:8b_inputdoc} and Fig.~\ref{fig:14b_inputdoc}.
We have the following observations:
\begin{enumerate}[leftmargin=*, topsep=0pt, itemsep=0pt]
    \item \textbf{More retrieved content improves performance, with \method consistently outperforming the baseline.} This indicates that \method is robust to variations in the number of input contents per retrieval and can maintain strong performance across different settings.
    
    \item \textbf{When the number of input contents per retrieval is small, \method cannot fully leverage the capabilities of LLMs. When the number is larger ($\ge 3$), it can achieve outstanding results.} This is mainly because \method’s training leverages PPR based on the co-occurrence mechanism. LLMs tend to focus on co-occurring data features, and when the number of input contents per retrieval is small, co-occurrence is difficult to achieve. This leads to a mismatch between inference and training, resulting in input distribution drift. When the number is sufficiently large, the information filtered by PPR has a higher information density, and at the same time, the LLM can exploit the co-occurrence in the data to achieve better performance.
    
    \item \textbf{\method is more scalable, with a more significant performance boost as the number of input contents per retrieval increases.} This is due to the construction of KAG and effective PPR filtering, which introduce triplets to increase information density and accuracy, thus reducing irrelevant content for each input. In contrast, the baseline method typically hits a performance bottleneck once the number of inputs reaches $3$, as traditional retrieval-and-rerank approaches still cannot avoid interference from irrelevant information in chunks.
\end{enumerate}

\subsubsection{Effects of Hyperparameter $\alpha$ in PPR}

\begin{figure*}[t]
\centering
\includegraphics[width=\textwidth]{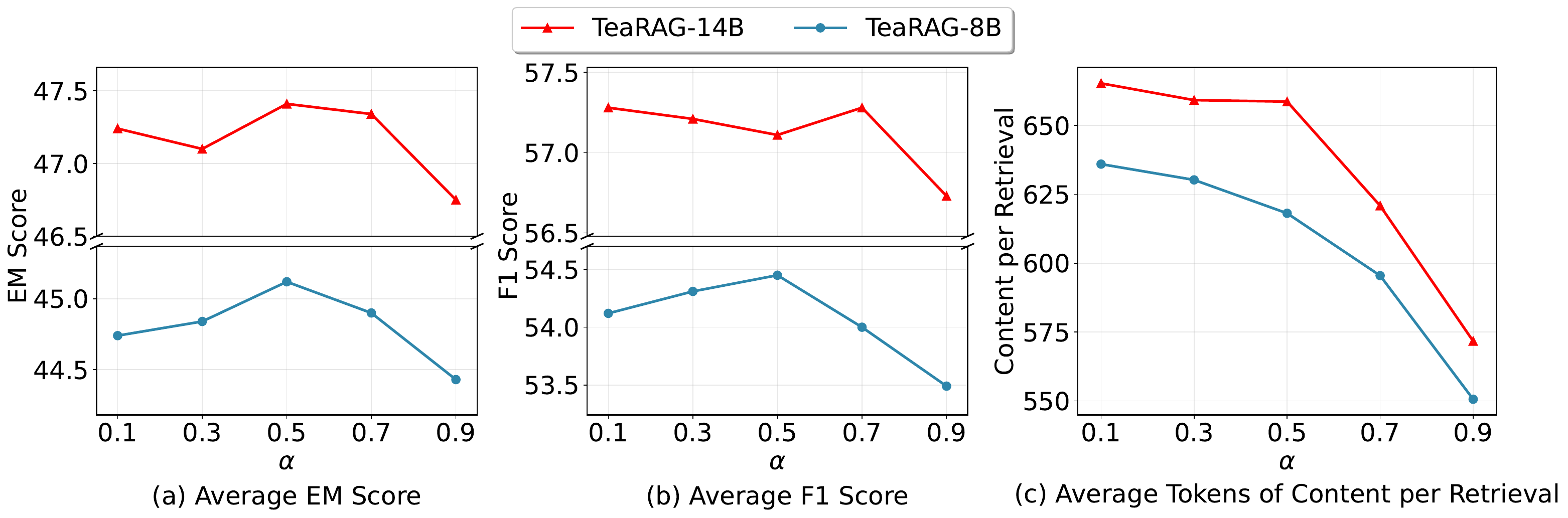}
\caption{Performance and the average number of content tokens per retrieval on six QA benchmarks for \method-8B and \method-14B across varying $\alpha$ in PPR.}
\label{fig:alpha_PPR}
\end{figure*}

In this section, we systematically investigate the influence of the PPR hyperparameter $\alpha$, which controls the focus of retrieved content. When $\alpha$ is small, PPR emphasizes the personalization vector, i.e., semantic information relevant to the query. When $\alpha$ is large, it prioritizes the structured information of co-occurrence links. The experimental results are shown in Fig.~\ref{fig:alpha_PPR}.
We have the following observations:
\begin{enumerate}[leftmargin=*, topsep=0pt, itemsep=0pt]
    \item \textbf{\method is robust to changes in $\alpha$.} The performance of \method varies little for $\alpha$ between $0.1$ and $0.7$. This robustness stems from KAG, which is constructed from the outputs of two retrieval methods, ensuring that its content is already highly relevant to the query.
    
    \item \textbf{\method achieves better performance by balancing query relevance and co-occurrence relations.} \method typically performs best when $\alpha$ is in the range $0.3$ to $0.7$. This indicates that PPR filtering effectively combines both factors to enhance information density and accuracy.

    \item \textbf{Greater emphasis on co-occurrence structure reduces the average number of content tokens per retrieval.} This is because, as co-occurrence structure receives more weight, information from triplet nodes connected to entities and chunks is more activated and ranked higher, thereby reducing the token count. This reveals a trade-off where despite fewer input tokens, missing important background leads to a significant performance drop, e.g., at $\alpha=0.9$.
\end{enumerate}

\subsubsection{Effects of Generation Temperature}

\begin{figure*}[!t]
    \centering
    \includegraphics[width=\textwidth]{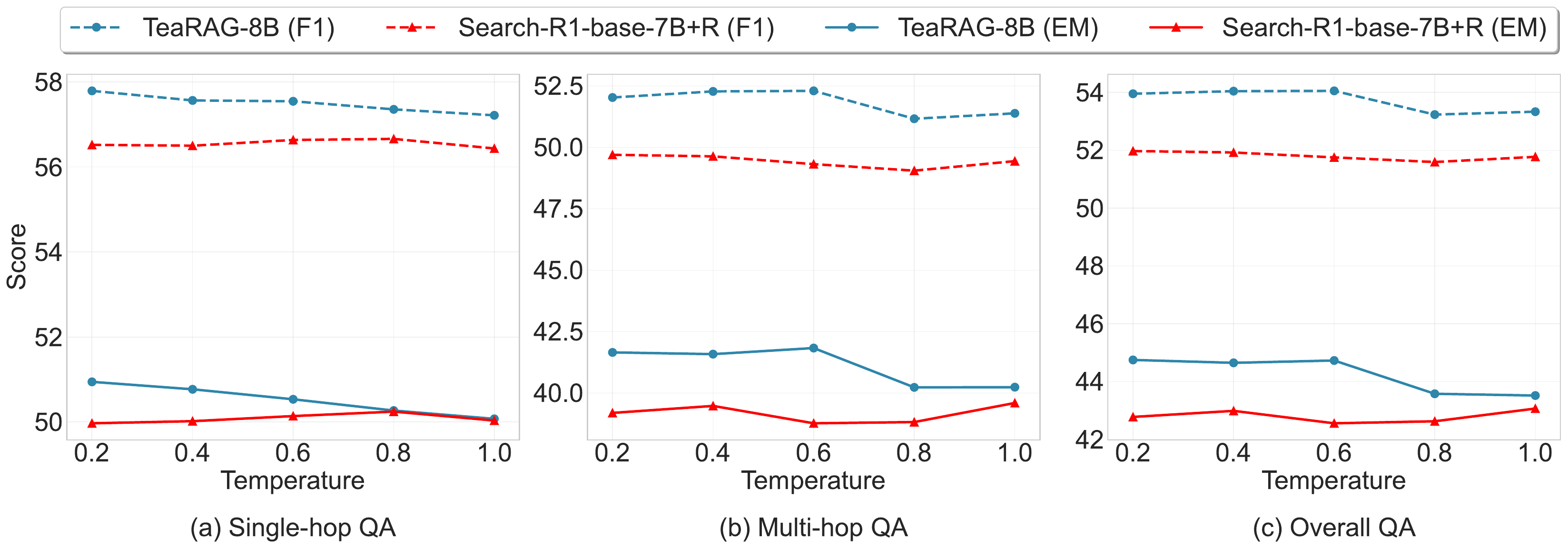}
    \caption{Performance of \method-8B with Search-R1-base-7B+R across generation temperatures.}
    \label{fig:8b_temp}
\end{figure*} 

 Because the sampling temperature has a pronounced impact on LLM performance~\cite{llmrl2025incorrect}, we vary the generation temperature for \method-8B and Search-R1-base-7B+R to probe robustness. The results are shown in Fig.~\ref{fig:8b_temp}.
We have the following observations:
\begin{enumerate}[leftmargin=*, topsep=0pt, itemsep=0pt]
    \item \textbf{Agentic RAG methods are not sensitive to temperature.} This is because agentic RAG not only requires the LLM to generate content, but also to ground its generation in retrieved external information, which reduces uncertainty.
    
    \item \textbf{\method-8B consistently outperforms Search-R1-base-7B+R across different temperatures, further demonstrating the effectiveness and robustness of \method.} We performed a two-sided paired t-test on the overall QA results, and $p<0.05$ indicates that our method is significantly better than Search-R1-base-7B+R.

    \item \textbf{As temperature increases, the performance of \method-8B exhibits a slight decline.} A likely reason is that higher temperature induces more diverse reasoning paths, which increases variance and can occasionally lead to deviations from the retrieved evidence or extra, unnecessary steps, thereby reducing answer accuracy and consistency.
    
\end{enumerate}

\subsection{Training and Inference Efficiency Analysis}

\begin{table}[t]
\caption{Training overhead across methods on 8 NVIDIA A100 (80G) GPUs.}
\label{tab:training}
\centering
\scalebox{0.56}{
\begin{tabular}{l|c|ccc|cc|c}
\Xhline{1.2pt}
\textbf{Method} & \textbf{Overall Training Time} & \textbf{Training Time} & \textbf{Inference Time} & \textbf{Reward Time} & \textbf{Training Steps} & \textbf{Time/Step} & \textbf{Memory Usage per GPU} \\
\Xhline{0.6pt}
\textbf{\method-8B} & 681m & 85m & 376m & 220m & 1,688 & 0.40m & 42G \\
\textbf{Search-R1-base-7B} & 2,944m & 1,320m & 1,624m & -& 1,005 & 2.93m & 79G \\
\cdashline{1-8}
\textbf{\method-14B} & 752m & 115m & 417m & 220m & 1,513 & 0.49m & 61G \\
\textbf{Search-R1-base-14B}& 4,958m  & 2,462m & 2,496m & - & 1,005 & 4.93m & 80G \\
\Xhline{1.2pt}
\end{tabular}}
\end{table}

In this section, we provide an in-depth analysis of the training and inference efficiency of \method. 
For the training overhead, the statistics are summarized in the Table~\ref{tab:training}. First, \method requires less total training time compared to Search-R1, primarily because the IP-DPO framework decouples the training and sampling stages, thereby enhancing overall training efficiency. Furthermore, \method achieves lower memory consumption through LoRA training and without reliance on auxiliary models such as PPO critics.

\begin{table}[t]
\caption{Inference efficiency on the 2WikiMultiHopQA dev dataset ($12,576$ questions).  }
\label{tab:inference}
\centering
\scalebox{0.65}{
\begin{tabular}{l|ccc|c|c|c}
\Xhline{1.2pt}
\textbf{Method} & \textbf{Semantic Retrieval} & \textbf{Graph Retrieval} & \textbf{KAG+PPR} & \textbf{Retrieval Time} & \textbf{Generation Time} & \textbf{Overall Time} \\
\Xhline{0.6pt}
\textbf{\method-8B (Sequential)} & 270s & 282s & 24s & 578s & 482s & 1,061s \\
\textbf{\method-8B (Parallel)} & 305s & 301s & 22s & 327s & 490s & 820s \\
\textbf{Search-R1-base-7B+R} & 601s & - & - & 601s &  1,641s &  2,243s \\
\cdashline{1-7}
\textbf{\method-14B (Sequential)} & 281s & 334s & 22s & 639s & 495s & 1,136s \\
\textbf{\method-14B (Parallel)} & 314s & 345s & 22s & 368s & 497s & 868s \\
\textbf{Search-R1-base-14B+R} & 525s & - & - & 525s & 2,655s & 3,181s \\
\Xhline{1.2pt}
\end{tabular}}
\end{table}

Furthermore, we evaluate \method's inference efficiency against Search-R1 on the 2WikiMultiHopQA dataset.
We implement two retrieval strategies for dual-path retrieval in \method-8B and \method-14B, namely sequential retrieval and parallel retrieval.
The inference time for each stage is presented in Table~\ref{tab:inference}.
Our analysis yields three key observations:
\begin{enumerate}[leftmargin=*, topsep=0pt, itemsep=0pt]
    \item \textbf{The overall inference time of \method is significantly lower than that of Search-R1.} This reduction can be attributed to the IP-DPO training method, which reduces the reasoning steps, thereby avoiding overthinking and redundant retrieval steps. Furthermore, the incorporation of knowledge triplets enhances information density and reduces token length. Consequently, the generation time is substantially reduced.
    
    \item \textbf{Parallel retrieval mitigates the latency overhead introduced by graph retrieval.} Since the semantic retrieval and graph retrieval processes are decoupled, they can be executed in parallel. Consequently, the overall retrieval latency is bounded by the maximum of the two individual branch latencies, rather than their sum. Parallel execution thereby introduces only marginal additional system-parallel overhead compared to a single-path retrieval baseline.

    \item \textbf{Both KAG construction and PPR filtering are efficient.} The two operations collectively account for a negligible fraction of the total inference latency, requiring on average only $1$ ms per reasoning step. This confirms that \method can efficiently leverage co-occurrence relationships to produce higher–information-density retrieval content.
    
\end{enumerate}

\begin{table*}[t]
\centering
\caption{Case comparison between \method-8B and Search-R1-base-7B+R.}
\label{tab:case_study}
\renewcommand\arraystretch{1}
\resizebox{0.85\textwidth}{!}{
\begin{tabular}{p{\textwidth}}
\hline
\textbf{Question:} Where was the place of death of Alexander Carl Otto Westphal's father? \\
\hline
\textbf{Ground Truth:} \textcolor{orange}{Kreuzlingen} \\
\hline
\textbf{\method-8B:} \\
\textit{Step 1:} \\
\textit{Important entity:} Alexander Carl Otto Westphal \\
\textit{Subquery:} Who was the father of Alexander Carl Otto Westphal? \\

\textcolor{red!80!black}{\textless{}Reference\textgreater{}}
\textcolor{blue}{Evidence 1:} Title: "Alexander Carl Otto Westphal" Text: number of significant scientists, including Otto Lowenstein (1889-1965), with whom he was co-author of "Experimentelle und klinische Studien zur Physiologie und Pathologie der Pupillenbewegungen, mit besonderer Berücksichtigung der Schizophrenie" ... \\

\textcolor{blue}{Evidence 2:} Title: "Alexander Carl Otto Westphal" Text: Alexander Carl Otto Westphal Alexander Carl Otto Westphal (18 May 1863, Berlin – 9 January 1941, Bonn) was a German neurologist and psychiatrist... \\
\textcolor{blue}{Evidence 3:} Triplet: Alexander Carl Otto Westphal father Carl Friedrich Otto Westphal \\
\textcolor{blue}{Evidence 4:} Triplet: Alexander Carl Otto Westphal son of Carl Friedrich Otto Westphal \\
\textcolor{blue}{Evidence 5:} Title: "Carl Friedrich Otto Westphal" Text: ... Carl Friedrich Otto Westphal Karl Friedrich Otto Westphal (23 March 1833, in Berlin – 27 January 1890, in \textcolor{orange}{Kreuzlingen}) was a German psychiatrist from Berlin. He was the son of Otto Carl Friedrich Westphal (1800–1879) and Karoline Friederike Heine and the father of Alexander Karl Otto Westphal (1863-1941)...
\textcolor{red!80!black}{\textless{}/Reference\textgreater{}}

\textit{Summary:} The fact Alexander Carl Otto Westphal's father Carl Friedrich Otto Westphal died in \textcolor{orange}{Kreuzlingen} logically implies the answer: \textcolor{orange}{Kreuzlingen}. \\
\textit{Final answer:} \textcolor{orange}{Kreuzlingen} \\
\hline
\textbf{Search-R1-base-7B+R:} \\
\texttt{<think>} I need to find the place of death of Alexander Carl Otto Westphal's father. I'll search for it. \texttt{</think>} \\
\texttt{<search>} Alexander Carl Otto Westphal's father \texttt{</search>} \\
\textcolor{red!80!black}{\texttt{<information>}} \textcolor{blue}{Doc 1} (Title: "Carl Friedrich Otto Westphal") Carl Friedrich Otto Westphal Karl Friedrich Otto Westphal (23 March 1833, in Berlin – 27 January 1890, in \textcolor{orange}{Kreuzlingen}) was a German psychiatrist from Berlin... \\
\textcolor{blue}{Doc 2} (Title: "Alexander Carl Otto Westphal") Alexander Carl Otto Westphal Alexander Carl Otto Westphal (18 May 1863, Berlin – 9 January 1941, Bonn) was a German neurologist and psychiatrist... \textcolor{blue}{[3 more docs...]} \textcolor{red!80!black}{\texttt{</information>}} \\

\texttt{<think>} I found out that Alexander Carl Otto Westphal's father is Carl Friedrich Otto Westphal. Now I need to find the place of death of his father, Otto Carl Friedrich Westphal. \texttt{</think>} \\

\texttt{<search>} place of death of Otto Carl Friedrich Westphal \texttt{</search>} \\

\textcolor{red!80!black}{\texttt{<information>}} \textcolor{blue}{Doc 1} (Title: "Carl Friedrich Otto Westphal") Carl Friedrich Otto Westphal Karl Friedrich Otto Westphal (23 March 1833, in Berlin – 27 January 1890, in \textcolor{orange}{Kreuzlingen}) was a German psychiatrist from Berlin. He was the son of Otto Carl Friedrich Westphal... 
\textcolor{blue}{[4 more docs...]}
\textcolor{red!80!black}{\texttt{</information>}} \\

\texttt{<think>} I found out that Otto Carl Friedrich Westphal died in Berlin. Now I need to find the place of death of his father, Otto Carl Friedrich Westphal. \texttt{</think>} \\

\texttt{<search>} place of death of Otto Carl Friedrich Westphal \texttt{</search>} \\

\textcolor{red!80!black}{\texttt{<information>}}
\textcolor{blue}{[5 docs with similar information...]}\textcolor{red!80!black}{\texttt{</information>}} \\
\texttt{<think>} I found out that Otto Carl Friedrich Westphal died in Berlin. Now I can provide the answer. \texttt{</think>} \\
\texttt{<answer>} Berlin \texttt{</answer>} \\
\hline
\end{tabular}
}
\end{table*}

Finally, we measure the GPU and main memory usage of the two retrieval methods. The semantic retriever requires $2.6$ GB of GPU memory and $5.8$ GB of main memory, whereas the graph retriever requires $2.6$ GB of GPU memory and $13.1$ GB of main memory. These results suggest that dual-path retrieval incurs only moderate resource overhead, which is acceptable given the reduction in the number of generated tokens and the resulting improvement in generation efficiency.

\subsection{Case Study}

In this section, we present a case study of \method-8B and Search-R1-base-7B+R to gain a clearer understanding of \method's advantages. As shown in Table~\ref{tab:case_study}, we observe two advantages of \method-8B. First, the content retrieved by \method-8B effectively increases information density by introducing knowledge triplets, and it further strengthens confidence in the correct information by leveraging the co-occurrence between chunks and triplets. In addition, \method-8B effectively reduces the number of iterations in the reasoning path, fully leveraging available information to arrive at the correct answer. In contrast, Search-R1-base-7B+R pulls in a large amount of document content, which distracts the model from effectively capturing key information to obtain the correct answer, and the model exhibits overthinking and redundant retrieval, leading to poor overall token efficiency.

\section{Conclusion and Future Work}
In this work, we explore a token-efficient agentic RAG framework that aims to improve the token utilization of reasoning paths while performing agentic RAG tasks. We observe that better token efficiency relies on increasing the information density of each retrieved content and reducing the number of reasoning iterations. To this end, we propose \method. \method provides a fully automated agentic pipeline, including important entity recognition, subquery generation, context retrieval, summary generation, and final answer generation. To raise the information density per retrieval step, we incorporate high-information-density knowledge triplets and adopt a hybrid retrieval strategy that combines semantic and graph retrieval. We further construct a KAG and apply co-occurrence-based PPR to preserve rich context and boost information density. To reduce the number of reasoning iterations, we introduce IP-DPO, which uses process-aware rewards to supervise the reasoning path, preventing overthinking and redundant retrieval. Extensive experiments demonstrate the effectiveness of our approach and its superior token efficiency.

To further broaden the framework's adaptability to domains lacking detailed evidence annotations, our future work will focus on extending process supervision beyond the reliance on human-annotated gold evidence. We plan to explore two directions to verify reasoning paths: (1) leveraging advanced LLMs as automated evaluators (LLM-as-a-Judge) to assess intermediate steps, and (2) training a specialized process reward model to provide step-by-step supervision.

\begin{acks}
This work was supported in part by the grants from National Science and Technology Major Project (No. 2023ZD0121104), and the Anhui Natural Science Foundation (No. 2508085ZD006).
Besides, this research was partially supported by National Natural Science Foundation of China (No.62502404), Hong Kong Research Grants Council (Research Impact Fund No.R1015-23, Collaborative Research Fund No.C1043-24GF, General Research Fund No. 11218325), Institute of Digital Medicine of City University of Hong Kong (No.9229503), Huawei (Huawei Innovation Research Program), Tencent (Tencent Rhino-Bird Focused Research Program, Tencent University Cooperation Project), Kuaishou (CCF-Kuaishou Large Model Explorer Fund No. 2025008, Kuaishou University Cooperation Project), Didi (CCF-Didi Gaia Scholars Research Fund), and Bytedance.
\end{acks}

\bibliographystyle{ACM-Reference-Format}
\bibliography{ref}

@inproceedings{jin2024flashrag,
  title={Flashrag: A modular toolkit for efficient retrieval-augmented generation research},
  author={Jin, Jiajie and Zhu, Yutao and Dou, Zhicheng and Dong, Guanting and Yang, Xinyu and Zhang, Chenghao and Zhao, Tong and Yang, Zhao and Wen, Ji-Rong},
  booktitle={Companion Proceedings of the ACM on Web Conference 2025},
  pages={737--740},
  year={2025}
}

@article{lewis2020retrieval,
  title={Retrieval-augmented generation for knowledge-intensive nlp tasks},
  author={Lewis, Patrick and Perez, Ethan and Piktus, Aleksandra and Petroni, Fabio and Karpukhin, Vladimir and Goyal, Naman and K{\"u}ttler, Heinrich and Lewis, Mike and Yih, Wen-tau and Rockt{\"a}schel, Tim and others},
  journal={NeurIPS},
  volume={33},
  pages={9459--9474},
  year={2020}
}

@inproceedings{bge_m3,
  title={M3-embedding: Multi-linguality, multi-functionality, multi-granularity text embeddings through self-knowledge distillation},
  author={Chen, Jianlyu and Xiao, Shitao and Zhang, Peitian and Luo, Kun and Lian, Defu and Liu, Zheng},
  booktitle={Findings of the association for computational linguistics: ACL 2024},
  pages={2318--2335},
  year={2024}
}

@article{wang2022text,
  title={Text embeddings by weakly-supervised contrastive pre-training},
  author={Wang, Liang and Yang, Nan and Huang, Xiaolong and Jiao, Binxing and Yang, Linjun and Jiang, Daxin and Majumder, Rangan and Wei, Furu},
  journal={arXiv preprint arXiv:2212.03533},
  year={2022}
}

@inproceedings{press2023measuring,
  title={Measuring and Narrowing the Compositionality Gap in Language Models},
  author={Press, Ofir and Zhang, Muru and Min, Sewon and Schmidt, Ludwig and Smith, Noah A and Lewis, Mike},
  booktitle={EMNLP}
}

@article{gao2023retrieval,
  title={Retrieval-augmented generation for large language models: A survey},
  author={Gao, Yunfan and Xiong, Yun and Gao, Xinyu and Jia, Kangxiang and Pan, Jinliu and Bi, Yuxi and Dai, Yi and Sun, Jiawei and Wang, Haofen and Wang, Haofen},
  journal={arXiv preprint arXiv:2312.10997},
  volume={2},
  year={2023}
}

@article{jiang2025s3,
  title={s3: You Don't Need That Much Data to Train a Search Agent via RL},
  author={Jiang, Pengcheng and Xu, Xueqiang and Lin, Jiacheng and Xiao, Jinfeng and Wang, Zifeng and Sun, Jimeng and Han, Jiawei},
  journal={arXiv preprint arXiv:2505.14146},
  year={2025}
}

@article{kalai2025language,
  title={Why language models hallucinate},
  author={Kalai, Adam Tauman and Nachum, Ofir and Vempala, Santosh S and Zhang, Edwin},
  journal={arXiv preprint arXiv:2509.04664},
  year={2025}
}

@article{gutierrez2025rag,
  title={From rag to memory: Non-parametric continual learning for large language models},
  author={Guti{\'e}rrez, Bernal Jim{\'e}nez and Shu, Yiheng and Qi, Weijian and Zhou, Sizhe and Su, Yu},
  journal={arXiv preprint arXiv:2502.14802},
  year={2025}
}

@inproceedings{jiang2023llmlingua,
  title={LLMLingua: Compressing Prompts for Accelerated Inference of Large Language Models},
  author={Jiang, Huiqiang and Wu, Qianhui and Lin, Chin-Yew and Yang, Yuqing and Qiu, Lili},
  booktitle={EMNLP}
}

@article{ma2024think,
  title={Think-on-graph 2.0: Deep and faithful large language model reasoning with knowledge-guided retrieval augmented generation},
  author={Ma, Shengjie and Xu, Chengjin and Jiang, Xuhui and Li, Muzhi and Qu, Huaren and Yang, Cehao and Mao, Jiaxin and Guo, Jian},
  journal={arXiv preprint arXiv:2407.10805},
  year={2024}
}

@inproceedings{sarmah2024hybridrag,
  title={Hybridrag: Integrating knowledge graphs and vector retrieval augmented generation for efficient information extraction},
  author={Sarmah, Bhaskarjit and Mehta, Dhagash and Hall, Benika and Rao, Rohan and Patel, Sunil and Pasquali, Stefano},
  booktitle={Proceedings of the 5th ACM International Conference on AI in Finance},
  pages={608--616},
  year={2024}
}

@article{wen2025hybridrag,
  title={HybridRAG-based LLM Agents for Low-Carbon Optimization in Low-Altitude Economy Networks},
  author={Wen, Jinbo and Su, Cheng and Kang, Jiawen and Nie, Jiangtian and Zhang, Yang and Tang, Jianhang and Niyato, Dusit and Yuen, Chau},
  journal={arXiv preprint arXiv:2506.15947},
  year={2025}
}

@inproceedings{wang2025richrag,
  title={RichRAG: Crafting Rich Responses for Multi-faceted Queries in Retrieval-Augmented Generation},
  author={Wang, Shuting and Yu, Xin and Wang, Mang and Chen, Weipeng and Zhu, Yutao and Dou, Zhicheng},
  booktitle={COLING},
  pages={11317--11333},
  year={2025}
}

@inproceedings{zhangevoking,
  title={Evoking User Memory: Personalizing LLM via Recollection-Familiarity Adaptive Retrieval},
  author={Zhang, Yingyi and Li, Junyi and Zhang, Wenlin and Jia, Pengyue and Li, Xianneng and Wang, Yichao and Xu, Derong and Wen, Yi and Guo, Huifeng and Liu, Yong and others},
  booktitle={The Fourteenth International Conference on Learning Representations}
}

@inproceedings{ma2023query,
  title={Query rewriting in retrieval-augmented large language models},
  author={Ma, Xinbei and Gong, Yeyun and He, Pengcheng and Zhao, Hai and Duan, Nan},
  booktitle={EMNLP},
  pages={5303--5315},
  year={2023}
}

@inproceedings{chenplan,
  title={Plan-on-Graph: Self-Correcting Adaptive Planning of Large Language Model on Knowledge Graphs},
  author={Chen, Liyi and Tong, Panrong and Jin, Zhongming and Sun, Ying and Ye, Jieping and Xiong, Hui},
  booktitle={NeurIPS}
}

@article{verma2024plan,
  title={Plan* rag: Efficient test-time planning for retrieval augmented generation},
  author={Verma, Prakhar and Midigeshi, Sukruta Prakash and Sinha, Gaurav and Solin, Arno and Natarajan, Nagarajan and Sharma, Amit},
  journal={arXiv preprint arXiv:2410.20753},
  year={2024}
}

@article{schulman2017proximal,
  title={Proximal policy optimization algorithms},
  author={Schulman, John and Wolski, Filip and Dhariwal, Prafulla and Radford, Alec and Klimov, Oleg},
  journal={arXiv preprint arXiv:1707.06347},
  year={2017}
}

@article{shao2024deepseekmath,
  title={Deepseekmath: Pushing the limits of mathematical reasoning in open language models},
  author={Shao, Zhihong and Wang, Peiyi and Zhu, Qihao and Xu, Runxin and Song, Junxiao and Bi, Xiao and Zhang, Haowei and Zhang, Mingchuan and Li, YK and Wu, Yang and others},
  journal={arXiv preprint arXiv:2402.03300},
  year={2024}
}

@article{xu2025align,
  title={Align-GRAG: Reasoning-Guided Dual Alignment for Graph Retrieval-Augmented Generation},
  author={Xu, Derong and Jia, Pengyue and Li, Xiaopeng and Zhang, Yingyi and Wang, Maolin and Liu, Qidong and Zhao, Xiangyu and Wang, Yichao and Guo, Huifeng and Tang, Ruiming and others},
  journal={arXiv preprint arXiv:2505.16237},
  year={2025}
}

@inproceedings{asai2023self,
  title={Self-rag: Learning to retrieve, generate, and critique through self-reflection},
  author={Asai, Akari and Wu, Zeqiu and Wang, Yizhong and Sil, Avirup and Hajishirzi, Hannaneh},
  booktitle={ICLR},
  year={2023}
}

@inproceedings{zhu-etal-2025-knowledge,
    title = "Knowledge Graph-Guided Retrieval Augmented Generation",
    author = "Zhu, Xiangrong  and
      Xie, Yuexiang  and
      Liu, Yi  and
      Li, Yaliang  and
      Hu, Wei",
    editor = "Chiruzzo, Luis  and
      Ritter, Alan  and
      Wang, Lu",
    booktitle = "NAACL",
    month = apr,
    year = "2025",
    address = "Albuquerque, New Mexico",
    publisher = "Association for Computational Linguistics",
    pages = "8912--8924",
    ISBN = "979-8-89176-189-6",
}

@article{guo2024lightrag,
  title={LightRAG: Simple and Fast Retrieval-Augmented Generation},
  author={Guo, Zirui and Xia, Lianghao and Yu, Yanhua and Ao, Tu and Huang, Chao},
  journal={arXiv preprint arXiv:2410.05779},
  year={2024}
}

@article{jiang2025ras,
  title={RAS: Retrieval-And-Structuring for Knowledge-Intensive LLM Generation},
  author={Jiang, Pengcheng and Cao, Lang and Zhu, Ruike and Jiang, Minhao and Zhang, Yunyi and Sun, Jimeng and Han, Jiawei},
  journal={arXiv preprint arXiv:2502.10996},
  year={2025}
}

@article{xu2024large,
  title={Large language models for generative information extraction: A survey},
  author={Xu, Derong and Chen, Wei and Peng, Wenjun and Zhang, Chao and Xu, Tong and Zhao, Xiangyu and Wu, Xian and Zheng, Yefeng and Wang, Yang and Chen, Enhong},
  journal={FCS},
  volume={18},
  number={6},
  pages={186357},
  year={2024},
  publisher={Springer}
}

@article{mackie2023grm,
  title={GRM: generative relevance modeling using relevance-aware sample estimation for document retrieval},
  author={Mackie, Iain and Sekulic, Ivan and Chatterjee, Shubham and Dalton, Jeffrey and Crestani, Fabio},
  journal={arXiv preprint arXiv:2306.09938},
  year={2023}
}

@inproceedings{lyu2024retrieve,
  title={Retrieve-Plan-Generation: An Iterative Planning and Answering Framework for Knowledge-Intensive LLM Generation},
  author={Lyu, Yuanjie and Niu, Zihan and Xie, Zheyong and Zhang, Chao and Xu, Tong and Wang, Yang and Chen, Enhong},
  booktitle={EMNLP},
  pages={4683--4702},
  year={2024}
}

@article{zhang2025process,
  title={Process vs. Outcome Reward: Which is Better for Agentic RAG Reinforcement Learning},
  author={Zhang, Wenlin and Li, Xiangyang and Dong, Kuicai and Wang, Yichao and Jia, Pengyue and Li, Xiaopeng and Zhang, Yingyi and Xu, Derong and Du, Zhaocheng and Guo, Huifeng and others},
  journal={arXiv preprint arXiv:2505.14069},
  year={2025}
}

@inproceedings{zhang2026personalize,
  title={Personalize before retrieve: Llm-based personalized query expansion for user-centric retrieval},
  author={Zhang, Yingyi and Jia, Pengyue and Xu, Derong and Wen, Yi and Li, Xianneng and Wang, Yichao and Zhang, Wenlin and Li, Xiaopeng and Gan, Weinan and Guo, Huifeng and others},
  booktitle={AAAI},
  volume={40},
  number={19},
  pages={16406--16414},
  year={2026}
}

@article{chen2025xiangqi,
  title={Xiangqi-r1: Enhancing spatial strategic reasoning in llms for chinese chess via reinforcement learning},
  author={Chen, Yuhao and Liu, Shuochen and Lyu, Yuanjie and Zhang, Chao and Shi, Jiayao and Xu, Tong},
  journal={arXiv preprint arXiv:2507.12215},
  year={2025}
}

@inproceedings{liu2026look,
  title={Look as You Think: Unifying Reasoning and Visual Evidence Attribution for Verifiable Document RAG via Reinforcement Learning},
  author={Liu, Shuochen and Luo, Pengfei and Zhang, Chao and Chen, Yuhao and Zhang, Haotian and Liu, Qi and Kou, Xin and Xu, Tong and Chen, Enhong},
  booktitle={AAAI},
  volume={40},
  number={38},
  pages={32159--32167},
  year={2026}
}

@article{lyu2025correction,
  title={From correction to mastery: Reinforced distillation of large language model agents},
  author={Lyu, Yuanjie and Wang, Chengyu and Huang, Jun and Xu, Tong},
  journal={arXiv preprint arXiv:2509.14257},
  year={2025}
}

@article{liu2026perma,
  title={Perma: Benchmarking personalized memory agents via event-driven preference and realistic task environments},
  author={Liu, Shuochen and Zhu, Junyi and Shu, Long and Lin, Junda and Chen, Yuhao and Zhang, Haotian and Zhang, Chao and Xu, Derong and Li, Jia and Tang, Bo and others},
  journal={arXiv preprint arXiv:2603.23231},
  year={2026}
}

@article{lyu2025crud,
  title={Crud-rag: A comprehensive chinese benchmark for retrieval-augmented generation of large language models},
  author={Lyu, Yuanjie and Li, Zhiyu and Niu, Simin and Xiong, Feiyu and Tang, Bo and Wang, Wenjin and Wu, Hao and Liu, Huanyong and Xu, Tong and Chen, Enhong},
  journal={TOIS},
  volume={43},
  number={2},
  pages={1--32},
  year={2025},
  publisher={ACM New York, NY}
}

@inproceedings{liang2025kag,
  title={Kag: Boosting llms in professional domains via knowledge augmented generation},
  author={Liang, Lei and Bo, Zhongpu and Gui, Zhengke and Zhu, Zhongshu and Zhong, Ling and Zhao, Peilong and Sun, Mengshu and Zhang, Zhiqiang and Zhou, Jun and Chen, Wenguang and others},
  booktitle={Companion Proceedings of the ACM on Web Conference 2025},
  pages={334--343},
  year={2025}
}

@inproceedings{xu2024multi,
  title={Multi-perspective Improvement of Knowledge Graph Completion with Large Language Models},
  author={Xu, Derong and Zhang, Ziheng and Lin, Zhenxi and Wu, Xian and Zhu, Zhihong and Xu, Tong and Zhao, Xiangyu and Zheng, Yefeng and Chen, Enhong},
  booktitle={Proceedings of the 2024 Joint International Conference on Computational Linguistics, Language Resources and Evaluation (LREC-COLING 2024)},
  pages={11956--11968},
  year={2024}
}

@inproceedings{xu2022relation,
  title={Relation-enhanced negative sampling for multimodal knowledge graph completion},
  author={Xu, Derong and Xu, Tong and Wu, Shiwei and Zhou, Jingbo and Chen, Enhong},
  booktitle={MM},
  pages={3857--3866},
  year={2022}
}

@article{yang2024qwen2,
  title={Qwen2.5 Technical Report},
  author={Yang, An and Yang, Baosong and Zhang, Beichen and Hui, Binyuan and Zheng, Bo and Yu, Bowen and Li, Chengyuan and Liu, Dayiheng and Huang, Fei and Wei, Haoran and others},
  journal={arXiv e-prints},
  pages={arXiv--2412},
  year={2024}
}

@article{hu2022lora,
  title={Lora: Low-rank adaptation of large language models.},
  author={Hu, Edward J and Shen, Yelong and Wallis, Phillip and Allen-Zhu, Zeyuan and Li, Yuanzhi and Wang, Shean and Wang, Lu and Chen, Weizhu and others},
  journal={ICLR},
  volume={1},
  number={2},
  pages={3},
  year={2022}
}

@article{rafailov2023direct,
  title={Direct preference optimization: Your language model is secretly a reward model},
  author={Rafailov, Rafael and Sharma, Archit and Mitchell, Eric and Manning, Christopher D and Ermon, Stefano and Finn, Chelsea},
  journal={NeurIPS},
  volume={36},
  pages={53728--53741},
  year={2023}
}

@article{edge2024local,
  title={From local to global: A graph rag approach to query-focused summarization},
  author={Edge, Darren and Trinh, Ha and Cheng, Newman and Bradley, Joshua and Chao, Alex and Mody, Apurva and Truitt, Steven and Metropolitansky, Dasha and Ness, Robert Osazuwa and Larson, Jonathan},
  journal={arXiv preprint arXiv:2404.16130},
  year={2024}
}

@inproceedings{trivedi2023interleaving,
  title={Interleaving Retrieval with Chain-of-Thought Reasoning for Knowledge-Intensive Multi-Step Questions},
  author={Trivedi, Harsh and Balasubramanian, Niranjan and Khot, Tushar and Sabharwal, Ashish},
  booktitle={ACL},
  year={2023}
}

@article{trivedi2022musique,
  title={♫ MuSiQue: Multihop Questions via Single-hop Question Composition},
  author={Trivedi, Harsh and Balasubramanian, Niranjan and Khot, Tushar and Sabharwal, Ashish},
  journal={TACL},
  volume={10},
  pages={539--554},
  year={2022},
  publisher={MIT Press One Broadway, 12th Floor, Cambridge, Massachusetts 02142, USA~…}
}

@article{song2025r1,
  title={R1-Searcher: Incentivizing the Search Capability in LLMs via Reinforcement Learning},
  author={Song, Huatong and Jiang, Jinhao and Min, Yingqian and Chen, Jie and Chen, Zhipeng and Zhao, Wayne Xin and Fang, Lei and Wen, Ji-Rong},
  journal={arXiv preprint arXiv:2503.05592},
  year={2025}
}

@article{wang2025stepsearch,
  title={StepSearch: Igniting LLMs Search Ability via Step-Wise Proximal Policy Optimization},
  author={Wang, Ziliang and Zheng, Xuhui and An, Kang and Ouyang, Cijun and Cai, Jialu and Wang, Yuhang and Wu, Yichao},
  journal={arXiv preprint arXiv:2505.15107},
  year={2025}
}

@article{sun2025zerosearch,
  title={Zerosearch: Incentivize the search capability of llms without searching},
  author={Sun, Hao and Qiao, Zile and Guo, Jiayan and Fan, Xuanbo and Hou, Yingyan and Jiang, Yong and Xie, Pengjun and Huang, Fei and Zhang, Yan},
  journal={arXiv preprint arXiv:2505.04588},
  year={2025}
}

@article{chen2024not,
  title={Do not think that much for 2+ 3=? on the overthinking of o1-like llms},
  author={Chen, Xingyu and Xu, Jiahao and Liang, Tian and He, Zhiwei and Pang, Jianhui and Yu, Dian and Song, Linfeng and Liu, Qiuzhi and Zhou, Mengfei and Zhang, Zhuosheng and others},
  journal={arXiv preprint arXiv:2412.21187},
  year={2024}
}

@article{he2024g,
  title={G-retriever: Retrieval-augmented generation for textual graph understanding and question answering},
  author={He, Xiaoxin and Tian, Yijun and Sun, Yifei and Chawla, Nitesh and Laurent, Thomas and LeCun, Yann and Bresson, Xavier and Hooi, Bryan},
  journal={NeurIPS},
  volume={37},
  pages={132876--132907},
  year={2024}
}

@article{zeng2024scaling,
  title={Scaling of search and learning: A roadmap to reproduce o1 from reinforcement learning perspective},
  author={Zeng, Zhiyuan and Cheng, Qinyuan and Yin, Zhangyue and Wang, Bo and Li, Shimin and Zhou, Yunhua and Guo, Qipeng and Huang, Xuanjing and Qiu, Xipeng},
  journal={arXiv preprint arXiv:2412.14135},
  year={2024}
}

@article{yu2025dapo,
  title={Dapo: An open-source llm reinforcement learning system at scale},
  author={Yu, Qiying and Zhang, Zheng and Zhu, Ruofei and Yuan, Yufeng and Zuo, Xiaochen and Yue, Yu and Dai, Weinan and Fan, Tiantian and Liu, Gaohong and Liu, Lingjun and others},
  journal={arXiv preprint arXiv:2503.14476},
  year={2025}
}

@article{jaech2024openai,
  title={Openai o1 system card},
  author={Jaech, Aaron and Kalai, Adam and Lerer, Adam and Richardson, Adam and El-Kishky, Ahmed and Low, Aiden and Helyar, Alec and Madry, Aleksander and Beutel, Alex and Carney, Alex and others},
  journal={arXiv preprint arXiv:2412.16720},
  year={2024}
}

@article{ouyang2022training,
  title={Training language models to follow instructions with human feedback},
  author={Ouyang, Long and Wu, Jeffrey and Jiang, Xu and Almeida, Diogo and Wainwright, Carroll and Mishkin, Pamela and Zhang, Chong and Agarwal, Sandhini and Slama, Katarina and Ray, Alex and others},
  journal={NeurIPS},
  volume={35},
  pages={27730--27744},
  year={2022}
}

@article{grattafiori2024llama,
  title={The llama 3 herd of models},
  author={Grattafiori, Aaron and Dubey, Abhimanyu and Jauhri, Abhinav and Pandey, Abhinav and Kadian, Abhishek and Al-Dahle, Ahmad and Letman, Aiesha and Mathur, Akhil and Schelten, Alan and Vaughan, Alex and others},
  journal={arXiv preprint arXiv:2407.21783},
  year={2024}
}

@article{brown2020language,
  title={Language models are few-shot learners},
  author={Brown, Tom and Mann, Benjamin and Ryder, Nick and Subbiah, Melanie and Kaplan, Jared D and Dhariwal, Prafulla and Neelakantan, Arvind and Shyam, Pranav and Sastry, Girish and Askell, Amanda and others},
  journal={NeurIPS},
  volume={33},
  pages={1877--1901},
  year={2020}
}

@article{yu2025graphrag,
  title={GraphRAG-R1: Graph Retrieval-Augmented Generation with Process-Constrained Reinforcement Learning},
  author={Yu, Chuanyue and Zhao, Kuo and Li, Yuhan and Chang, Heng and Feng, Mingjian and Jiang, Xiangzhe and Sun, Yufei and Li, Jia and Zhang, Yuzhi and Li, Jianxin and others},
  journal={arXiv preprint arXiv:2507.23581},
  year={2025}
}

@article{Agent4Ranking,
author = {Li, Xiaopeng and Su, Lixin and Jia, Pengyue and Cheng, Suqi and Wang, Junfeng and Yin, Dawei and Zhao, Xiangyu},
title = {Agent4Ranking: Semantic Robust Ranking via Personalized Query Rewriting Using Multi-agent LLMs},
year = {2025},
publisher = {Association for Computing Machinery},
address = {New York, NY, USA},
issn = {1046-8188},
url = {https://doi.org/10.1145/3749099},
doi = {10.1145/3749099},
note = {Just Accepted},
journal = {TOIS},
month = jul
}

@article{yu2024rankrag,
  title={Rankrag: Unifying context ranking with retrieval-augmented generation in llms},
  author={Yu, Yue and Ping, Wei and Liu, Zihan and Wang, Boxin and You, Jiaxuan and Zhang, Chao and Shoeybi, Mohammad and Catanzaro, Bryan},
  journal={NeurIPS},
  volume={37},
  pages={121156--121184},
  year={2024}
}

@article{webglm,
author = {Lai, Hanyu and Liu, Xiao and Yu, Hao and Xu, Yifan and Iong, Iat Long and Yao, Shuntian and Zeng, Aohan and Du, Zhengxiao and Dong, Yuxiao and Tang, Jie},
title = {WebGLM: Towards an Efficient and Reliable Web-Enhanced Question-Answering System},
year = {2025},
issue_date = {September 2025},
publisher = {Association for Computing Machinery},
address = {New York, NY, USA},
volume = {43},
number = {5},
issn = {1046-8188},
url = {https://doi.org/10.1145/3729421},
doi = {10.1145/3729421},
journal = {ACM Trans. Inf. Syst.},
month = jul,
articleno = {122},
numpages = {43}
}

@article{luo2025graph,
  title={Graph-R1: Towards Agentic GraphRAG Framework via End-to-end Reinforcement Learning},
  author={Luo, Haoran and Chen, Guanting and Lin, Qika and Guo, Yikai and Xu, Fangzhi and Kuang, Zemin and Song, Meina and Wu, Xiaobao and Zhu, Yifan and Tuan, Luu Anh and others},
  journal={arXiv preprint arXiv:2507.21892},
  year={2025}
}

@article{cuadron2025danger,
  title={The danger of overthinking: Examining the reasoning-action dilemma in agentic tasks},
  author={Cuadron, Alejandro and Li, Dacheng and Ma, Wenjie and Wang, Xingyao and Wang, Yichuan and Zhuang, Siyuan and Liu, Shu and Schroeder, Luis Gaspar and Xia, Tian and Mao, Huanzhi and others},
  journal={arXiv preprint arXiv:2502.08235},
  year={2025}
}

@inproceedings{kirkunderstanding,
  title={Understanding the Effects of RLHF on LLM Generalisation and Diversity},
  author={Kirk, Robert and Mediratta, Ishita and Nalmpantis, Christoforos and Luketina, Jelena and Hambro, Eric and Grefenstette, Edward and Raileanu, Roberta},
  booktitle={The Twelfth International Conference on Learning Representations}
}

@article{shi2025search,
  title={Search and Refine During Think: Autonomous Retrieval-Augmented Reasoning of LLMs},
  author={Shi, Yaorui and Li, Sihang and Wu, Chang and Liu, Zhiyuan and Fang, Junfeng and Cai, Hengxing and Zhang, An and Wang, Xiang},
  journal={arXiv preprint arXiv:2505.11277},
  year={2025}
}

@article{jin2025empirical,
  title={An Empirical Study on Reinforcement Learning for Reasoning-Search Interleaved LLM Agents},
  author={Jin, Bowen and Yoon, Jinsung and Kargupta, Priyanka and Arik, Sercan O and Han, Jiawei},
  journal={arXiv preprint arXiv:2505.15117},
  year={2025}
}

@article{song2025r1plus,
  title={R1-Searcher++: Incentivizing the Dynamic Knowledge Acquisition of LLMs via Reinforcement Learning},
  author={Song, Huatong and Jiang, Jinhao and Tian, Wenqing and Chen, Zhipeng and Wu, Yuhuan and Zhao, Jiahao and Min, Yingqian and Zhao, Wayne Xin and Fang, Lei and Wen, Ji-Rong},
  journal={arXiv preprint arXiv:2505.17005},
  year={2025}
}

@article{hu2025reinforce++,
  title={Reinforce++: An efficient rlhf algorithm with robustness to both prompt and reward models},
  author={Hu, Jian and Liu, Jason Klein and Xu, Haotian and Shen, Wei},
  journal={arXiv preprint arXiv:2501.03262},
  year={2025}
}

@article{wang2025acting,
  title={Acting Less is Reasoning More! Teaching Model to Act Efficiently},
  author={Wang, Hongru and Qian, Cheng and Zhong, Wanjun and Chen, Xiusi and Qiu, Jiahao and Huang, Shijue and Jin, Bowen and Wang, Mengdi and Wong, Kam-Fai and Ji, Heng},
  journal={arXiv preprint arXiv:2504.14870},
  year={2025}
}

@article{jin2025search,
  title={Search-r1: Training llms to reason and leverage search engines with reinforcement learning},
  author={Jin, Bowen and Zeng, Hansi and Yue, Zhenrui and Wang, Dong and Zamani, Hamed and Han, Jiawei},
  journal={arXiv preprint arXiv:2503.09516},
  year={2025}
}

@article{yuan2023scaling,
  title={Scaling Relationship on Learning Mathematical Reasoning with Large Language Models},
  author={Yuan, Zheng and Yuan, Hongyi and Li, Chengpeng and Dong, Guanting and Lu, Keming and Tan, Chuanqi and Zhou, Chang and Zhou, Jingren},
  journal={arXiv e-prints},
  pages={arXiv--2308},
  year={2023}
}

@article{xiong2025minimalist,
  title={A minimalist approach to llm reasoning: from rejection sampling to reinforce},
  author={Xiong, Wei and Yao, Jiarui and Xu, Yuhui and Pang, Bo and Wang, Lei and Sahoo, Doyen and Li, Junnan and Jiang, Nan and Zhang, Tong and Xiong, Caiming and others},
  journal={arXiv preprint arXiv:2504.11343},
  year={2025}
}

@inproceedings{khaki2024rs,
  title={RS-DPO: A Hybrid Rejection Sampling and Direct Preference Optimization Method for Alignment of Large Language Models},
  author={Khaki, Saeed and Li, JinJin and Ma, Lan and Yang, Liu and Ramachandra, Prathap},
  booktitle={Findings of NAACL},
  pages={1665--1680},
  year={2024}
}

@article{ivison2024unpacking,
  title={Unpacking dpo and ppo: Disentangling best practices for learning from preference feedback},
  author={Ivison, Hamish and Wang, Yizhong and Liu, Jiacheng and Wu, Zeqiu and Pyatkin, Valentina and Lambert, Nathan and Smith, Noah A and Choi, Yejin and Hajishirzi, Hanna},
  journal={NeurIPS},
  volume={37},
  pages={36602--36633},
  year={2024}
}

@article{tu2025enhancing,
  title={Enhancing LLM Reasoning with Iterative DPO: A Comprehensive Empirical Investigation},
  author={Tu, Songjun and Lin, Jiahao and Tian, Xiangyu and Zhang, Qichao and Li, Linjing and Fu, Yuqian and Xu, Nan and He, Wei and Lan, Xiangyuan and Jiang, Dongmei and others},
  journal={arXiv preprint arXiv:2503.12854},
  year={2025}
}

@article{pang2024iterative,
  title={Iterative reasoning preference optimization},
  author={Pang, Richard Yuanzhe and Yuan, Weizhe and He, He and Cho, Kyunghyun and Sukhbaatar, Sainbayar and Weston, Jason},
  journal={NeurIPS},
  volume={37},
  pages={116617--116637},
  year={2024}
}

@article{fang2025kirag,
  title={KiRAG: Knowledge-Driven Iterative Retriever for Enhancing Retrieval-Augmented Generation},
  author={Fang, Jinyuan and Meng, Zaiqiao and Macdonald, Craig},
  journal={arXiv preprint arXiv:2502.18397},
  year={2025}
}

@article{gao2025beyond,
  title={Beyond Ten Turns: Unlocking Long-Horizon Agentic Search with Large-Scale Asynchronous RL},
  author={Gao, Jiaxuan and Fu, Wei and Xie, Minyang and Xu, Shusheng and He, Chuyi and Mei, Zhiyu and Zhu, Banghua and Wu, Yi},
  journal={arXiv preprint arXiv:2508.07976},
  year={2025}
}

@article{pal2024smaug,
  title={Smaug: Fixing failure modes of preference optimisation with dpo-positive},
  author={Pal, Arka and Karkhanis, Deep and Dooley, Samuel and Roberts, Manley and Naidu, Siddartha and White, Colin},
  journal={arXiv preprint arXiv:2402.13228},
  year={2024}
}

@inproceedings{sunthink,
  title={Think-on-Graph: Deep and Responsible Reasoning of Large Language Model on Knowledge Graph},
  author={Sun, Jiashuo and Xu, Chengjin and Tang, Lumingyuan and Wang, Saizhuo and Lin, Chen and Gong, Yeyun and Ni, Lionel and Shum, Heung-Yeung and Guo, Jian},
  booktitle={ICLR}
}

@inproceedings{yao2023react,
  title={React: Synergizing reasoning and acting in language models},
  author={Yao, Shunyu and Zhao, Jeffrey and Yu, Dian and Du, Nan and Shafran, Izhak and Narasimhan, Karthik and Cao, Yuan},
  booktitle={ICLR},
  year={2023}
}

@article{wang2024survey,
  title={A survey on large language model based autonomous agents},
  author={Wang, Lei and Ma, Chen and Feng, Xueyang and Zhang, Zeyu and Yang, Hao and Zhang, Jingsen and Chen, Zhiyuan and Tang, Jiakai and Chen, Xu and Lin, Yankai and others},
  journal={FCS},
  volume={18},
  number={6},
  pages={186345},
  year={2024},
  publisher={Springer}
}

@inproceedings{gutierrez2024hipporag,
  title={Hipporag: Neurobiologically inspired long-term memory for large language models},
  author={Guti{\'e}rrez, Bernal Jim{\'e}nez and Shu, Yiheng and Gu, Yu and Yasunaga, Michihiro and Su, Yu},
  booktitle={NeurIPS},
  year={2024}
}

@article{kwiatkowski2019natural,
  title={Natural questions: a benchmark for question answering research},
  author={Kwiatkowski, Tom and Palomaki, Jennimaria and Redfield, Olivia and Collins, Michael and Parikh, Ankur and Alberti, Chris and Epstein, Danielle and Polosukhin, Illia and Devlin, Jacob and Lee, Kenton and others},
  journal={TACL},
  volume={7},
  pages={453--466},
  year={2019},
  publisher={MIT Press One Rogers Street, Cambridge, MA 02142-1209, USA journals-info~…}
}

@inproceedings{mallen2023not,
  title={When Not to Trust Language Models: Investigating Effectiveness of Parametric and Non-Parametric Memories},
  author={Mallen, Alex and Asai, Akari and Zhong, Victor and Das, Rajarshi and Khashabi, Daniel and Hajishirzi, Hannaneh},
  booktitle={ACL},
  pages={9802--9822},
  year={2023}
}

@inproceedings{yang2018hotpotqa,
  title={HotpotQA: A Dataset for Diverse, Explainable Multi-hop Question Answering},
  author={Yang, Zhilin and Qi, Peng and Zhang, Saizheng and Bengio, Yoshua and Cohen, William and Salakhutdinov, Ruslan and Manning, Christopher D},
  booktitle={EMNLP},
  pages={2369--2380},
  year={2018}
}

@inproceedings{ho2020constructing,
  title={Constructing A Multi-hop QA Dataset for Comprehensive Evaluation of Reasoning Steps},
  author={Ho, Xanh and Nguyen, Anh-Khoa Duong and Sugawara, Saku and Aizawa, Akiko},
  booktitle={COLING},
  pages={6609--6625},
  year={2020}
}

@inproceedings{yueinference,
  title={Inference Scaling for Long-Context Retrieval Augmented Generation},
  author={Yue, Zhenrui and Zhuang, Honglei and Bai, Aijun and Hui, Kai and Jagerman, Rolf and Zeng, Hansi and Qin, Zhen and Wang, Dong and Wang, Xuanhui and Bendersky, Michael},
  booktitle={ICLR}
}

@article{li2025search,
  title={Search-o1: Agentic search-enhanced large reasoning models},
  author={Li, Xiaoxi and Dong, Guanting and Jin, Jiajie and Zhang, Yuyao and Zhou, Yujia and Zhu, Yutao and Zhang, Peitian and Dou, Zhicheng},
  journal={arXiv preprint arXiv:2501.05366},
  year={2025}
}

@inproceedings{karpukhin2020dense,
  title={Dense Passage Retrieval for Open-Domain Question Answering.},
  author={Karpukhin, Vladimir and Oguz, Barlas and Min, Sewon and Lewis, Patrick SH and Wu, Ledell and Edunov, Sergey and Chen, Danqi and Yih, Wen-tau},
  booktitle={EMNLP},
  pages={6769--6781},
  year={2020}
}

@misc{llmrl2025incorrect,
title={Incorrect Baseline Evaluations Call into Question Recent LLM-RL Claims},
author={Nikhil Chandak and Shashwat Goel and Ameya Prabhu},
year={2025},
howpublished={\url{https://safe-lip-9a8.notion.site/Incorrect-Baseline-Evaluations-Call-into-Question-Recent-LLM-RL-Claims-2012f1fbf0ee8094ab8ded1953c15a37?pvs=4}},
note={Notion Blog}
}

@article{xu2023recomp,
  title={Recomp: Improving retrieval-augmented lms with compression and selective augmentation},
  author={Xu, Fangyuan and Shi, Weijia and Choi, Eunsol},
  booktitle={ICLR},
  year={2024}
}

@article{alonso2024mixture,
  title={Mixture-of-PageRanks: Replacing Long-Context with Real-Time, Sparse GraphRAG},
  author={Alonso, Nicholas and Millidge, Beren},
  journal={arXiv preprint arXiv:2412.06078},
  year={2024}
}


\end{document}